\documentclass[12pt]{scrartcl}
\usepackage[utf8]{inputenc}
\usepackage[T1]{fontenc}
\usepackage{lmodern}
\usepackage{caption}
\usepackage[english]{babel}
\usepackage[authoryear]{natbib}
\usepackage{abstract}
\usepackage{amsmath, amssymb, amsfonts, amsthm}
\usepackage{graphicx}
\usepackage{siunitx}
\usepackage{subcaption}
\usepackage{xr-hyper}
\usepackage[toc,page]{appendix}
\usepackage[hidelinks]{hyperref}
\usepackage{bm}
\usepackage{bbm}
\usepackage{colortbl}
\usepackage{longtable}
\usepackage{threeparttable}
\usepackage{rotating}
\usepackage[normalem]{ulem}
\usepackage[a4paper, left=1in, right=1in, top=1in, bottom=1in]{geometry}
\usepackage{setspace}
\usepackage{flafter}
\usepackage{placeins}
\usepackage{morefloats}
\usepackage{array}
\usepackage{tabularx,booktabs}\newcolumntype{b}{X}
\newcolumntype{r}{>{\hsize=.2\hsize}X}
\newcolumntype{s}{>{\hsize=.1\hsize}X}
\makeatletter
\newcommand\notsotiny{\@setfontsize\notsotiny{7.5pt}{8pt}}
\makeatother
\usepackage{authblk}
\usepackage{mathtools}
\usepackage{bm}
\usepackage{lscape}
\usepackage{enumitem}
\usepackage{multirow}
\usepackage{xcolor,colortbl}

\definecolor{Gray}{gray}{0.85}
\definecolor{LightCyan}{rgb}{0.88,1,1}
\newcolumntype{a}{>{\columncolor{Gray}}c}
\newcolumntype{b}{>{\columncolor{white}}c}
\newtheorem{assumption}{Assumption}
\newtheorem{theorem}{Theorem}
\newtheorem{property}{Property}
\newtheorem{definition}{Definition}
\newtheorem{proposition}{Proposition}
\newtheorem{algorithm}{Algorithm}

\newtheorem{corollary}{Corollary}
\newtheorem{lemma}{Lemma}

\def\spacingset#1{\renewcommand{\baselinestretch}%
{#1}\small\normalsize} \spacingset{1}

\title{Detecting Grouped Local Average Treatment Effects and Selecting True Instruments\\ With an Application to Estimating the Effect of Prison on Recidivism}

\author[1]{Nicolas Apfel}
\author[2]{Helmut Farbmacher}
\author[2]{Rebecca Groh}
\author[3]{Martin Huber}
\author[4]{Henrika Langen}
\affil[1]{University of Innsbruck, Austria}
\affil[2]{TU Munich, Germany}
\affil[3]{University of Fribourg, Switzerland}
\affil[4]{IHS, Vienna, Austria\thanks{Corresponding address: nicolas.apfel@uibk.ac.at. Nicolas Apfel gratefully acknowledges funding through ESRC grant EST013567/1.}}
\newcommand\cites[1]{\citeauthor{#1}'s\ (\citeyear{#1})}

\begin{document}
	
	\maketitle
%\spacingset{2} % DON'T change the spacing!
\doublespacing
\pagenumbering{gobble}
\newpage
\begin{abstract} \noindent
Considering an endogenous treatment with heterogeneous effects and multiple instruments, we propose a method to identify instruments entailing identical local average treatment effects while simultaneously eliminating instruments that violate identifying assumptions. Our two-step procedure involves clustering instruments with identical treatment propensities (first step) and identical reduced forms given the propensity (second step). This permits detecting instruments that satisfy the identifying assumptions under the condition that they form the largest group among all instruments with the same propensity. We demonstrate consistency of our method and apply it to estimate the impact of imprisonment on recidivism, using judge assignments as instruments.

\end{abstract}

\pagenumbering{arabic}	
%\newpage
\section{Introduction}

\iffalse
	IV general introduction with assumptions – Question (P,R)
	What we do in a nutshell (A)
	More details on what we do (A)
	Where is it relevant? And how do we contribute to that (P)
	Literature we contribute to
	Remainder of paper
\fi

\noindent Instrumental variables (IV) analysis is a well-established method for estimating the effect of a treatment on an outcome in the presence of unobserved confounding. In models with a single endogenous treatment and multiple IVs, different instruments can identify heterogeneous local average treatment effects (LATE), provided they satisfy the standard LATE assumptions: relevance, exogeneity, exclusion, and monotonicity \citep[see, e.g.,][]{Imbens+94, Angrist+96}. 
Recent advances in parametric statistics have focused on distinguishing valid instruments from those that violate exclusion. However, these approaches often overlook the case of heterogeneous treatment effects.

\noindent In this paper, we propose a method to detect instrumental variables (IVs) that satisfy the LATE assumptions in heterogeneous treatment effect models, even when the set of candidate instruments includes invalid IVs that violate these assumptions. The method consists of two steps.

\noindent In the first step, we apply Agglomerative Hierarchical Clustering \citep[AHC,][]{Ward1963Hierarchical} to assign IVs into clusters - referred to as \textit{clubs} - based on identical conditional treatment probabilities, henceforth called \textit{propensity scores}. These clubs are used to generate homogeneous complier groups when pairing IVs. The intuition behind this approach is that if two distinct pairs of IVs exhibit the same higher and lower propensity scores, then the complier groups associated with these pairs must share identical unobserved characteristics, provided the LATE assumptions hold. Thus, our approach enables clustering IVs into club-pairs with homogeneous LATEs under valid IV restrictions. A critical assumption in this step is that the propensity scores defining the first-stage effect of the instrument on treatment are clustered in the population. 
In the second step, we distinguish valid IVs from invalid IVs that violate one or more LATE assumptions. This is achieved using the plurality assumption, which posits that the largest groups of IV values within clubs satisfy the LATE assumptions and share the same LATEs within group pairs. In this framework, a \textit{group} refers to a set of IVs with identical reduced form estimands, defined as the expected outcomes conditional on the IVs. Valid IVs are identified by hierarchically clustering the reduced form estimates within clubs (defined by first-stage propensity scores) and selecting the cluster with the relative majority of IVs.  %The approaches suited for the selection of valid IVs vary with the asymptotic setting that the researcher has in mind. %COMMENT: Please say more here about the settings in a couple of sentence?

%Our two-step procedure broadly fits into the fast evolving literature on the integration of machine learning methods in causal inference, by applying unsupervised machine learning for clustering instrumental variables based on their first stage estimates and selecting IVs that satisfy the identifying assumptions, respectively. 
\noindent We show that, under certain conditions, the first-step selection procedure consistently classifies propensity scores into clubs, provided such clubs exist. Similarly, the second-step selection procedure consistently identifies IVs that satisfy the LATE assumptions, assuming the plurality assumption holds. Additionally, we evaluate the finite sample performance of our method through a simulation study. The results indicate that the method performs well in both first- and second-step clustering, even with a relatively moderate number of observations per instrument, and when a subset of instruments violates the exclusion restriction.

%%%

\noindent Our study contributes to the growing literature on detecting and selecting valid and invalid IVs \citep{Kang2016Instrumental, Windmeijer2019Use, Guo2018Confidence, Windmeijer2021Confidence, Apfel2021Agglomerative, Fan2024Endogenous} that works under the assumption that at least a subset of instruments is valid. This literature builds on the foundational work of \citet{Andrews1999Consistent}. A key limitation of existing IV selection methods is their reliance on the assumption of homogeneous treatment effects, which prevents them from distinguishing violations of identifying assumptions from effect heterogeneity. 

\noindent Our approach shares similarities with \citet{Apfel2021Agglomerative}, who also use Agglomerative Hierarchical Clustering (AHC) to identify valid IVs under homogeneous treatment effects by clustering IV-specific estimates of the treatment effect. Notably, they highlight that AHC can detect heterogeneity arising from LATEs when all IVs are valid. However, their method - and others like it - cannot address cases involving validity violations and treatment effect heterogeneity, a gap our paper aims to fill. Moreover, existing approaches generally do not accommodate simultaneous exclusion restriction violations and IV exogeneity violations \citep{Apfel2022Falsification}. They also fail to account for violations of treatment monotonicity in the instrument, as discussed in \citet{Imbens+94}. We are the first in the IV selection literature to address both violations of IV assumptions and the heterogeneity of treatment effects. 

\noindent Another unresolved question in the IV selection literature is how to effectively preselect relevant IVs. Our approach naturally finds IVs with a non-zero first stage by pairing values with differing propensity scores. A further difference relative to this literature is one of scope: while \citet{Apfel2021Agglomerative} and related papers can be applied to general multiple-regressor settings, our framework considers a single discrete (or discretizable) instrument expanded into dummies. \\
\noindent A paper that departs from the homogeneous treatment effects assumption is \citet[SW]{Sun2025Pairwise}, who propose a technique that exploits testable conditions for instrument validity to remove invalid variation in the instruments. This approach differs from ours in three main respects. \textit{First}, the testable conditions in SW (drawing on \citealt{Kitagawa2015Test}, \citealt{Mourifie2017Testing}, and \citealt{Sun2023Instrument}) are necessary but not sufficient, so their procedure may retain invalid IVs. By contrast, we establish consistent selection of the validity set under the plurality assumption. \textit{Second}, SW maintain the first-stage assumption for all pairs of IV values, which is a strong requirement when the number of judges is large. In our framework, the existence of non-singleton clubs corresponds to a violation of this first-stage assumption for at least one pair, and the larger the share of pairs with equal propensity scores, the fewer club-pairs there are. In the limiting case where all judges have distinct propensity scores, our method reduces to a setting with singleton clubs, where the paired-comparison logic of SW is the natural alternative. \textit{Third}, our approach mitigates the high-dimensionality problem that arises when comparing a large number of IV values by leveraging clustering to reduce dimensionality and estimating club-pair-specific LATEs. We view the two approaches as complementary: our method is better-suited to settings with few detectable clubs of judges with similar stringency, while SW's pairwise approach is more natural when all judges have distinct propensity scores, e.g. in settings when there are many clubs, each with few judges. We provide a more detailed discussion of the links and differences between the approaches in Appendix \ref{app:lit}.

\noindent As an empirical contribution, we apply our method to the well-studied question of whether incarceration reduces offenders' likelihood of reoffending. This question is highly policy-relevant, given that incarceration rates in the United States reach 1 percent of the population annually. Following the seminal work of \citet{Kling2006}, a common estimation strategy in this literature is to use the random assignment of judges to cases as instrumental variables. These IVs are assumed to influence incarceration decisions but not directly affect the outcome. 
While many studies report zero or even crime-inducing effects of incarceration, these effects may vary across geographic and institutional contexts (e.g., prison conditions). For instance, \citet{Bhuller2020Incarceration} find that imprisonment reduces crime in Norway. Additionally, effects may vary within a single country, potentially even by court. Each IV, defined by a pair of judges differing in stringency, enables the estimation of a potentially distinct LATE. %The LATE corresponds to the average effect of incarceration on recidivism among those offenders who comply with the IV in the sense that they are incarcerated under the more stringent judge, but not under the less stringent judge when considering a specific pair of judges as IV. 
Since these LATEs correspond to different complier groups, which typically differ in unobservables, a conventional two-stage least squares (2SLS) approach using all IVs simultaneously may obscure important heterogeneity in the effects of incarceration.

\noindent Another concern with judge-based IVs is that they may fail to satisfy the LATE assumptions. For instance, they might directly affect the outcome, thereby violating the exclusion restriction, or they may be associated with unobserved characteristics that influence the outcome, violating IV exogeneity. Additionally, \citet{sigstad2023monotonicity} provides empirical evidence suggesting that a significant proportion of judge IVs may violate monotonicity. This can occur, for example, if a generally stringent judge (i.e., has a high treatment propensity score) is more lenient toward certain types of offenses. 
Our method addresses these challenges by allowing for violations of the LATE assumptions across multiple judges while simultaneously investigating effect heterogeneity.  %For example, judges could violate the exclusion restriction by varying in terms of the use of alternative measures to incarceration, such as electronic monitoring \citep{Loeffler2021Impact}. Furthermore, the monotonicity assumption may be violated, which imposes that more stringent judges are stricter across all types of offenses and offenders than less stringent judges such that any offender incarcerated under a less stringent judge would also be incarcerated under a more stringent judge.  

\noindent We analyze a newly collected dataset on minor crimes in Minnesota from 2009 to 2017. To benchmark our results against conventional methods, we first apply ordinary least squares (OLS), which indicates crime-inducing effects of incarceration. Similarly, linear IV regression using 2SLS with all judge IVs simultaneously also suggests crime-inducing effects. In contrast, our first-step clustering method for judge-based IVs reveals significant variation in LATE estimates across club-pairs. Estimates range from negative to positive and high to low but are mostly statistically insignificant. These results highlight substantial heterogeneity in the effects of imprisonment on recidivism across judicial stringency levels - this heterogeneity is obscured by 2SLS.%Within a certain club-pair (which consists of several judges), constructing LATE estimators based on multiple judge-specific IVs, taking one club as basis, permits testing the overidentifying restriction that all LATE estimates are equal. Running joint tests on an overidentifying set of IVs within a club-pair mostly rejects the overall validity of IVs. %, which is corroborated by the fact that we find substantial heterogeneity in LATE estimates within club-pairs. 
%For this reason, we also apply our second-step procedure for selecting those instruments that satisfy the LATE assumption under the plurality rule. After applying the second-step selection, the overidentification tests no longer reject the LATE assumptions for the remaining IVs. %In a next step %COMMENT: add a discussion on whether we also test for the validity of IVs in the clusters and what the results are.

\noindent The remainder of this paper is structured as follows. Section \ref{sec:ModelAndAssumptions} introduces the treatment effects model, the LATE framework, and the identifying assumptions. We also show that under these assumptions, the complier populations implied by two pairs of judges with identical propensity scores share the same LATE. In Section \ref{sec:Method}, we present our two-step procedure: (i) detecting club-pairs of IVs (based on judges) with equal lower and higher propensity scores and (ii) selecting IVs that satisfy the LATE assumptions under the plurality condition, which states that the largest group of IVs satisfies these assumptions. Section \ref{sec:consistency} discusses consistency and asymptotic normality of our estimation procedure. 
Section \ref{sec:Application} applies our method to evaluate the effect of incarceration on recidivism in Minnesota. Finally, Section \ref{sec:Conclusion} concludes.
% COMMENT: the section sec:Motivation should be integrated in the empirical application and no longer be a section on its own. Some arguments of the section may go into the section sec:ModelAndAssumptions to give an illustration of the model considered. So we don't need the following sentence any more: In Section \ref{sec:Motivation}, we introduce the empirical problem. We want to estimate the effect of incarceration on offender recidivism.

%\input{sections/motivation.tex}
\section{Model and Assumptions}\label{sec:ModelAndAssumptions}
 
\subsection{Notation}\label{notation}

\noindent For modeling the causal effects of interest and discussing the assumptions required for their identification, we use the potential outcome framework, see, for instance, \citet{Neyman23} and \citet{Rubin74}. To this end, we denote by $D$ a binary treatment, e.g., \ incarceration, which can take values $d \in \{0,1\}$ and a discrete or continuous outcome $Y$, in our empirical problem, a binary indicator on recidivism.  %In general, we will refer to random variables by capital letters and specific values thereof by lower case letters. We are interested in the causal effect of the treatment on a discrete or continuous outcome $Y$, in our empirical problem on a binary indicator on recidivism. Furthermore, w
We denote by $Z$ an instrumental variable (IV), which needs to satisfy certain conditions outlined below, which is presumably multivalued and discrete. That is, $Z$ may take values $z \in \{1, ..., J\}$, with $J$ corresponding to the total number of values of the instrument. Let $\mathcal{J} = \{1, ..., J\}$ be the full set of possible IV values. $n_z$ denotes the number of observations with a specific value $z$ and $\sum_{z} n_z = n$ is the overall number of observations in the sample. In our empirical application, the instrument values $z$ indicate the assignment to alternative judges who differ in incarceration rates, i.e., the likelihood of issuing prison sentences. Furthermore, $n_z$ stands for the number of cases per judge. $D(z)$ corresponds to the potential treatment state when setting the IV to a hypothetical value $Z=z$ in support of $Z$. In contrast, $Y(d,z)$ is the potential outcome under  treatment value $d$ (0 or 1) and IV value $z$.  Denote $\mid \cdot \mid$ as the cardinality of a set. Let $\mathbbm{1}(\cdot)$ denote the indicator function, which is equal to one if its argument is satisfied and zero otherwise. Bold letters are reserved for matrices and vectors, while uppercase letters not in bold are random variables. We denote matrices in upper case and bold, $\mathbf{X}$, and vectors in lower case and bold, $\mathbf{x}$. In particular, $\mathbf{Z}$ is the matrix of dummies for observations where the instrument takes value $z$. %Denote the sign function of a difference of two scalar values $a$ and $b$ as $\sigma(a,b)=\operatorname{sgn}(b-a)= \frac{b-a}{|b-a|}$. %Calligraphic letters such as $\mathcal{A}$ denote a union of sets of values. 

\subsection{LATE assumptions}\label{lateassumptions}

\noindent Following \citet{Angrist+96}, individuals can be classified into compliance types as a function of how someone's treatment state depends on or reacts to switching the value of the instrument. 
In the following, we assume that the values of the instrument \( Z \) can be ordered in a meaningful way, such that \( z \geq z' \) implies, for example, a weakly higher judge stringency of judge \( z \) relative to judge \( z' \). This ordering allows us to formally state the IV assumptions, which are assumed to hold globally, i.e., for any possible IV pair \( z \) and \( z' \). 
%Let us, to this end, consider two distinct IV values, $z$ and $z'$, which can presumably be ordered in a sensible way, e.g., \ two judges with a higher and lower incarceration rate, respectively.
Individuals satisfying $(D(z) = 1, D(z') = 0)$ are compliers in the sense that they are treated, i.e., incarcerated under the higher value of the IV ($Z = z$), when assigned to a more stringent judge, while remaining non-treated under the lower value ($Z = z'$), when assigned to a more lenient judge.  %Never takers neither receive the treatment under a higher, nor under a lower IV value, thus satisfying $(D(z)=D(z')=0)$, while always takers are treated under either IV value, i.e.\ $(D(z)=D(z')=1)$. Finally, defiers counteract the instrument assignment by being treated under the lower IV value implying a more lenient judge, while remaining untreated under the higher IV value implying a stricter judge, such that $(D(z)=0,D(z')=1)$ holds. 
It is important to note that different choices of values $z$ and $z'$ generally entail different proportions of complier types, see, e.g.,\ the discussion in \citet{Froe02a}. For instance, the share of compliers is likely larger when considering two judges that greatly differ in terms of their stringency rather than two judges, one of which is slightly more stringent than the other.

\noindent \citet{Imbens+94} discuss assumptions under which an IV-based evaluation approach nonparametrically identifies a LATE on the compliers under heterogeneous treatment effects. For two values $z>z'$, this LATE is formally defined as
\begin{equation}\label{def:LATE}
	\Delta_{z,z'}=E[Y(1)-Y(0)|D(z)=1,D(z')=0].
\end{equation}
%In terms of assumptions, the instrument must be as good as randomly assigned, which rules out statistical associations with background characteristics affecting the outcome, and must not have a direct effect on the outcome other than through the treatment. Furthermore, defiers must not exist, while compliers must exist in the population. 
Consequently, we can recover the LATE for a specific pair of instrument values \( z \) and \( z' \).

%In our empirical application, quasi-random assignment appears only plausible when controlling for observed pre-treatment covariates, henceforth denoted by $X$. For this reason, we follow \citet{Abadie00} and \citet{Froe02a} and invoke IV assumptions that hold conditional on the covariates.

\begin{assumption}\label{ass:Validity}Validity
	$$\textrm{(i) $Z \perp (D(z),Y(z,d), D(z'), Y(z',d))$ and (ii) $Y(z,d) = Y(z',d)=Y(d)$}.$$
\end{assumption}

\begin{assumption}\label{ass:Monotonicity}Monotonicity
	$$\Pr(D(z) \geq D(z')) = 1.$$
\end{assumption}

\begin{assumption}\label{ass:FirstStage}First Stage
	$$E(D|Z=z) - E(D|Z=z') \neq 0.$$
\end{assumption}%

\noindent Under Assumptions \ref{ass:Validity} to \ref{ass:FirstStage}, the LATE on compliers who are responsive to IV shifts from $z'$ to $z$ corresponds to a so-called \citet{Wald40}-estimand. The latter consists of the (reduced form) average effect of the instrument shift on the outcome scaled by the (first stage) effect of the instrument shift on the treatment:
\begin{eqnarray}\label{Wald}
	\frac{E(Y| Z=z)-E(Y| Z=z')}{\Pr(D=1| Z=z)-\Pr(D=1| Z=z')}
\end{eqnarray}
\noindent For the case of a discretely distributed IV with mass points at $z$ and $z'$, \eqref{Wald} is equivalent to the probability limit of a 2SLS in which only observations with $Z \in \{z,z'\}$ are considered. It is also worth noting that $\Pr(D=1| Z=z)-\Pr(D=1| Z=z')$ identifies the complier share $\Pr(D(z)=1,D(z')=0)$.

%\citet{Froe02a} also argues that instead of directly controlling for $Z,X$, one may alternatively use the conditional treatment probability given $Z$ and $X$ as instrument to obtain the conditional LATE. This is feasible because this conditional treatment probability, also known as propensity score, fully captures the association of $D$ and $Z$. Denoting the propensity score by $p(Z,X)=\Pr(D=1|Z,X)$, this implies that the conditional LATE corresponds to
%\begin{eqnarray}\label{identX2}
%\Delta_{z,z'}(x)=\frac{E[Y| p(Z,X)=p(z',X),X]-E[Y| p(Z,X)=p(z,X),X]}{E[D| p(Z,X)=p(z',X),X]-E[D| p(Z,X)=p(z,X),X]}.
%\end{eqnarray}

\noindent An important fact relevant to our proposed method is that \citet{Vy02} proves that Assumption \ref{ass:Monotonicity}, when imposed globally across all possible values of the ordered instrument, is equivalent to assuming a so-called threshold-crossing model for the treatment. In this model, the treatment is an additively separable function of the instrument and an unobserved term. It takes the value one whenever the function on the instrument is larger than or equal to the unobserved term. Formally,\footnote{We have abstracted from covariates for simplicity in the main text and discuss them in Appendix \ref{app:covariates}.}
\begin{equation}\label{eq:vyt_equivalence}
	D=\mathbbm{1}(\psi(Z)\geq V),
\end{equation}
where $V$ is a scalar (index of) unobservable(s), and $\psi(Z)$ is a nonparametric function of $Z$. This allows us to characterize the compliers (and other compliance types) in terms of the distribution of $V$. Because $D(z)=1$ implies that $\psi(z)\geq V$ in \eqref{eq:vyt_equivalence} and $D(z')=0$ implies that $\psi(z')< V$, it is easy to see that the distribution of $V$ among compliers satisfies $\psi(z)\geq V > \psi(z')$. For this reason, it follows that
\begin{equation}\label{LATE}
	\Delta_{z,z'}=E[Y(1)-Y(0)|v\geq  V > v'],
\end{equation}
for values $v=\psi(z)$ and $v'=\psi(z')$ for the unobservable $V$.

\noindent Let us now assume that there exists another pair of instrument values $z^*,z''$, in our case, two further judges, which also satisfies Assumptions \ref{ass:Validity} to \ref{ass:FirstStage} and generates the same compliance behavior as the previous pair $z,z'$. Formally,  $v=\psi(z^*)=\psi(z)$ and $v'=\psi(z')=\psi(z'')$. It follows that
\begin{equation}\label{eq:SameLATE}
	\Delta_{z^*,z''}=E[Y(1)-Y(0)|v \geq V > v']=\Delta_{z,z'}
\end{equation}
As both pairs of instruments refer to the very same complier group regarding the distribution of unobservables $V$, the LATE identified by these pairs is identical. This in turn implies that the Wald estimand $\frac{E(Y| Z=z^*)-E(Y| Z=z'')}{\Pr(D=1| Z=z^*)-\Pr(D=1| Z=z'')}$  is equivalent to that in \eqref{Wald} and that both denominators $\Pr(D=1| Z=z)-\Pr(D=1| Z=z')$ and $\Pr(D=1| Z=z^*)-\Pr(D=1| Z=z'')$ identify the share of compliers satisfying $\Pr(v \geq V  > v')$.

\subsection{Grouped propensity scores}

To simplify notation, we denote the treatment propensity score as a function of the instrument by $p_z=\Pr(D=1 | Z=z) = E(D | Z=z)$, where the second equality holds because $D$ is binary. In the next step, we assume that propensity scores follow a specific cluster structure, such that the propensity scores within a cluster are homogeneous, even across different instrument values (e.g.,\ judges). %This assumption is crucial in our context.
To this end, we introduce some further notation. Let $C_k$ denote some cluster $k$ of propensity scores $p_z$ and $p_k^0$ the constant propensity score of all instruments in this cluster. $K^0$ is the true number of clusters of IVs with the same propensity score, and without the zero superscript, $K$ is a number of clusters, but not necessarily the correct one. Formally, we make the following cluster assumption, in analogy to \citet{Su2016Identifying} (who look at panel data models more generally):
\begin{assumption}{Existence of Clubs}\label{ass:Clubs}
$$p_z = \sum_{k=1}^{K^0} p_k^0  \mathbbm{1}(z \in C_k).$$
We have cluster $C_k$ such that $p_k^0 \neq p_{\ell}^0$ for any $k \neq \ell$. The clusters do not overlap, and the union of all clusters is the full set of IVs, $C_k \bigcap C_{\ell} = \emptyset$ for any $k \neq \ell$ and $\bigcup\limits_{k=1}^{K^0} C_k = \{1, ..., J\}$. The number of clubs is larger than one and smaller than the number of IV values, $1 < K^0 < J$. %An equivalent way to state this assumption is that there exist distinct $z, \,\, z^*$ such that $p_z - p_{z^*} = 0$.
\end{assumption}
\noindent Here, $k$ defines the club identity. Assumption \ref{ass:Clubs} implies that there exist sets of $z$ with the very same propensity score, which we call \textit{clubs}, in line with the growth literature, which studies convergence clubs with clustering. When $K^0=J$, each judge belongs to its own club. Another way to look at the assumption is that there exists at least one representation with $J-1$ dummies and an intercept included where a judge $z$ serves as the reference category and the first-stage effect of at least one of the dummies of the remaining judges, say $z'$, is zero, implying the same propensity score for judge pair $z$, $z'$. In other words, the club assumption with $K^0<J$ is equivalent with a violation of the first-stage assumption. %This assumption seems restrictive at first, but as we will see soon, it 
Yet another way to state the assumption is that the function $\psi(z)$ is not one-to-one. 
This assumption is known in the literature as \textit{local irrelevance condition}. In \citet{Pan2024Locally}, for example, it is implied by assumption 2 (ii). \citet{DHaultfoeuille2024Testing} rely on the local irrelevance condition for the instrument (their assumption 4) to test the exclusion restriction. Similar assumptions can also be found in \citet{Torgovitsky2015Identification} and \citet{DHaultfoeuille2015Identification}. 

\noindent This assumption is not strong as there is a way to verify it in the data, and it can be relaxed to a case where each judge belongs to its own club, i.e., we have singleton clubs and $K_0=J$. To strengthen the argument when using our method, researchers should provide a priori reasons for the feasibility of the club assumption. We provide arguments for the judge IV example in the application section. 
The existence of several clubs permits forming pairs of clubs with distinct propensity scores $p_k^0$ and $p_{\ell}^0$ to construct IVs with a non-zero first stage for LATE evaluation. IV methods relying on any judge in club $k$ and any judge in a different club $\ell$ identify the very same LATE if assumptions \ref{ass:Validity} to \ref{ass:FirstStage} are satisfied, which follows from (\ref{eq:SameLATE}). %We will henceforth refer to the total of IVs constructed from a specific pair of clubs $k$ and $\ell$ as a union, which is indexed by $u$. That is, instruments belong to this union if they satisfy $z \in C_k$ and $z' \in C_{\ell}$. Based on this definition,

\noindent The following proposition states the identification result for the LATE in a setting with clubs, adding assumption \ref{ass:Clubs}. It follows directly from equation \ref{eq:SameLATE}. 
\begin{proposition}\label{th:MainResult}
For every pair $\{z, \,\, z'\}$, with $z \in C_k$ and $z' \in C_{\ell}$, $k\neq \ell$,  that satisfy \ref{ass:Validity} to \ref{ass:FirstStage}, the corresponding LATEs are the same.
\end{proposition}
%\textcolor{red}{2. What about controls? Can we just make everything conditional on observables?}

\subsection{Violations of the LATE assumptions}\label{sec:violations}

In the next step, we refine our approach to allow for the possibility that some instruments violate any of the LATE assumptions. We discuss the bias of the Wald estimand resulting from violating the assumptions in detail in Appendix \ref{app:violation_bias}.

\noindent We define the validity set, $\mathcal{V}$, as follows:
\begin{definition}[Validity set]\label{def:validityset}
	The validity set $\mathcal{V} \subseteq \mathcal{J}$ is the set of all IV values $z \in \mathcal{J}$ such that there exists some $z' \in \mathcal{J}$ belonging to a different club than $z$, and the pair $(z, z')$ satisfies Assumptions 1 to 3. We denote by $V_k = \mathcal{V} \cap C_k$ the subset of validity-set values belonging to club $k$, so that $\mathcal{V} = \bigcup_{k=1}^{K^0} V_k$.
\end{definition}
\noindent In words, the validity set contains all instrument values that form at least one valid pair with a value from a different club. Note that $z$ remains in the validity set as long as there exists some $z'$ in a different club such that the pair $(z, z')$ satisfies Assumptions 1 to 3, even if $z$ also forms invalid pairs with other instrument values. The validity set depends on assumption \ref{ass:Clubs} in the sense that only if $K^0>1$, assumption \ref{ass:FirstStage} can be fulfilled for some pair $z,\,z'$.  The validity set is analogous to the validity value set in Definition 3.2 of \citet{Sun2025Pairwise}. Note that the latter concentrates on the use of the validity pair set, i.e., the set of pairs that fulfill the LATE assumptions, while we directly use the validity set. In our example, the validity set is the subset of judges that identify judge-pair or club-pair LATEs when paired together. Judges not in the validity set could influence recidivism through other dimensions of their decision than imprisonment, or for pairs where these judges are involved, monotonicity might be violated because a usually severe judge does not incarcerate individuals who would have been incarcerated under a more lenient judge.

\noindent To find single judges satisfying the IV assumptions, we consider expected recidivism by judge, denoted by $r_z=E(Y|Z=z)$. %, which corresponds to reduced form coefficients as a function of the judge IV. 
%To cluster the reduced form coefficients, 
Next, we define \textit{groups}:

\begin{definition}{Group}\label{def:group}\\
A group $R_k^g$ is a set of IV values $z$ with the same reduced form coefficients:
\begin{equation}
R_k^g = \{z \in C_k: r_z = g\} \quad \text{ with some constant } g.
\end{equation}
\end{definition}
\noindent A group $R_k^g$ therefore corresponds to a set of IV values $z$ with the same mean outcome $g$ within a club $k$ as defined by Assumption \ref{ass:Clubs}. In our case, these are judges with the same incarceration and recidivism rates. %Let us denote by $R = \bigcup\limits_{k=1}^{K} R_k^g$ the union of groups of reduced form averages across all clubs and note that $R_k^g \subseteq C_k$.
% Intuitively, given that there is no effect of moving from $z$ to $z^*$ on $D$ inside of a club, the effect of moving from $z$ to $z^*$ on $Y$ can be attributed to a violation of one of the remaining LATE assumptions, for example, the exclusion restriction. Therefore, we want to limit our analysis to sections of the reduced form functions where $r_z=r_{z*}$ for $z\neq z^*$. That is the case in \cites{Li2024Discordant} assumption E.3, where the potential outcomes, $Y_{dz}$, are equal for all values $z \leq t$, which is visualized by a flat portion of the potential outcome function $Y_{dz}$, in dependence of $z$.  
Intuitively, since moving from \(z\) to \(z^*\) does not affect \(D\) inside a club, any effect of shifting from \(z\) to \(z^*\) on \(Y\) can be attributed to a violation of one of the remaining LATE assumptions, such as the exclusion restriction. Therefore, we limit our analysis to sections of the reduced form functions where \(r_z = r_{z^*}\) for \(z \neq z^*\). The setting we have in mind is also similar to the one implied by assumption E.3 in \citet{Li2024Discordant}, where the potential outcomes \(Y_{dz}\) are equal for all values of \(z \leq t\). This is represented by a flat segment of the potential outcome function \(Y_{dz}\) in relation to \(z\).

\noindent Next, we define the reduced form-based group with the highest number of IV values within a propensity score-based club as $R_k^M = R_{k}^g$, s.t. $|R_{k}^g| > \underset{g' \neq g}{max}|R_k^{g'}|$ and the union of such dominant groups across clubs by $\mathcal{R}^M = \bigcup\limits_{k=1}^{K^0}R_k^M$.\footnote{In our example $R_k^M$ corresponds to the largest number of judges with the same recidivism rate within a club $k$.} 
% To identify LATEs based on such dominant sets of IVs, we assume that the latter correspond to the validity set provided in Definition \ref{def:validityset}, which imposes an IV plurality assumption:
%\begin{assumption}{Plurality}\label{ass:plurality}: $\mathcal{V} = R^M$
%\end{assumption}
By proposition \ref{th:MainResult} (and the discussion leading to it), instrument values within a club that share the same reduced form coefficient are precisely those that yield identical Wald estimands when paired with values from another club. Plurality then asserts that the union of largest reduced-form groups within each club is the validity set:

\begin{assumption}[Plurality]\label{ass:plurality}
	For each club $k$, $V_k = R_{k}^M$.
\end{assumption}

\noindent Equivalently, we write $\mathcal{V} = \mathcal{R}^{M}$. Plurality implies that the respective largest groups of judges with the same recidivism rates within any club of homogeneous incarceration rates satisfy the LATE assumptions and, thus, belong to the validity set. Using the group with the maximal size is analogous to the plurality rule in \citet{Guo2018Confidence} and \citet{Windmeijer2021Confidence}.

\noindent To better understand plurality in the context of our paper consider the following toy example, which is visualized in table \ref{tab:withinplurality} in Appendix \ref{app:example}. Assume there are eleven judges, indexed from 1 to 11, with different imprisonment and recidivism rates in the cases that they consider: the four judges 1 to 4 have imprisonment rates of 10 percent, judges 5 to 8 have an imprisonment rate of 40 percent and the remaining judges are strict judges with 80 percent imprisonment rate. Hence, there are three clubs. In club 1, two judges have a recidivism rate of 20 percent, one has a rate of 40 percent and one 50 percent. Hence, this club’s largest group of judges has the same recidivism rate. We select now also the largest groups inside the other two clubs. The plurality assumption says that judges who are part of the largest group inside of each club belong to the validity set, i.e., if we pair them together, they create valid IVs. Using the largest group is a credible strategy when violations of the key assumptions are expected to be the exception and not the rule. This aligns naturally with many empirical applications, where researchers begin from the working assumption that all IVs are valid and seek a mild relaxation that permits excluding only a small subset of potentially invalid groups.

\noindent If the largest group is not unique, one option is to randomly select one of the maximal groups, as suggested in \citet{Windmeijer2021Confidence}. A more conservative alternative is to treat the tie as evidence against plurality and disregard the affected club. More generally, the size of the second-largest group relative to the largest serves as a useful diagnostic: when the two are similar in size, results should be interpreted cautiously and we recommend reporting estimates from both groups. When the next-largest group is much smaller, or when an absolute majority holds, the results are more trustworthy. In our application, this is clearly the case. In practice, when faced with clubs formed by a singleton $z$ we cannot verify whether they are valid or not and we therefore discard them. 

%Additionally, even in the absence of violations, we can effectively proceed with the method while disregarding assumption \ref{ass:plurality} and the second step of group clustering. Focusing on club-pairs reduces the number of comparisons from $\binom{J}{2}$ to $\binom{K}{2}$,  which can be a considerable reduction. Importantly, this approach also eliminates comparisons where $p_z=p_{z*}$, which would lead to weak IVs.
\noindent Even when the plurality assumption is in doubt, the first-stage clustering step alone delivers value. Skipping Assumption \ref{ass:plurality} and the within-club group clustering, the method reduces the number of pairwise comparisons from $\binom{J}{2}$ to $\binom{K}{2}$ and automatically excludes pairs with $p_z = p_{z^*}$, which would yield weak IVs. The first-stage clustering is therefore useful even as a standalone procedure.

\section{Method}\label{sec:Method}

\subsection{Intuition and general procedure}

% General idea
% Two-steo procedure general
% Step 1
% Step 2
% Postclass estimation

%Before presenting the methodological details of our method, we briefly discuss the general idea of our approach. We recall that by Proposition \ref{th:MainResult}, pairs $z, z'$ and $z*, z''$ which satisfy the LATE assumptions and yield the same propensity scores $p_z=p_z^*$ and $p_z'=p_z^{''}$ identify an identical LATE (for the very same complier group in terms of unobservables). We write the Wald estimator, $\beta$, as
%\begin{equation}
%\hat{\beta}_{z,z'} = \frac{\hat{r}_z - \hat{r}_z'}{\hat{p}_z - \hat{p}_z'} = \hat{\beta}_{z*,z^{''}}.
%\end{equation}
\noindent The key idea of our paper is to first detect clubs with equal treatment propensity scores $p$ and then isolate groups with equal reduced form outcome means $r$ in the data. We may then pair such groups to estimate LATEs. Because the quantities $p$ and $r$ in each group are homogeneous, so are the LATEs of specific group-pairs. The approach is illustrated in Figure \ref{app:fig:Illustration} in the Appendix. %As discussed in more detail in the following sections, we suggest a two-step clustering procedure for detecting first stage-based clubs as well as instruments that satisfy the LATE assumptions. Before clustering, we obtain the propensity scores from the coefficients of a regression of the treatment on all judge dummies. In the first step, we cluster propensity score estimates into clubs. In the second step, we use clustering within each club to determine the largest group of judge-based instruments with homogeneous estimates of the reduced form outcome. After this, we pair such largest groups with distinct propensity scores to obtain LATE estimates for specific club-pairs. This approach is illustrated in Figure \ref{app:fig:Illustration}.

%\begin{figure}[ht]
%	\begin{center}
%		%\includegraphics[scale=0.23]{../../ClubsAndGroups_Visualization2.pdf}
%		\includegraphics[scale=0.23]{ClubsAndGroups_Visualization2.pdf}
%		\includegraphics[scale=0.23]{ClubsAndGroups_Visualization3.pdf}
%		%\includegraphics[scale=0.23]{../../ClubsAndGroups_Visualization3.pdf}
%	\end{center}
%	\caption{Illustration of the estimation approach\label{app:fig:Illustration}}
%	\scriptsize
%	\textit{Note:} The first step of the procedure (left graph) detects clubs with comparable propensity scores. This provides us with two clubs in the given example. The second step detects groups with comparable reduced form estimates of judge- (or IV-) specific mean outcomes and selects the largest groups within clubs as the validity set (right graph). After this two-step classification, pairing the largest groups across different clubs (or propensity score values) permits estimating the LATE.
%\end{figure}

\noindent We use the Agglomerative Hierarchical Clustering (AHC) algorithm of \citet{Ward1963Hierarchical} to perform both steps of our procedure. To further improve the clustering, we also combine AHC with the clustering algorithm in regression via data-driven segmentation \citep[CARDS,][]{Ke2015Homogeneity}. We report the results of the CARDS approach in Appendix \ref{app:CARDS:sim}. We cannot achieve better results using the CARDS approach when compared to AHC. Furthermore, CARDS is not computationally efficient, while AHC is much faster and, therefore, is preferable overall. 
In a previous version of the paper, we also considered the Classifier-Lasso \citep{Su2016Identifying} for estimating club identities. However, again, we could not improve the results compared to AHC.

\subsection{Club classification}

\noindent The first step for estimating LATEs consists of detecting clubs of treatment propensity scores based on Assumption \ref{ass:Clubs}. We estimate club membership based on the AHC. This method was first applied to select valid instruments by \citet{Apfel2021Agglomerative} in the context of over-identified parametric models where the number of valid instruments exceeds the number of regressors and plurality is fulfilled. Here, we use it to cluster the treatment propensity scores as a function of the instruments. We denote by $\mathcal{C}$ a partition into clusters with $C_k \in \mathcal{C}$ and by $\hat{\mathcal{C}}$ the estimated partition of the data. The true partition is denoted by $\mathcal{C}^0$. In the following, we describe the algorithm, which is closely related to the one given in \citet{Apfel2021Agglomerative}: 
\begin{algorithm}\label{algo:ward}
	Agglomerative Hierarchical Clustering
	\begin{enumerate}
		\item \textbf{Input:} Calculate the IV-specific propensity scores, $\hat{p}_z=\frac{1}{N_z}\sum_{i=1}^{N}D_i\mathbbm{1}(i: Z = z)$, as the estimated coefficients of a regression of the treatment on all binary IVs. Calculate the Euclidean distance between them, stored in a dissimilarity matrix.
		\item \textbf{Initialize:} At the beginning, each propensity score is assigned to its own cluster, and therefore, the total number of clusters at the initialization step is $J$.
		\item \textbf{Join:} The two closest clusters in terms of their weighted squared Euclidean distance are merged into a new cluster. Formally, the weighted squared Euclidean distance is defined as $\frac{|C_k||C_l|}{|C_k| + |C_l|}||\bar{C}_k - \bar{C}_l||^2$, with $|C_k|$ denoting the number of propensity scores in cluster $k$ and $\bar{C}_k$ the arithmetic mean of propensity scores in cluster $k$.
		\item \textbf{Iterate:} Repeat the joining step until all propensity scores are in one cluster.
	\end{enumerate}
\end{algorithm}
\noindent 
We note that distinct weighting schemes in the Euclidean distance correspond to distinct objective functions to be minimized. We follow the classical approach in \citet{Ward1963Hierarchical} and define the sum of within-cluster variance as the objective function. As discussed in \citet{Apfel2021Agglomerative}, the results do not rely on this specific choice of the dissimilarity metric. When running the algorithm, we obtain a path of $S = J-1$ steps, with clusters of size $|\hat{C}_k| \in \{1, ..., J\}$ at each step. Note that each number of clusters, $K$, entails a different estimated partition $\hat{\mathcal{C}}(K)$.

\noindent To apply Algorithm \ref{algo:ward} in practice, a central question is at which step one should stop the merging process of clusters. In this context, the number of clusters $K$ is the penalty parameter to choose. %Note that we reduced the number of penalty parameters to choose from two to only one.
We propose the following procedure based on F-tests for equality of parameters to optimally select $K$.
Starting with $K=1$, we test whether all propensity scores are equal. If the test rejects the null, we proceed to the step of the AHC with $K=2$ and test for equality of all propensity scores within each cluster (or club). 

\begin{algorithm}\label{algo:F-test}
	Selecting the number of clusters via F-test
	\begin{enumerate}
		\item Select a critical value $\xi_n$
		\item Set $K=1$
		\item Perform an F-test, with $H_0:$ Propensity scores within any cluster are equal
		\item \textit{Updating:} If rejected, increase $K$ by 1
		\item Iterate \textit{testing} and \textit{updating} until F-test does no longer reject $H_0$.
	\end{enumerate}
\end{algorithm}
\noindent $\xi_n$ is a sequence of critical values. If it diverges, but not too fast, consistent classification is ensured.\footnote{For details on selection consistency, see Section \ref{sec:consistency}.} The asymptotic F-statistic for the test applied in Algorithm \ref{algo:F-test} is defined as\footnote{For simplicity, we use the test statistic under homoskedasticity but using a heteroskedasticity-robust version would be straightforward.}
\begin{equation}\label{eq:Ftest}
	(J-K)F(\hat{\mathbf{p}}, \mathcal{C}) = (\mathbf{R}(\mathcal{C})\hat{\mathbf{p}}- \mathbf{0}_J)^T\left[s^2\mathbf{R}(\mathcal{C})(\mathbf{Z}^T\mathbf{Z})^{-1} \mathbf{R}(\mathcal{C})^T \right]^{-1}(\mathbf{R}(\mathcal{C})\hat{\mathbf{p}}- \mathbf{0}_J),
\end{equation}
\noindent where $\mathbf{R}(\mathcal{C})$ is a hypothesis matrix with dimension $(J-K) \times J$. Denote $\hat{\mathbf{p}}=(\hat{p}_1,...,\hat{p}_J)'$ the vector of coefficients of the regression of the treatment variable on the matrix of dummies which turn to one when $Z_i=z$. This is equivalent to calculating the $z$-specific means. $s^2$ is the estimated error variance of that regression. For the IV of the first club to be tested for equality, the matrix entry is one, and for the remaining IVs, it is zero, except for one other IV in the club, whose entry is -1.\footnote{Note that in each row there are only  two non-zero entries.} In this way, we test the equality of all coefficients. $\mathbf{0}_J$ is a $(J \times 1)$ vector of zeros. Under the $H0$, the test statistic follows a $\chi^2$-distribution with $J-K$ degrees of freedom. %If the test is consistent and the critical values are chosen such that they diverge not too fast, see e.g.\ \citet{Windmeijer2021Confidence}, the classification via Algorithms \ref{algo:ward} and \ref{algo:F-test} is consistent in the sense that the club classification assigns all judges to their respective correct clubs in the limit.

\subsection{Group classification}

\noindent After clustering on the first-stage propensity score, $p_z$, any remaining heterogeneity in the LATEs must come from violating one or several LATE assumptions. Put differently, if all judge-specific IVs constructed within a club-pair satisfy the LATE assumptions, then IV estimates based on the various judge IVs should yield comparable LATEs and tests of overidentifying restrictions (e.g. Hansen-Sargan) in IV models should not reject. %For instance, one may apply a Hansen-Sargan-type test to multiple IVs created from two paired clubs of homogeneous propensity scores. In classical parametric IV models which impose homogeneous treatment effects and do not rely on first-stage clustering for generating clubs, such tests may not only have power against violations of IV assumptions, but also against a violation of effect homogeneity. In our framework allowing for heterogenous treatment effects, however, the application of Hansen-Sargan-type tests within club-pairs tailors it to testing the LATE assumptions rather than functional form restrictions. %If the test rejects, the estimate in a union is hence deemed uninformative.

\noindent To allow for a violation of the LATE assumptions, we proceed to the second step of our procedure, a clustering approach for determining the validity set $\mathcal{V}$ as provided in Definition \ref{def:validityset}. To this end, we apply Algorithms \ref{algo:ward} and \ref{algo:F-test} once more, but now for detecting groups within a club that are homogeneous in terms of reduced form estimates of judge-specific mean outcomes $r_z$. This requires first estimating the reduced form parameters by a linear regression of the outcome $Y$ on IV dummies for all judges (but no constant) within a club with homogeneous propensity scores. Therefore, the instrument $Z$ satisfies $Z \in C^k$ in club $k$. Then, the identical cluster procedure as previously used for detecting  first-stage clubs is applied to the reduced form estimates of $r_z$, $\hat{r}_z=\frac{1}{n_z}\sum_{i=1}^{N}Y_i\mathbbm{1}(i: Z_i = z)$, in order to classify them into homogeneous groups, $\hat{R}_k^g$. Finally, among these, we select the groups with the largest number of instruments within a club, $\hat{R}_k^M=\hat{R}_k^g,\,\, s.t. |\hat{R}_k^g| > \underset{g'\neq g}{max}|\hat{R}_k^{g'}|$. as the validity set, relying on the plurality assumption introduced earlier. The union of these sets is selected as the validity set, $\hat{\mathcal{V}} = \bigcup\limits_{k=1}^{K} \hat{R}_k^M$.\footnote{If Algorithm \ref{algo:F-test} proceeds until the last step for the propensity scores, this indicates that there are singleton clubs in the data. Then, group selection is not possible, because the plurality can only be fulfilled if $\mathcal{J}=\mathcal{V}$, i.e. the set of IV values is identical to the validity set.}

\subsection{Post-classification estimation}\label{sec:Postclassification}

\noindent Clustering first-stage clubs and reduced-form groups within clubs provides us with an estimate of clubs $\hat{\mathcal{C}}$ as well as the validity set $\hat{\mathcal{V}}$, which consists of the largest club-specific groups. Based on these estimates, we compute the LATE by the sample analog of equation \eqref{Wald}  using observations in the validity set that come from two different clubs (and, thus, differ in propensity scores).

\noindent The Group-Pair IV (GPIV) estimator for the ordered pair $(k,\ell)$ with $k<\ell$ is
\begin{equation}\label{eq:gpiv-scalar}
\hat{\beta}_{k\ell}^{\text{GPIV}} = \frac{\hat{r}_k - \hat{r}_{\ell} }{\hat{p}_k - \hat{p}_{\ell} } 
\end{equation}
with 
\begin{equation}
	\hat{p}_{k} = \frac{1}{n_k^M}\sum_{i=1}^n  D_i \mathbbm{1}(i: \,Z_i \in \hat{R}_k^{M}) \ , \ \
	\hat{r}_k = \frac{1}{n_k^M}\sum_{i=1}^n Y_i \mathbbm{1}(i: \,Z_i \in \hat{R}_k^{M}) \ ,\ \ 
	n_k^M = \sum_{i=1}^n \mathbbm{1}(i: \,Z_i \in \hat{R}_k^{M}) \ ,
\end{equation}
and similarly for group $\ell$. The groupwise incarceration ($\hat{p}_k$) and recidivism ($\hat{r}_k$) rates are based on the number of cases ($n_k^M$) decided by the judges in each group.

% ---------------------------------------------------------------------
% 4. Vector of all GPIV estimators
% ---------------------------------------------------------------------
\noindent For $K$ clubs, the full set of GPIV estimates contains all of these ordered pairs:
\begin{equation}\label{eq:gpiv-vector}
\hat{\boldsymbol{\beta}}^{\text{GPIV}} = \bigl( \hat{\beta}_{12}^{\text{GPIV}},\; \hat{\beta}_{13}^{\text{GPIV}},\; \dots,\; \hat{\beta}_{(K-1), K}^{\text{GPIV}} \bigr)'.
\end{equation}

\section{Consistency and asymptotic normality \label{sec:consistency}}

\subsection{Consistent classification}\label{sec:consistentclassification}

\noindent In the following, we prove that our procedure consistently classifies propensity scores and selects the validity set. Then, we show that the GPIV estimates using the selected groups are asymptotically normal. All proofs can be found in the appendix.

\noindent The following assumption is introduced to demonstrate that the algorithm covers the true partition.
\begin{assumption}\label{ass:LLN} We assume that
	\begin{enumerate}[label=\roman*.]
		\item A LLN holds such that $\forall z$ \\ $\frac{1}{n_z} \sum\limits_{i=1}^{n} D_i\mathbbm{1}(i: \, Z_i = z) \overset{p}{\rightarrow} E(D_i | Z_i = z)$  and $\frac{1}{n_z} \sum\limits_{i=1}^{n} Y_i\mathbbm{1}(i: \, Z_i = z) \overset{p}{\rightarrow} E(Y_i | Z_i = z)$
		\item $\frac{n_z}{n} \rightarrow c_z>0$, $\forall z$, where $c_z$ is a constant that may vary by $z$. 
	\end{enumerate}
\end{assumption}
\noindent The following results will establish consistent classification. 
\begin{corollary}{The true partition is on the selection path\label{coro:OnPath}}.\\
	Under assumptions \ref{ass:Clubs} and \ref{ass:LLN}, as $n \rightarrow \infty$ and $K=K^0$: $\underset{n \rightarrow \infty}{lim}\, P(\hat{\mathcal{C}}(K)=\mathcal{C}^0)=1$.
\end{corollary}
\noindent This follows from Lemma \ref{lemma:SameClubFirst} in Appendix \ref{sec:SameClubFirst}. Intuitively, we start with $K=J$ and then merge only clusters that contain propensity scores from the same club. Only when all $p_z$ from each club are in a cluster, respectively, the algorithm starts to merge clusters with members of different clubs.

\begin{theorem}Consistent classification \label{th:ConsistentClassification}\\
Let $\xi_n$ be the critical value for the F-test in Algorithm \ref{algo:F-test}. Let $\hat{\mathcal{C}}$ be the partition selected from Algorithms \ref{algo:ward} and \ref{algo:F-test}. Under Assumptions \ref{ass:Clubs} and \ref{ass:LLN}, for $\xi_n \rightarrow \infty$ and $\xi_n = o(n)$, as $n\rightarrow \infty$
$$\underset{n \rightarrow \infty}{lim} P (\hat{\mathcal{C}} = \mathcal{C}^0) = 1.$$
\end{theorem}
\noindent 
This theorem shows that asymptotically, Algorithms \ref{algo:ward} and \ref{algo:F-test} correctly select the club identities, i.e., the correct partition $\mathcal{C}^0$. The following corollary states that the analogous algorithms for the maximal group selection inside each club will select the validity set. 
\begin{corollary}Consistent group selection \label{coro:ConsistentGroupSelection}\\	
Let $\hat{\mathcal{R}}^{M}$ be the partition selected when applying Algorithms \ref{algo:ward} and Algorithm \ref{algo:F-test} to $\hat{r}_z$. Under Assumptions \ref{ass:plurality} and \ref{ass:LLN}, for $\xi_n \rightarrow \infty$ and $\xi_n = o(n)$, as $n \rightarrow \infty$
	$$\underset{n \rightarrow \infty}{lim} P (\hat{\mathcal{R}}^{M} = \mathcal{V}) = 1.$$
\end{corollary}
\noindent The proof of this result follows the ones of Theorem \ref{th:ConsistentClassification} in this paper and of Theorem 1 in \citet{Apfel2021Agglomerative} closely and is hence omitted. The main difference is that the correct selection of the validity set does not rely on correctly assigning each judge to its group. As long as the invalid judges are not in the largest group, the algorithms still correctly select the validity set, even when the algorithm partitions judges into the wrong number of groups. 
By Theorems \ref{th:ConsistentClassification} and Corollary \ref{coro:ConsistentGroupSelection} we now have that as $n \rightarrow \infty$, $\hat{\mathcal{C}} = \mathcal{C}^0$ and $\hat{\mathcal{R}}^{M} = \mathcal{V}$ with probability approaching one. 

\subsection{Asymptotic normality of the GPIV estimates}\label{sec:asynorm}
Now we proceed to show asymptotic normality of the GPIV estimates. To do so, we first introduce the \textit{oracle estimator} and its asymptotic distribution. We denote the partitions based on the true clubs, $\mathcal{C}^0$, and the validity set, $\mathcal{V}$, as \textit{oracle} partitions and the GPIV estimator based on these ideal partitions as oracle estimator. 

\noindent Let \(V_1,\dots,V_{K^0}\) be the true validity value sets (disjoint subpopulations) for the \(K^0\) clubs.  
For each \(k=1,\dots,K^0\), define

\[
p_k = E[D\mid Z\in V_k] \ , \ \
r_k = E[Y\mid Z\in V_k] \ , \ \
\pi_k =P(Z\in V_k) > 0 \ .
\]

\noindent Clubs are ordered so that \(p_1 > p_{2} > \dots > p_{K^0}\) (strictly decreasing treatment propensity). This ensures that for any \(k<\ell\) the forward Wald ratio has a positive denominator.

\noindent The groupwise incarceration and recidivism means based on the oracle classification, i.e. the sample means for the true clubs and the validity sets, are
\begin{equation}\label{eq:oraclemeans}
\tilde{p}_k = \frac{1}{\tilde{n}_k}\sum_{i=1}^n D_i \mathbbm{1}(i: Z_i\in V_k) \ , \ \
\tilde{r}_k = \frac{1}{\tilde{n}_k}\sum_{i=1}^n Y_i \mathbbm{1}(i: Z_i\in V_k) \ ,\ \
\tilde{n}_k = \sum_{i=1}^n \mathbbm{1}(i: Z_i\in V_k).
\end{equation}
\iffalse
\noindent Then, consider the \(2K^0\)-vector of means
\[
\bar{\mathbf{X}}_n = \bigl( \tilde{p}_1,\; \tilde{r}_1, \; \tilde{p}_2,\; \tilde{r}_2, \; \dots,\; \tilde{p}_{K^0},\; \tilde{r}_{K^0} \; \bigr)'
\]
\noindent with the corresponding population means collected in $\boldsymbol{\mu}_X$. 
\fi
\noindent Define the vector of oracle subsample means $\tilde{\boldsymbol{\mu}} = (\tilde{r}_1,\tilde{p}_1,\dots, \tilde{r}_{K^0},\tilde{p}_{K^0})' \, $, and let \(\boldsymbol{\mu} = (r_{1},p_{1},\dots,r_{K^0},p_{K^0})'\), the corresponding population means. To show asymptotic normality of the oracle estimator, we introduce the following Lemma. 

\begin{lemma}[Asymptotic distribution of \(\tilde{\boldsymbol{\mu}}\)]\label{lem:mu_asymnorm}
	Under Assumption \ref{ass:multiCLT}, $\tilde{\boldsymbol{\mu}}$ satisfies
	\[
	\sqrt{n}\bigl( \tilde{\boldsymbol{\mu}} - \boldsymbol{\mu} \bigr) \xrightarrow{d} N(0,\mathbf{\Sigma}_{\boldsymbol{\mu}}).
	\]
	where the asymptotic covariance matrix \(\mathbf{\Sigma}_{\boldsymbol{\mu}}\) is block‑diagonal. 
	\begin{align}\label{eq:symmetricmu}
		\mathbf{\Sigma}_{\boldsymbol{\mu}} &  \;=\; \bigoplus_{k=1}^{K^0} \frac{1}{\pi_k}
		\begin{pmatrix}
			\operatorname{Var}(Y_i\mid V_{ki}=1) & \operatorname{Cov}(Y_i,D_i\mid V_{ki}=1) \\
			\operatorname{Cov}(Y_i,D_i\mid V_{ki}=1) & \operatorname{Var}(D_i\mid V_{ki}=1)
		\end{pmatrix}\\ \nonumber
		& \;=\; \operatorname{diag}\left( \frac{1}{\pi_1}\mathbf{\Sigma}_{YD\mid V_1},\;
		\frac{1}{\pi_2}\mathbf{\Sigma}_{YD\mid V_2},\; \dots,\;
		\frac{1}{\pi_{K^0}}\mathbf{\Sigma}_{YD\mid V_{K^0}} \right)
	\end{align}
\end{lemma}
\noindent where \(V_{ki} = \mathbbm{1}(i: \,Z_i\in V_k)\), an indicator that turns to one for all cases assigned to a judge in club $k$. The $2 \times 2$ matrix in brackets is implicitly denoted as $\mathbf{\Sigma}_{YD\mid V_k}$. Assumption \ref{ass:multiCLT} means to assume that the numerators and denominators of the ratios in $\tilde{\mathbf{\mu}}$ follow a CLT.\footnote{This holds for example under iid sampling. Because this assumption is almost trivial and would introduce additional notation, we introduce it before the proof of Lemma \ref{lem:mu_asymnorm} in Appendix \ref{app:oracle_distribution}}

\noindent The oracle estimator for a specific group comparison is:\[\hat{\beta}_{k\ell}^{\text{or}} = \frac{\tilde{r}_k - \tilde{r}_{\ell}}{\tilde{p}_k - \tilde{p}_{\ell}}\]and the vector of oracle Wald estimators, for the set of ordered pairs $(k,\ell)$ with $k<\ell$ is \begin{equation}\label{eq:oracle}	\hat{\bm{\beta}}^{or} = \left( \hat{\beta}^{or}_{1 2},\hat{\beta}^{or}_{13}, ..., \hat{\beta}^{or}_{(K^0-1),K^0} \right)'\end{equation}

\noindent The asymptotic distribution of the oracle estimator is derived in the next proposition. 
%The oracle estimator for a specific group comparison is:
%\[
%\hat{\beta}_{k\ell}^{\text{or}} = \frac{\tilde{r}_k - \tilde{r}_{\ell}}{\tilde{p}_k - \tilde{p}_{\ell}}
%\]
%and the vector of oracle Wald estimators, for the set of ordered pairs $(k,\ell)$ with $k<\ell$ is 
%\begin{equation}\label{eq:oracle}
%	\hat{\bm{\beta}}^{or} = \left( \hat{\beta}^{or}_{1 2},\hat{\beta}^{or}_{13}, ..., \hat{\beta}^{or}_{(K^0-1)K} \right)'
%\end{equation}
Define $\Delta_{k,\ell}=p_k-p_{\ell}>0$, for all $k < \ell$. For two ordered group pairs \((k,\ell)\) and \((k',\ell')\) with \(k<\ell\) and \(k'<\ell'\) that share at most one club, denote the index of the shared club \(\{s\} = \{k,\ell\} \cap \{k',\ell'\}\). We can have the case that the shared club is in the same position, $k=k'$ or $\ell=\ell'$, the shared club is not in the same position, $k=\ell$ or $k'=\ell'$, or there is no shared club at all, $s=\{\emptyset\}$. 

%The corresponding entry of the asymptotic covariance matrix \(\boldsymbol{\Omega}_{\text{or}}\) is

\begin{proposition}[Asymptotic normality of \(\hat{\boldsymbol{\beta}}^{\text{or}}\)]\label{prop:oracle_asymnorm}
	Under Assumptions \ref{ass:LLN} and \ref{ass:multiCLT}, the vector $\hat{\boldsymbol{\beta}}^{\text{or}}$
	satisfies
	\[
	\sqrt{n}\bigl( \hat{\boldsymbol{\beta}}^{\text{or}} - \boldsymbol{\beta} \bigr) \xrightarrow{d} N(0,\mathbf{\Omega}_{\text{or}}),
	\]
	where \(\boldsymbol{\beta}\) is the vector of LATEs \(\beta_{k,\ell}=\frac{r_k-r_{\ell}}{p_k-p_{\ell}}\) and \(\mathbf{\Omega}_{\text{or}}\) ($\binom{K^0}{2} \times \binom{K^0}{2}$) has entries:
	
	\begin{enumerate}
		\item Diagonal entries (estimator variances): 
		$
		[\mathbf{\Omega}_{\text{or}}]_{k\ell,k\ell} = \frac{1}{\Delta_{k\ell}^2}\left( \frac{\operatorname{Var}(Y-\beta_{k,\ell}D\mid V_k)}{\pi_k} + \frac{\operatorname{Var}(Y-\beta_{k,\ell}D\mid V_{\ell})}{\pi_{\ell}} \right),
		$
		\item Covariances: \[
		[\boldsymbol{\Omega}_{\text{or}}]_{(k,\ell),(k',\ell')}
		=
		\begin{cases}
			\dfrac{1}{\Delta_{k\ell}\Delta_{k'\ell'}} \cdot
			\dfrac{\operatorname{Cov}\bigl(Y - \beta_{k\ell}D,\; Y - \beta_{k'\ell'}D \mid V_s\bigr)}{\pi_s}
			& \text{if } k = k' \text{ or } \ell = \ell', \\[1.5em]
			-\dfrac{1}{\Delta_{k\ell}\Delta_{k'\ell'}} \cdot
			\dfrac{\operatorname{Cov}\bigl(Y - \beta_{k\ell}D,\; Y - \beta_{k'\ell'}D \mid V_s\bigr)}{\pi_s}
			& \text{if } k = \ell' \text{ or } \ell = k', \\[1.5em]
			0 & \text{if } \{k,\ell\}\cap\{k',\ell'\} = \emptyset.
		\end{cases}
		\]
		\iffalse
		\item Covariance between two estimators sharing one club:
		\[
		[\mathbf{\Omega}_{\text{or}}]_{k\ell,k\ell'} = \frac{\gamma_{(k,\ell),(k',\ell')} }{\Delta_{k,\ell}\Delta_{k,\ell'}} \cdot
		\frac{\operatorname{Cov}\bigl(Y - \beta_{k,\ell}D,\; Y - \beta_{k,\ell'}D \mid V_k\bigr)}{\pi_k},
		\]
		
		\item Covariance between two estimators that share no clubs (disjoint pairs):
		\[
		[\mathbf{\Omega}_{\text{or}}]_{k\ell,k' \ell'}=0 \quad\text{if }\{k,\ell\}\cap\{k', \ell'\}=\emptyset.\]
		\fi
	\end{enumerate}
\end{proposition}
\noindent The covariance entries are distinguished for three cases: when the shared club is in the same position, when it is not in the same position and when there is no shared club at all. %The second case, when two estimators share one club appears for example when \((k,\ell)\) and \((k,{\ell'})\) with \(\ell\neq \ell'\). 
Building on this result and on consistent classification as shown in Theorem \ref{th:ConsistentClassification} and Corollary \ref{coro:ConsistentGroupSelection}, we can now derive the asymptotic distribution of the GPIV estimator.

	\begin{theorem}[Asymptotic normality of the GPIV estimator]\label{th:gpiv_asymnorm}
	Under Assumptions \ref{ass:Clubs}-\ref{ass:multiCLT}, for $\xi_n \rightarrow \infty$, $\xi_n = o(n)$ and $n \rightarrow \infty$
	\[
	\sqrt{n}\bigl( \hat{\boldsymbol{\beta}}^{\text{GPIV}} - \boldsymbol{\beta} \bigr) \xrightarrow{d} N(0,\mathbf{\Omega}_{\text{or}}).
	\]
\end{theorem}

\iffalse
The matrix \(\Omega_{\text{or}}\) can be consistently estimated by replacing population quantities (\(r_k,p_k,\pi_k,\beta_{k\ell},\operatorname{Var}(\cdot\mid V_k),\operatorname{Cov}(\cdot,\cdot\mid V_k)\)) with their sample analogs based on the oracle groups.

\begin{theorem}Asymptotic normality \label{th:AsymptoticNormality}\\
Under Assumptions \ref{ass:Clubs} to \ref{ass:CLT}, for $\xi_{n} \rightarrow \infty$ and $\xi_{n} = o(n)$,
\begin{equation}\label{eq:AsymptoticNormality}
	\sqrt{n} (\hat{\mathbf{\beta}}(\hat{\mathcal{C}}, \hat{R}_{max},k,\ell) - \Delta(k,\ell) ) \overset{d}{\rightarrow} N(0, \omega_{k,\ell})
\end{equation}
for all pairs $k,\ell$.
\end{theorem}
\fi
%\footnote{We're interested in the club-pair estimators, but if a weighted combination of the club-pair estimators is of interest, where weights would be provided by the econometrician as in \citet{Sun2024Pairwise}, one can denote $\hat{\mathbf{\beta}}_{k,\ell} \equiv \hat{\mathbf{\beta}}(\hat{\mathcal{C}}, \hat{R},k,\ell)$ and introduce the $\frac{K(K-1)}{2}$-dimensional vector $\hat{\bm{\beta}} = \{\hat{\mathbf{\beta}}_{1,2}, ..., \hat{\mathbf{\beta}}_{K-1,K}\}$, to then use Theorem B.1 in \citet{Sun2024Pairwise} to show convergence to a normal distribution. Hypothesis testing using the distribution of this vector is then straightforward.}

%Simulation Results 
\subsection{Simulation results}
\noindent In simulations,f presented in Appendix \ref{sim}, we investigate the performance of our method in finite samples. We create one setting with and one without violations of the LATE assumptions. In the setting without invalidity, after the first-stage clustering, the Hansen \citep{Sargan1958Estimation, Hansen1982Large} p-values are above any conventional significance level, correctly indicating no evidence for invalid instruments, and therefore, our procedure stops. In the setting with invalidity after the first-stage clustering on the propensity scores, the Hansen test\footnote{Note that the Hansen test statistic is obtained from an overidentified specification by setting one club equal to the reference category and using all other judge-IVs as separate dummies. The test is not used as part of the algorithm but to illustrate that there is still heterogeneity in the treatment effects.} leads to rejecting the null, which could be due to effect heterogeneity and/or invalidity. As judges have been classified into clubs with the same imprisonment rates, we can rule out effect heterogeneity, which leaves violations of the remaining LATE assumptions as the reason for rejection. To address invalidity, we apply the second step of our procedure, after which the Hansen-test no longer rejects. 
The results show that the method delivers reliable results in various settings, retrieving clubs and groups that fulfill the LATE assumptions. The simulations also show that the performance improves with the number of cases per judge. We also analyze the impact of violating Assumption \ref{ass:Clubs} by adding Gaussian noise to the judges' propensity scores, finding that our method performs well under minor deviations but deteriorates with larger deviations.

\section{Application}\label{sec:Application}

\subsection{Assessing the effect of incarceration on recidivism}\label{sec:assessing}

% Idea
% Equation

% Problem
% Solution
% Literature
% Violations
% New assumptions

%Incarceration is thought to have an effect on crime by deterring it but also through experiencing incarceration, which might have a chastening effect. But jail can also contribute to rehabilitate inmates and prepare them for reintegration into society. A crime-increasing effect is possible when the time in jail lets human capital depreciate or fosters integration into criminal networks.

\noindent In recent years, several studies have addressed the question of whether incarceration affects the likelihood of recidivism (future criminal behavior) or, in other words, whether sentencing offenders to prison affects the probability of them relapsing into criminal behavior. Incarceration could decrease the likelihood of recidivism if prisons help rehabilitate offenders and reintegrate them into society. If, however, the time of imprisonment integrates them into criminal networks or leads to a loss of human capital, incarceration might increase the likelihood of recidivism. To answer this question, authors typically aim to estimate the following empirical model:
\begin{equation}\label{eq:EmpiricalModel}
Y_{jc} = D_{jc} \beta + \mathbf{x}_{jc} \theta + \varepsilon_{jc}
\end{equation}
where $Y_{jc}$ denotes the outcome, an indicator for recidivism, i.e., re-indictment or re-incarceration in a certain time window after conviction, $j$ is the judge-index and $c$ is the case-index. $D_{jc}$ is the treatment variable with the coefficient of interest $\beta$, indicating whether a judge has sentenced the convict to a term of imprisonment or has decided on a non-prison sentence, such as probation. $\mathbf{x}_{jc}$ are observed covariates potentially affecting the probability of recidivism, with the coefficient vector $\theta$. \citet{Nagin2009Imprisonment} suggest using prior record, offense type, age, race, and sex as key control variables. $\varepsilon_{jc}$ reflects unobserved characteristics affecting the outcome.

\noindent Estimating equation \eqref{eq:EmpiricalModel} by OLS is generally biased and inconsistent if the association between outcome and right-hand side variables is non-linear and/or unobserved confounders jointly affect $D$ and $Y$ even after controlling for $X$. For instance, the judge might have information (not available to the researcher) about the defendant's previous offenses or possible addictions, which may simultaneously influence the judge's sentencing and the likelihood of recidivism. Likewise, the sentence may be influenced by the defendant's behavior in court, which may shed light on the defendant's potential to recidivate. To address this issue of unobserved confounders, several studies have leveraged the fact that in some judicial systems, cases are randomly assigned to judges whose use of prison sentences varies systematically. They have considered dummies for being assigned to a particular judge or the incarceration rate by the judge as instrumental variables.  %Furthermore, we refer to the probability of receiving a prison sentence given the instrument as propensity score. The practice of random judge assignment has been introduced to avoid so-called judge-shopping, i.e.\ filing numerous lawsuits in the hope of being eventually assigned to a favorable judge.
Following this literature, we also use a set of judge IV dummies. 

\noindent Earlier evidence on the effect of imprisonment is mixed, but studies that find crime-increasing and null effects are in the majority. \citet{Loeffler2021Impact} provide a review of 13 published IV-based studies on the effect of incarceration on recidivism, concluding that those based on data from U.S. courts mainly find either an insignificant or a significant recidivism-increasing effect. It appears that U.S. prisons are ineffective in terms of rehabilitation and resocialization and, thus, do not serve to fight crime beyond general deterrence effects. In contrast, a study by \citet{Bhuller2020Incarceration} finds that incarceration in Norwegian prisons has a significant recidivism-reducing effect and a positive impact on employment, pointing to considerable heterogeneity in the effect of incarceration across regions, possibly due to differences in prison infrastructure and support services, such as labor market training opportunities.\footnote{Many studies in this literature focus on US data. But there is also some research on data from other countries, such as Chile \citep{Cortes2019juvenile}.  Most studies examine the incarceration effect only among convicts, while some, such as \citet{Cortes2019juvenile} and \citet{Leslie2017unintended}, assess the impact of pre-trial detention among all individuals accused of a crime.}

\noindent Effect heterogeneity can also arise within the same institutional context, in that defendants sentenced to prison by different judges generally differ in their background characteristics (such as personality traits), and these, in turn, can influence the effect of incarceration. For this reason, the LATE, i.e., the effect of imprisonment on recidivism among individuals who would be incarcerated under a stricter but not under a more lenient judge, might generally differ across distinct pairs of judges (or, relatedly, distinct propensities of incarceration). Such effect heterogeneity would be masked when applying 2SLS simultaneously to all judge IVs because the 2SLS estimator is a weighted combination of just-identified IV estimators, which estimate LATEs \citep[e.g.][]{angrist2009mostly, andrews2019structure}. This is one motivation for using our approach.

\noindent Further, the instruments might fail to satisfy the identifying assumptions \ref{ass:Validity} to \ref{ass:FirstStage}. For instance, judges could differ in terms of their use of alternative measures to imprisonment, such as electronic monitoring \citep{Loeffler2021Impact}, which would violate the exclusion restriction postulated in Assumption \ref{ass:Validity}. Further, the IVs might violate  Assumption \ref{ass:Monotonicity}, requiring monotonicity of the treatment in the instrument, see for instance the discussion in \citet{Frandsen2023Judging}. Depending on how different judges weigh the individual aspects of a case, a rather stringent judge with a relatively high rate of prison sentences could refrain from incarcerating a particular defendant who would have been imprisoned by a judge with a lower incarceration rate. i.e., their (potential) sentences could contradict the judges' order of (average) severity. Such a situation entails the existence of defiers and, thus, a violation of weak monotonicity of the treatment in the instrument.

\noindent To consistently estimate the LATEs of interest, we invoke the existence of clubs of judges with a similar propensity to incarcerate, as well as plurality, such that the largest group of judges constitutes IVs fulfilling the LATE assumptions. Here, we use conditional LATE assumptions, because we condition on observed covariates, as described in Section \ref{main_analysis}. The cluster assumption appears realistic as long as judges can be plausibly categorized into a limited number of judge types, each with a distinct rate of prison sentences.

\noindent In the political science and legal literature, the idea that judges belong to different types and can be clustered as such is familiar. 
\citet{Bennett1989Styles} mentions that it ``is commonplace to characterize judges [...] as conservative or liberal, activist or restrained'' and lists a number of other categories. \citet{Cross2008Judging} are motivated by law scholars often discussing judge typologies. The types of judges mentioned in this work are ``innovators'', ``realists'', and ``interpreters'', and the two types of judicial entrepreneurs and minimalists are introduced. The authors perform cluster analysis using three quality criteria of 58 judges and find nine clusters. This analysis uses \cites{Ward1963Hierarchical} agglomerative hierarchical clustering, the same algorithm our paper proposes in the context of LATE estimation.

\noindent Judges might have been educated and practiced at the same institutions, establishing cultures of higher or lower use of prison sentences. Moreover, in the US, the severity of judges might reflect their allegiance to political parties. \citet{Yung2012Typology} uses AHC on four metrics for the following dimensions: judicial activism, ideology, independence, and partisanship for 178 judges and finds nine judge-type clusters. Using multinomial logistic regression, the author shows that factors such as attending law school and appointing the president's political party are predictive of the judge type.

\noindent While these clustering exercises do not search for clusters of judge stringency, they motivate why the idea of clubs can be sensible. In the context of our example, we treat the existence of clubs as an empirical problem. 
As an example, \citet{Green2010Using} exploit variation in judicial calendars, where it could be the case that judges assigned to the same calendar affect each other's decisions. In some applications, such as in \cites{Eren2021Juvenile} Fig. 1, the distribution of judge stringency displays multiple peaks, such that the club assumption seems justified. As mentioned above, the club assumption can be verified in the data, and if there is no club, the algorithm will run until each judge is in their own cluster. Finally, some judges have moved between courts and reappeared in the data with different judge IDs, which might lead to similar stringency.
%Dobbie2018
%Agan \& Dobbie
%Bald2022Causal, Fig. 2
%Silver and Zhang 2022? Fig 1
%Lee 23? Fig 1

\subsection{Data}\label{sec:Data}

\noindent The model developed in Section \ref{sec:ModelAndAssumptions} is applied to a data set of offenders in the U.S. state of Minnesota, which is composed of data from two primary sources: the first one is an extract from the Minnesota Judicial Branch case database containing information on the offender, including their full name, date of birth and place of residence, for all criminal cases from 2009 to 2020; the second source is a data set from the Minnesota Sentencing Guidelines Commission on all adult offender cases in Minnesota between 2001 and 2017, with information on the type and severity of the offense as well as the judge hearing the case. We link these two datasets using the case number as an identifier to obtain a data set ranging from 2009 to 2017 on all criminal offense cases, including information on offender, judge, offense, and conviction. We note that the data do not include judge-specific covariates, which limits our analysis, as these characteristics could be related to both judicial stringency and sentencing practices.

\noindent For each offender in our data set, we identify all criminal cases in Minnesota in which they were involved between 2009 and 2017, using the offenders' full names and dates of birth as identifiers. Despite also having information on the offenders' place of residence, we do not include this information to identify recidivism to account for changes of residence, which occur frequently, especially after returning from jail or prison. This way, we accept the low risk of incorrectly linking the cases of two individuals with the same name and date of birth while reducing the risk of not detecting recidivism. In addition, we cannot detect recidivism if an offender commits a crime in another state, has moved to another state/country, or has changed their name.

\noindent For estimating the effect of incarceration on offender recidivism, we consider recidivism within three years of sentencing as the outcome. Therefore, to observe the three-year post-conviction period of every offender, we must reduce our dataset to the years 2009 to 2014. According to the Minnesota Order for Assignment of Cases, all criminal cases in Minnesota are randomly assigned to a judge with jurisdiction in the county where the crime is tried. After being assigned to a case, a judge in active service must preside over it until its resolution, i.e., random judge assignment, as required for our IV approach, is guaranteed by Minnesota state case assignment rules. This ensures that the assignment of defendants to individual judges is as good as random within each court, which provides the basis for using judge dummies as candidate instruments.

\noindent As senior judges are permitted to opt-out from hearing a case, we remove all cases heard by a senior judge to ensure random judge assignment. Then, although most judges in Minnesota remain in the same court throughout their tenure, some judges in the sample have changed court during our observation period. These judges are assigned a different ID for each court, such that the terms at different courts are treated as if they belong to different judges to account for differences in county crime profiles, which are reflected in the judges' sentencing practices. 

\noindent To ensure that the vast majority of offenders have been able to re-offend in the data within the three years following sentencing, i.e., are not in prison for the entire three years during which we observe potential recidivism, we need to reduce the data set to minor crimes. An analysis of the recidivism effect based on the entire dataset would require strict control for crime and offender profiles to avoid bias in the estimated effect caused by the inclusion of observations for which recidivism is highly unlikely due to long-term incarceration. At the same time, including offenses that result in a lengthy prison sentence would not improve the quality of the estimator for the recidivism effect. Recent studies on the effect of incarceration have reduced their datasets based on different rules: \citet{Loeffler2013does} have concentrated on cases of the three lowest charge classes, \citet{Green2010Using} on drug offenses. We follow \citet{Bhuller2020Incarceration} by reducing the dataset based on the sentence lengths that an independent institution - in our case, the Minnesota Sentencing Guidelines Commission - recommends for each observed offense given the type of crime, the severity of the offense and the offender's criminal history. We reduce our dataset to cases with presumptive sentences of up to three years (a detailed list of the included offenses can be found in Appendix \ref{app:crimelist}). The resulting sample contains 48,849 cases involving 38,874 unique offenders. Only some 10 percent of the offenses in our sample did not result in incarceration.

\noindent Some 82 percent of the offenses in the final dataset have resulted in a sentence of up to one year in county jails, while in about 14 percent of the observed cases, offenders were sentenced to one to two years in state prison, meaning in 96 percent of the cases offenders had at least one year after their official release date to recidivate. In only some 0.9 percent of the cases, the offender was convicted of three or more years in state prison. Given that usually only two-thirds of a sentence is served in prison and the rest is on probation, the share of offenders having at least one year to recidivate is even higher than 96 percent. Table \ref{app:tab:desc} in Appendix \ref{app:desc} provides some descriptive statistics for our data.

\noindent We restrict the data to judges that heard at least 300 randomly assigned cases between 2009 and 2017 and, in addition, to cases tried in counties where no fewer than two of these judges are stationed in any given year. The resulting sample contains %44,520, 23,958 and 
11,219 cases. In the Appendix, we provide additional exercises to this application in which we look at judges with at least 200 cases (\ref{app:add}) as well as tackling the problem of controlling for observed covariates in an alternative way by considering a subset of the data that is similar in terms of observables (\ref{sec:Subset}).

\subsection{Balance check and placebo test}\label{sec:balance_placebo}

\noindent  Because our method is explicitly designed for settings in which some judges may violate the LATE assumptions, some imbalance in pre-sentence covariates across judges is expected. 
We check the covariate balance among clubs, because we perform the IV analyses at the club-pair level. The balance table \ref{tab:club_balance} can be found in Appendix \ref{app:application} and shows that the clubs are roughly balanced on most covariates such as gender, age at sentencing or severity of the crime, but for race and offense, there is some imbalance. For example, the percentage of white defendants for judges in clubs 1 and 2 is close to 40\% but it is close to 55\% for judges in club 3. 
To tackle this issue, any systematic diﬀerences in observable defendant characteristics across judges and clubs are addressed through the covariate partialling described in Section \ref{main_analysis}, ensuring our estimates account for these diﬀerences. 

\noindent As a further check on the plausibility of random assignment, we perform a placebo
test using a predetermined variable related to the outcome. Specifically, we regress each offender's number
of prior cases at the time of judgment on all club (or judge)
dummies, controlling for year fixed effects. We choose prior cases as an outcome variable, because it is fixed before the current
case and therefore cannot be affected by the assigned judge. We include year effects because
prior-case counts accumulate mechanically over the sample window, so that the joint
test asks whether assignment predicts prior cases \textit{beyond} this mechanical time
trend. Under random assignment, the dummies should not be jointly predictive of this
predetermined variable. The results are reported in Table~\ref{tab:placebo}, with
standard errors clustered by offender. At the club level (the level at which our
analysis is conducted) the club dummies are jointly insignificant
($F = 0.08$, $p = 0.93$), indicating that assignment does not predict prior criminal
history. This is notable because, although the clubs differ somewhat in their covariates (Table~\ref{tab:club_balance}), they are balanced on prior
record.%
\footnote{We prefer the club-level test, since our analysis is conducted at the
	club-pair level, but we also report the test using all judge dummies. There, the
	dummies are jointly significant at the $5\%$ level ($F = 1.57$, $p = 0.047$). With an $F$-statistic just above one we argue that this may be reflect small differences across judges, which average out at the club level. Conclusions are unchanged under classical and HC1
	standard errors. We therefore read the placebo results as supportive of the
	identifying assumptions.}

\subsection{Main analysis} \label{main_analysis}
 
\noindent We start with our first-step AHC and choose the significance level for the F-test as 0.1/log(N), following \citet{Windmeijer2021Confidence}. In Table \ref{tab:firststage}, we show the results of the first stage clustering for the sample with at least 300 cases per judge. We first partial out the controls from the outcome, treatment and IVs and then run the first-stage regression of the treatment (imprisonment) on all judge dummies. With that, we get five clubs of sizes 8, 12, 3, 1 and 1. We exclude the two singleton clubs since the second stage selection would be uninformative with a singleton club, and we cannot be sure the judge is valid. Both singleton clubs have propensity scores outside the range of the retained clubs, whose means lie between 0.63 and 0.77. One is a clear outlier at 0.13, and the other at 0.51 is substantially below the nearest club's range (0.61-0.67). The minimum distance from each singleton club to its nearest neighbor (roughly 0.5 and 0.12, respectively) substantially exceeds the within-club ranges of the retained clubs (0.032-0.061), confirming these represent genuinely distinct judges rather than borderline cases. The propensity scores for the non-singleton clubs range from 0.63 to 0.77.

\begin{table}[!htbp]
	\caption{First-stage clubs, 300}
	\label{tab:firststage}
 	\begin{center}
 	% latex table generated in R 4.2.2 by xtable 1.8-4 package
% Wed Jun  3 13:43:41 2026
\begin{tabular}{p{1.5cm}p{1.5cm}p{1.5cm}}
  \toprule
Club & Mean & Nr \\ 
  \midrule
1 & 0.73 & 8 \\ 
  2 & 0.77 & 12 \\ 
  3 & 0.63 & 3 \\ 
  4 & 0.13 & 1 \\ 
  5 & 0.51 & 1 \\ 
   \bottomrule
\end{tabular}
	
 	\end{center}
	\textit{Note: Clustering results when minimum case count per judge is 300. Nr stands for the number of judges in each cluster.}
\end{table}

\noindent In Table \ref{tab:postahc}, we run an OLS regression and a 2SLS estimation, using judge dummies as instrumental variables, as baseline regressions. This IV approach is close to \citet{Green2010Using}, who use dummies for judicial calendars as instruments. We include race-, gender-, offense-type, year-dummies, severity, age, the squares of the latter two and race-gender interaction dummies. In practice, we partial these controls out of the outcome and the treatment variable. This way, our LATE estimates focus on the remaining variation, unaﬀected by observable diﬀerences in defendant characteristics across judge clubs.

\noindent The OLS estimate is 0.06, and it is estimated with precision. The 2SLS estimate is 0.03, but it is insignificant, with an F-statistic at 76.50. The results of the baseline OLS and 2SLS estimations are in line with the IV literature on the effect of incarceration on recidivism \citep[see, e.g.][]{Loeffler2021Impact}.

\begin{table}[!htb]
	\caption{Effect of Imprisonment on Recidivism, 300}
	\label{tab:postahc}
	\begin{center}
%	\scriptsize
%	\hline
	%\input{../../Stata/postahc200.tex}
	\begin{tabularx}{0.6\textwidth}{cXXXXX}
          &\multicolumn{1}{c}{OLS}&\multicolumn{1}{c}{TSLS}&\multicolumn{1}{c}{1-2}&\multicolumn{1}{c}{1-3}&\multicolumn{1}{c}{2-3}\\

\midrule \multicolumn{6}{p{0.6\textwidth}}{\centering Panel A: Post Club Clustering}\\  \midrule D&     0.06&     0.03&      0.5&     -0.1&     0.03\\
          &   (0.01)&   (0.03)&    (0.3)&    (0.2)&    (0.1)\\

J         &         &         &       20&       11&       15\\
N         &    11219&    11219&     4444&     2259&     3373\\
P         &         &     0.00&     0.00&     0.49&     0.00\\
F         &         &    76.50&    47.20&    52.38&   113.70\\
\end{tabularx}

	\begin{tabularx}{0.6\textwidth}{cXXXXX}

\midrule \multicolumn{6}{p{0.6\textwidth}}{\centering Panel B: Post Valid IV Selection} \\ \midrule D&         &         &      0.7&     -0.1&     0.08\\
          &         &         &    (0.3)&    (0.2)&    (0.1)\\

J         &         &         &       19&       11&       14\\
N         &         &         &     4288&     2259&     3217\\
P         &         &         &     0.01&     0.49&     0.03\\
F         &         &         &    47.68&    52.38&   114.63\\
\hline 
%\multicolumn{6}{p{0.6\textwidth}}{Note: Sample restricted so that minimum number of cases per judge is 300. Standard errors are clustered by Offender-ID. Significance level in testing procedure: 0.1/log(N). D: coefficient for LATE estimate when comparing judge clubs or groups as detailed in top line. Panel A: Results after club-clustering, without valid IV selection, Panel B: Results after valid IV selection. J: Number of judges, N: number of observations, P: p-value of Hansen-test from an over-identified specification.} 
\end{tabularx}

	\end{center}
\textit{Note: Sample restricted so that minimum number of cases per judge is 300. Standard errors are clustered by Offender-ID. Significance level in testing procedure: 0.1/log(N). D: coefficient for LATE estimate when comparing judge clubs or groups as detailed in top line. Panel A: Results after club-clustering, without valid IV selection, Panel B: Results after valid IV selection. J: Number of judges, N: number of observations, P: p-value of Hansen-test from an over-identified specification.}
\end{table}

\noindent The three further columns in Table \ref{tab:postahc} show the estimates for the pairwise comparisons between the three clubs. In Panel A, we do not run an IV selection but use all of the judges within the clubs to build an IV that compares two clubs. The coefficients range from -0.1 to 0.5. Using comparison 1-2 as an example, this means that the subpopulation of individuals sent to jail under a judge from club 1 but who would not be sent to jail under a judge from club 2 has a 50 percentage point higher probability of committing another crime. These effects are not statistically significant. The number of judges involved here lies between 11 and 20. F-statistics are mostly high and range between 47 and 114. A red flag is that even though we found different estimates with relevant first stages, two of the three tests of overidentifying restrictions still reject clearly, with p-values close to zero for two of the three comparisons. 

\noindent %{\color{red} \sout{If we look for the largest group of reduced form coefficients and reduce the number of IVs further via the second-step AHC, the first-stage F-statistics are still in a similar range as before.} 
In Panel B, we report the estimates after selecting the valid IVs via the second-step AHC. Only in one of the three clubs has a judge been selected as invalid. The largest group comprises 11 of 12 judges, and the second-largest is therefore a singleton, lending credibility to the plurality selection. Even after the selection of valid IVs, the first-stage F-statistics are still in a similar range as before. The Hansen-Sargan tests increase for the two cases where they were very close to 0. The estimates now range from -0.1 to 0.7. The estimate for the comparison between club 1 and club 2 is 0.7 and it is now significant, suggesting that there is indeed a very strong effect of imprisonment on recidivism.\footnote{Note that we have used sample-splitting, selecting clubs and groups on one part of the sample and estimating on the other one.} 

\noindent To check whether this selection outcome is typical, we have resampled and reestimated 10,000 times and find that most judges are selected as valid in more than 90\% of the repetitions. The results can be found in Table \ref{tab:app_resamples} in the appendix. The one judge that is selected as invalid in our application is also the one with the lowest frequency of being selected as valid across samples (64.1\%). We read this as supporting the robustness of our analysis.

\begin{figure}[!htbp]
	\begin{center}
		\includegraphics[scale=0.75]{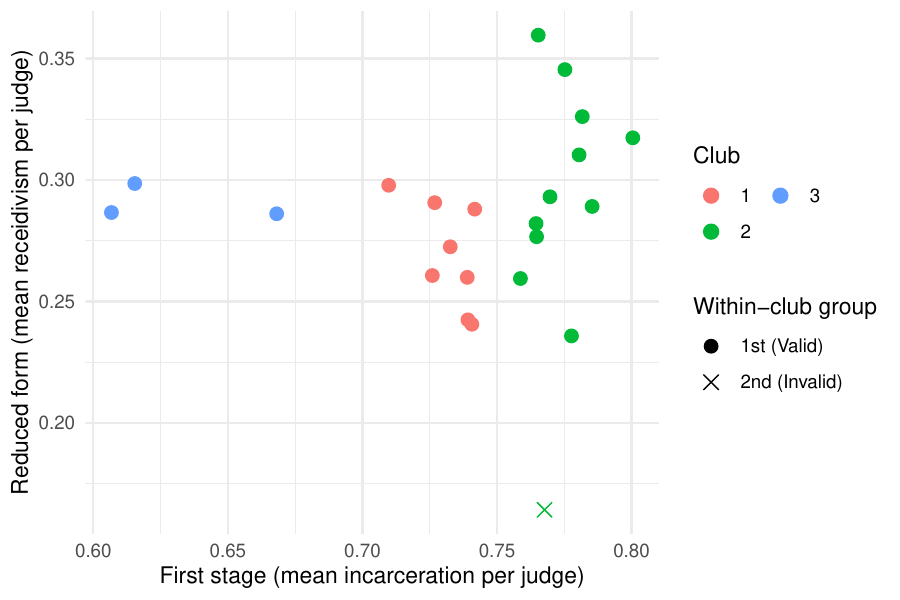}
	\caption{Reduced-form clustering}\label{fig:App_RF_300}
	\end{center}
%	\\
	\textit{Note: Minimum number of cases per judge: 300. On vertical axis: mean outcomes per judge. Crosses denote judges selected as invalid, dots denote the selected validity set. Different colors denote different clubs. }
\end{figure}

\noindent The first-stage clustering and the validity set selection via clustering inside each club is presented in Figure \ref{fig:App_RF_300}. We can see three clusters of propensity scores, with the lowest and smallest being most clearly separated from the other two. Judge-specific propensity scores, ordered by mean propensity score, are also shown in Figure \ref{fig:App_FS_300} in the Appendix. In club 2, one judge is selected as invalid, and their reduced form appears to be a clear outlier compared to the $r_z$ of the remaining judges.
%The first-stage clustering result for the six clubs is visualized in figure \ref{fig:App_FS_300}. By visual inspection we can deduct that there are jumps in the propensity scores, suggesting the existence of clubs and clubs are more clearly separated as mean propensities become smaller. The validity set selection via clustering inside each club is presented in figure \ref{fig:App_RF_200}. In the third club from the left and the two rightmost clubs, multiple judges are selected as invalid. 

\noindent The key takeaway of this application is that our method can discover different LATEs from a large set of IVs and provides a list of LATEs that would otherwise be collapsed into a single 2SLS estimate. Using the second-step AHC, we find subsets of IVs in the club-pairs that seem more likely to fulfill the exclusion restriction. 
% Maybe: if we look at the entire map -> loss of power.
Using the first-step AHC, we can find different clubs of propensity scores which yield very high first-stage F-statistics. Whether there is a monotonously increasing effect of the disaggregation of the 2SLS estimate into several LATEs on the first-stage F-statistic is unclear and is left for future research. 
Moreover, one might think about even more flexible ways to control for observables, such as through random forests or other machine learning methods.
% takeaways

%\subsection{Additional analyses}

%In the Appendix, we have added two additional exercises to this application. First, we have repeated the analysis, but now looking at judges with at least 200 cases. When doing this, we find five clusters, of which one singleton cluster (Table \ref{tab:fs-200} and Figure \ref{fig:App_FS_200}). Invalid judges are now found in two of the clubs (\ref{fig:App_RF_200}). When inspecting pairwise LATEs, qualitative results are similar to before in that the positive 2SLS estimate seems to be driven by a few positive, large and significant LATEs. In section \ref{sec:Subset}, we tackle the problem of controlling for observed covariates in an alternative way, by considering a subset of the data that is similar in terms of observables. 

%\FloatBarrier 
\section{Conclusion}\label{sec:Conclusion}

\noindent In this paper, we proposed a method based on hierarchical clustering for (i) estimating grouped LATEs under the condition that multiple instruments have the same firsts stage effect on the treatment and (ii) detecting instruments that satisfy the LATE assumptions under the plurality condition that those assumptions hold for a relative majority of instruments with the same first stage. The method first groups IV values based on their treatment propensity scores given the instrument and then groups the reduced form estimates (the mean outcomes given the instruments) within groups of propensity scores to determine the largest cluster of instruments as the one satisfying the LATE assumptions. This approach consistently selects clusters of identical first stages (and thus, complier groups) and valid instruments to estimate the grouped LATEs under a plurality condition. Simulation results suggested our method’s decent finite sample performance, even under a relatively moderate number of observations per instrument and with deviations from the club assumptions. 

\noindent Finally, we applied the procedure to assess the heterogeneous effects of incarceration on recidivism when using randomly assigned judges as (binary) instruments, where it appears likely that the LATE assumptions fail for multiple judge-pairs.

\noindent A direction for future work would be to apply the methodology to data that includes judge-level administrative records. Access to such data would help interpret the sources of heterogeneity in the LATEs (across complier groups) and lend additional credibility to the identified clubs and invalid judges. Professional characteristics such as experience or tenure, or demographic attributes such as age or race, may drive systematic diﬀerences across judges. If these are correlated with the judges' incarceration behavior, our procedure may partly capture those diﬀerences rather than pure heterogeneity in LATEs. With data providing such information, one could partial out observable judge characteristics prior to clustering, as we did for case-level characteristics. This would allow researchers to assess whether the recovered club structure genuinely reflects heterogeneity in LATEs or instead mirrors observable diﬀerences across judges. 

\noindent Our method is proposed for discrete or discretizable IVs and hence, it could be extended to settings with multiple continuous instruments, as in Mendelian randomization, where genetic markers serve as potential instruments.
\newpage

% Next steps:
% Assumptions on first-stage
% Show that true clustering is on path
% Conditions for correct selection of K for IC
% Conditions for correct selection of K for F-test
	
\renewcommand\appendix{\par
	\setcounter{section}{0}%
	\setcounter{table}{0}%
	\setcounter{figure}{0}%
	\renewcommand\thesection{\Alph{section}}%
	\renewcommand\thetable{\Alph{section}.\arabic{table}}}
\renewcommand\thefigure{\Alph{section}.\arabic{figure}}
\clearpage

\begin{appendix}
	\numberwithin{equation}{section}
	\noindent \textbf{\LARGE Appendices - For online publication only}
	
	\section{Methodological Appendix}

\subsection{Covariates}\label{app:covariates}
\noindent In our exposition, we have abstracted from covariates. There are multiple possibilities to control for covariates, and we will assume that instrument values can be ordered in a meaningful way conditional on $X$. Therefore, \( z \geq z' \) implies, for example, a weakly higher judge stringency of judge \( z \) relative to judge \( z' \) conditional on $X$. 

\noindent First, we can restrict the dataset to observations with similar values of $X=x$ so that LATEs are now conditional on $X$ and identify
\begin{equation}
	\Delta_{z,z'}(x)=E[Y(1)-Y(0)|v\geq  V > v', X=x].
\end{equation}
\noindent Our method can then be applied to subsets of the data. We follow this approach in an additional analysis in appendix \ref{sec:Subset}.

%A second approach would be to cluster on the predicted values of the propensity score and reduced form equation directly instead of using the first-stage coefficients, hence finding groups of $E(D| Z=z,X=x)$, so that homogeneity of these fitted values can be due to both $X$ or $Z$. To use this approach, however, an extension of the stopping rule is not straightforward. For example, we could create dummies based on the clusters but then it is not clear how to test that individual-specific propensity scores at a certain step of the clustering algorithm are indeed equal. An alternative would be to evaluate at each step via an information criterion. We leave this for future work.

\noindent Second, we can residualize $D$ and $Y$ with respect to observed factors $X$ using linear regressions and then use the same clustering approach as explained before. In this way, we can tackle confounding due to $X$. We are effectively estimating the regression $D_i=\mathbf{Z}'_i\bm{\pi}+ \mathbf{X}_i'\bm{\theta} + \varepsilon_i$ and clustering on $\hat{\bm{\pi}}$. We perform the analogous clustering approach with the reduced form estimates conditional on $X$: $Y_i=\mathbf{Z}'_i\bm{\Gamma}+ \mathbf{X}_i'\bm{\psi} + u_i$. This approach allows the identification of equal propensity scores once controlling for $X$. %One caveat of this approach is that we impose linearity with respect to $X$ in the first stage and the reduced form. 
In this case, the Wald estimand simplifies to $\frac{E(Y| Z=z_1, \,\,X=x)-E(Y| Z=z_2, \,\, X=x)}{\Pr(D=1| Z=z_1, \,\, X=x)-\Pr(D=1| Z=z_2, \,\, X=x)}=\frac{\Gamma_1-\Gamma_2}{\pi_1-\pi_2} = \frac{\Delta\Gamma}{\Delta\pi}$, where $\Gamma_1$ and $\Gamma_2$ (and $\pi_1$ and $\pi_2$ accordingly) are reduced form means, conditional on $X$, for two binary IVs 1 and 2 and  $\Delta\Gamma$ and $\Delta\pi$ are the differences between these means. These can be estimated via the reduced form and the first stage coefficients when taking one of the IVs as the reference category.

\subsection{Comparison with literature}\label{app:lit}

In this section we discuss differences between our approach and the ones taken by the previous literature on valid IV selection in \citet{Kang2016Instrumental} and subsequent papers as well as \citet[SW]{Sun2025Pairwise}.

\noindent In the introduction we have already mentioned the key difference between the setting considered in \citet{Kang2016Instrumental}, \citet{Windmeijer2019Use}, \citet{Guo2018Confidence} or \citet{Apfel2021Agglomerative} and the approach taken here. These papers consider a case with a homogeneous treatment effect. It is interesting to consider in which case these two approaches coincide to better understand how they are linked. If there are only two clubs with a single judge 1 in the basis category and all other judges $j\in \{2,...,J\}$ in the second club, such that $\pi_1 \neq \pi_j\,\, \forall j$ so that $\Delta\pi$ is homogeneous for all pairs, we can select valid and invalid IVs based on the plurality assumption via $\frac{\Delta\Gamma}{\Delta\pi}$, or directly via $\Delta\Gamma$. Only in such a special case with $K^0=2$ are the approaches from the previous literature directly applicable to our setting with clubs. However, note that the first-stage assumption in this setting depends on the correct choice of the basis judge (from the singleton club) and there are $J-1$ incorrect choices (from the second, larger club) and therefore this example is more interesting as a pedagogical construct and less as a practical tool. Arguably, such a structure is implausible in real data applications.%In the judge example it would be implausible to assume such a club structure with all judges having the same stringency and only one being an outlier. 

\noindent One advantage that the homogeneous treatment effect literature has over our approach is that it can be extended to the multiple regressor case as \citet{Apfel2021Agglomerative} demonstrate. The current paper uses a somewhat more specific setting where we consider a single discrete or discretizable IV. The method is still applicable if combinations of IV values of two or more discrete IVs define a set of binary IVs. 
%It is clear that for the ``traditional'' approach to work, this setting is restrictive: this very specific club structure must be given and comparisons with the reference judge cannot be systematically invalid.

\noindent Next, we discuss differences between our approach and the one taken by \citet[SW]{Sun2025Pairwise}. We argue that the two approaches use some of the same language but consider different settings. The conditions that they use are different and they lead to different answers. SW use the testable conditions for instrument validity from \citet{Kitagawa2015Test}, \citet{Mourifie2017Testing} and \citet{Sun2023Instrument} to discard invalid IVs. 

\noindent In SW, the first-stage condition (\ref{ass:FirstStage} in our paper) is upheld for all pairs of IV values, as per their Assumption 3.2. This condition can be quite strong, as the number of pairs considered grows quadratically in $J$. In our context, this first-stage assumption is assumed to be violated for at least one pair, when the number of clubs is lower than the number of judges. This violation of the first-stage assumption creates a non-singleton club in our setting such that $K^0 < J$. See also the discussion below assumption \ref{ass:Clubs}.

\noindent The high-dimensionality that arises from the large number of pairs is therefore addressed differently in the two approaches. SW aggregate all non-discarded pair-specific estimates using researcher specified weights, while our paper looks at club-pair estimates, preserving some of the heterogeneity while avoiding estimation of pair-specific LATE estimates with only few observations. We could also aggregate the club-pair LATEs to one parameter when a set of weights is chosen. 

\noindent Moreover, the conditions that SW use are the conditions from \citet{Kitagawa2015Test}, \citet{Mourifie2017Testing} and \citet{Sun2023Instrument}. These conditions are necessary but not sufficient as also mentioned in SW. \citet{Kitagawa2015Test} also discusses this, and states it in part i) of his proposition 1.1. There are two difficulties with this approach: first, it could be that some of the retained pairs are in fact invalid and second, the (asymptotic) power of this type of test typically decreases when the fraction of compliers increases, meaning that when the first stage is strong, the power of the test can become low, leading to a situation where neither low nor high strength provide a useful setting. Hence, the performance of methods that rely on testing these necessary conditions, including methods such as \citet{Huber2015Testing} or \citet{Farbmacher2022Instrument} can be somewhat limited in practice. In contrast, our method uses sufficient conditions for validity in a setting with clubs. When plurality holds and we concentrate on comparing the largest groups, the involved pairs are valid. The caveat of our approach then is that we may miss some valid IV value pairs where the values do not fall into the largest groups. 

\noindent Finally, in line with the different approaches used, they are concerned with different objects: SW relies on consistent selection of the largest validity \textit{pair} set, while this paper relies on consistent selection of the validity \textit{value} set. 

\noindent Taken together, these differences suggest that our method and SW are complementary rather than competing. The two approaches are well-suited to different empirical settings, and the choice between them can in practice be informed by the output of our first-stage clustering. When the clustering procedure detects a moderate number of clubs, that is, when subsets of IV values share similar propensity scores, our method is appropriate, since the club structure provides the leverage needed to exploit the relaxed first-stage condition and to identify the validity value set under plurality. Conversely, when the clustering procedure produces predominantly small or even singleton clubs, our framework has little structure to exploit, and SW's pairwise approach becomes the natural alternative. In intermediate cases, both methods may be informative, with our procedure delivering club-pair-specific LATEs that preserve heterogeneity across distinct complier groups, and SW providing a single aggregated estimate using researcher-specified weights.

\subsection{Proofs}\label{sec:proofs}

\subsubsection{Lemma \ref{lemma:SameClubFirst} and its proof}\label{sec:SameClubFirst}
We establish a Lemma from which the correct assignment in Corollary \ref{coro:OnPath} follows. 
Its proof is very close to the one in \citet*{Apfel2021Agglomerative}, but we reproduce it here to suit the setting and notation of the first-stage propensity scores.

\begin{lemma}{(Clusters with propensity scores from the same club $C_k$ are merged first\label{lemma:SameClubFirst})}\\
	Let $q$ and $q'$ denote two clusters such that all propensity scores in the two clusters both satisfy $p_z \in C_k$ (i.e., the propensity scores in the two clusters belong to the same single club). Let $q''$ be a different cluster s.t. $\exists p_z: p_z \in \hat{C}_{q''}$ and $p_z \in C_{\ell}$ but $\ell \neq k$ (i.e., there is at least one propensity score in this cluster which does not belong to the club from which the previous two clusters come). Under Assumptions \ref{ass:Clubs} and \ref{ass:LLN}, clusters $q$ and $q'$ are merged with probability converging to 1, as $n\rightarrow\infty$ in the first $J-K^0$ steps of algorithm \ref{algo:ward}.
\end{lemma}

%\textcolor{red}{Check notation in proof and move to appendix}\\
\noindent We subsequently demonstrate on a high level that the probability that an estimated cluster $\hat{C}_q$ with elements from the true club $C_{k}$ is merged with another estimated  cluster with elements from the same club $\hat{C}_{k}$ goes to 1.
The proof closely follows the one in \citet*{Apfel2021Agglomerative} and is divided into three parts for proving the subsequent statements:

\begin{enumerate}
	\item The means of clusters that include propensity scores from the same club also converge to the same value as each propensity score in the club.
	\item The algorithm merges the two clusters that have minimal distance.
	\item Clusters with propensity scores from the same club have a distance of zero and clusters with propensity scores from more than one club have a non-zero distance, with the probability going to one.
\end{enumerate}

\begin{proof}
	\textit{Part 1}:
	Consider
	\begin{align*}
		\begin{split}
			z, z' \in C_k \,\quad %,\quad k \\
			z'' \in C_{\ell} \, %,\quad \ell,
			\quad k \neq \ell.\\
		\end{split}
	\end{align*}
	Under Assumptions \ref{ass:Clubs} and \ref{ass:LLN} (i), it follows that
	\begin{align}
		\begin{split}
			plim(\hat{p}_{z}) = plim(\hat{p}_{z'}) = p_k^0\\
			plim(\hat{p}_{z''}) = p_{\ell}^0
		\end{split}
	\end{align}
	Let $\hat{C}_q$ and $\hat{C}_{q'}$ be estimated clusters with propensity scores from the same club: $\hat{C}_q$, $\hat{C}_{q'} \subset C_{k}$ and $\hat{C}_{q''} \subset C_{\ell}$. Let $\bar{\hat{C}}_k$ denote the arithmetic mean of the propensity scores in an estimated cluster. 
	
	\begin{equation}\label{eq:conv}
		plim \,\, \bar{\hat{C}}_q = plim \frac{\sum\limits_{\hat{p}_{z} \in \hat{C}_q} \hat{p}_{z}}{|\hat{C}_q|} = \frac{|\hat{C}_q|p_k^0}{|\hat{C}_q|} = p_k^0 .
	\end{equation}
	
	\noindent
	\textit{Part 2:}
	Consider the case that the algorithm decides whether to merge two estimated clusters, $\hat{C}_q$ and $\hat{C}_{q'}$, containing propensity scores from the same club or to merge two estimated clusters containing propensity scores from more than one underlying club, $C_q$ and $C_{q''}$. The two closest clusters in terms of their weighted Euclidean distance are merged first. Hence, we need to consider the distances between  $\hat{C}_q$ and $\hat{C}_{q'}$,  $\hat{C}_q$ and $\hat{C}_{q''}$,  as well as $\hat{C}_{q'}$ and $\hat{C}_{q''}$.
	
	$\hat{C}_q$ is merged with a cluster with elements of its own club $C_{k}$ iff
	$\frac{|C_q||C_{q'}|}{|C_q| + |C_{q'}|}||\bar{\hat{C}}_q - \bar{\hat{C}}_{q'}||^2 < \frac{|C_q||C_{q''}|}{|C_q| + |C_{q''}|}||\bar{\hat{C}}_q - \bar{\hat{C}}_{q''}||^2$. Therefore, the following two are equivalent
	
	\begin{equation*}
		lim \, P (\hat{C}_q \cup \hat{C}_{q'} = \hat{C}_{q,q'} \subseteq C_k) = 1
	\end{equation*}
	\begin{equation}\label{eq:Lemma1ToShow}
		\Leftrightarrow \quad  lim \, P(\frac{|\hat{C}_q||\hat{C}_{q'}|}{|\hat{C}_q| + |\hat{C}_{q'}|}||\bar{\hat{C}}_q - \bar{\hat{C}}_{q'}||^2 < \frac{|\hat{C}_q||\hat{C}_{q''}|}{|\hat{C}_q| + |\hat{C}_{q''}|}||\bar{\hat{C}}_q - \bar{\hat{C}}_{q''}||^2) = 1
	\end{equation}
	where $\hat{C}_{q,q'}$ is the merged cluster. 
	
	\textit{Part 3}: Our objective is to prove \eqref{eq:Lemma1ToShow}. We can then prove $lim \, P(\frac{|\hat{C}_q||\hat{C}_{q'}|}{|\hat{C}_q| + |\hat{C}_{q'}|}||\bar{\hat{C}}_q - \bar{\hat{C}}_{q'}||^2 < \frac{|\hat{C}_{q'}||\hat{C}_{q''}|}{|\hat{C}_{q'}| + |\hat{C}_{q''}|}||\bar{\hat{C}}_{q'} - \bar{\hat{C}}_{q''}||^2) = 1$ by changing the subscripts. 
	First, define $a := \frac{|\hat{C}_q||\hat{C}_{q'}|}{|C_q| + |\hat{C}_{q'}|}||\bar{\hat{C}}_q - \bar{\hat{C}}_{q'}||^2$ , $b:=\frac{|\hat{C}_q||\hat{C}_{q''}|}{|\hat{C}_q| + |\hat{C}_{q''}|}||\bar{\hat{C}}_q - \bar{\hat{C}}_{q''}||^2$ and $c:=\frac{|C_q||C_{q''}|}{|C_q| + |C_{q''}|}(p_k^0-p_{\ell}^0)^2$. 
	Under \eqref{eq:conv}
	\begin{align*}
		\begin{split}
			plim(a) =  \bm{0}, \quad 	plim(b) = c\\
			%plim(\frac{|\mathcal{S}_j||\mathcal{S}_k|}{|\mathcal{S}_j| + |\mathcal{S}_k|}||\bar{\mathcal{S}}_j - \bar{\mathcal{S}}_k||^2 - \frac{|\mathcal{S}_j||\mathcal{S}_l|}{|\mathcal{S}_j| + |\mathcal{S}_l|}||\bar{\mathcal{S}}_j - \bar{\mathcal{S}}_l||^2)& =  - \frac{|\mathcal{S}_j||\mathcal{S}_l|}{|\mathcal{S}_j| + |\mathcal{S}_l|} (\mathbf{q}-\mathbf{r})' (\mathbf{q}-\mathbf{r})\\
		\end{split}
	\end{align*}
	\noindent The remainder of this proof is identical to the one in \citet*{Apfel2021Agglomerative} and it is therefore not repeated here. Note that a merge of two clusters with elements from the same club is possible only when clusters with elements from the same club are available, i.e. for the first $J-K^0$ steps of Algorithm \ref{algo:ward}. 
	
	\iffalse
	It needs to be shown that $\underset{n \rightarrow \infty}{\lim}\, P (a<b)=1$, which is obtained by a proof by contradiction, by showing that $\underset{n \rightarrow \infty}{\lim}\, P(b<a)\neq0$ entails a contradiction. To ease notation, $\lim$ stands for $\underset{n \rightarrow \infty}{\lim}$ in the subsequent discussion.
	\noindent
	By the definitions of convergence in probability, it follows that
	\begin{equation}\label{eq:plim-a}
		\lim \, P(a < \varepsilon) = 1
	\end{equation}
	and
	\begin{equation}\label{eq:plim-b-c}
		\lim \, P(|b-c|< \varepsilon)=1 \text{.}
	\end{equation}
	for any $\varepsilon$.
	Therefore, $\lim \, P(b<a)\neq0$ and $\lim \, P(a < \varepsilon)=1$ imply $\lim \, P(b<\varepsilon)\neq0$.
	
	Now, consider $\varepsilon < \frac{1}{2}c$. Then,
	\begin{equation}\label{eq:lemmaproofcontradict}
		\lim \, P(b<\frac{1}{2}c)\neq0
	\end{equation}
	Because of the absolute value of $b-c$, consider two cases, $b<c$ and $b>c$.
	If $b<c$, then $\lim \, P(c-b< \frac{1}{2}c)=1 \, \Leftrightarrow \, \lim \, P(c-b > \frac{1}{2}c)=0$.
	$\Rightarrow \, \lim \, P(b<\frac{1}{2}c)=0$, which contradicts \eqref{eq:lemmaproofcontradict}.
	If $b\geq c$, then $a<\varepsilon<\frac{1}{2}c<c \leq b$ and hence
	$\lim \, P(a<b)=1 \, \Leftrightarrow \, \lim \, P(b\leq a)=0$, which is a contradiction, too.
	\fi
\end{proof}

%\subsubsection{Proof of Corollary \ref{coro:OnPath}}

%Here, we prove that $\mathcal{C}^0$ is on the selection path, i.e. for some $K$, Algorithm \ref{algo:ward} assigns judges to their respective clubs. 

\subsubsection{Proof of Theorem \ref{th:ConsistentClassification}}

%Results in the inference section needs more explanations and discussions. For example, I personally would like to see Assumptions 6-8 stated in the  main text. I would also like to see a complete definition of the asymptotic variance of Theorem 2 and discussions about it. It would also be helpful if the authors could explain the critical value sequence $\epsilon_n$ used for
%the Algorithm 2 F test, as is specified in Theorem 1 and Corollary 2. I understand it's required for consistency of classification. But in Section 3.2 when Algorithm 2 was introduced, this notation is not introduced.

The proof of this theorem follows \citet{Apfel2021Agglomerative} closely but is reproduced here for completeness. 
\begin{proof}
	The proof of Theorem \ref{th:ConsistentClassification} is structured as follows:
	\begin{enumerate}
		\item We show that asymptotically the selection path generated by Algo. \ref{algo:ward} contains $\mathcal{C}^0$.
		\item We show that Algorithm \ref{algo:F-test} can recover $\mathcal{C}^0$ from the selection path of Algo. \ref{algo:ward}.
	\end{enumerate}
	\textit{Part 1} follows from Corollary \ref{coro:OnPath} directly.
	
	\noindent \textit{Part 2}: Firstly, we establish the properties of the F-test. 
	The F-test statistic is given by equation \ref{eq:Ftest} in the main text. The asymptotic F-statistic has the following properties: 
	\iffalse
	\begin{property}Properties of the F-statistic
		\begin{enumerate}
			\item When $\mathcal{I}=\hat{\mathcal{I}}$: $S(\hat{\bm{\theta}}_\mathcal{\hat{\mathcal{I}}}) \overset{d}{\rightarrow} \chi^2_{|\mathcal{J}| - |\mathcal{I}| - P}$
			\item When $\mathcal{I}\neq\hat{\mathcal{I}}$ with $|\mathcal{\hat{I}}| < |\mathcal{I}|$: $S(\hat{\bm{\theta}}_\mathcal{\hat{\mathcal{I}}})=O_p(n)$.
		\end{enumerate}
	\end{property}
	\fi
	
	\begin{property}Properties of the F-statistic\label{prop:Fstatistic}
		\begin{enumerate}
			\item For all $\mathcal{C}$ such that each $C_q$ belongs to a club from the true partition $\mathcal{C}^0$:\\ $(J-K)F(\hat{\mathbf{p}},\mathcal{C}) \overset{d}{\rightarrow} \chi^2(J-K)$
			\item For all $\mathcal{C}$ such that at least one $C_k$ belongs to more than one cluster from the true partition $\mathcal{C}^0$: $(J-K)F(\hat{\mathbf{p}},\mathcal{C})=O_p(n)$.
		\end{enumerate}
	\end{property}
	\noindent The first property follows from standard arguments. The second property can be seen as follows. Note that under the alternative hypothesis $(plim \,\mathbf{R}(\mathcal{C})\hat{\mathbf{p}} \neq \mathbf{0}_J)$, under Assumption 7 (i), because the hypothesis matrix computes differences between $\hat{p}_z$ belonging to different clubs and instead $\mathbf{R}(\mathcal{C})\hat{\mathbf{p}} \overset{p}{\rightarrow} \bm{\delta} \neq \bm{0}$, $s^2 \overset{p}{\rightarrow} \sigma^2$, $ \mathbf{Z}^T\mathbf{Z}/n = diag(\frac{n_1}{n},..., \frac{n_J}{n}) \overset{p}{\rightarrow} \mathbf{S}_{\mathbf{Z}^T\mathbf{Z}} = diag(\pi_1,..., \pi_J)$ so that 
	\begin{align}
		\begin{split}
			\frac{(J-K)F(\hat{\mathbf{p}},\mathcal{C})}{n}  & = (\mathbf{R}(\mathcal{C})\hat{\mathbf{p}}- \mathbf{0}_J)'\left[s^2\mathbf{R}(\mathcal{C})(\frac{\mathbf{Z}^T\mathbf{Z}}{n})^{-1} \mathbf{R}(\mathcal{C})' \right]^{-1}(\mathbf{R}(\mathcal{C})\hat{\mathbf{p}}- \mathbf{0}_J)\\  & \overset{p}{\rightarrow}\bm{\delta}^T\left[\sigma_0^2\mathbf{R}(\mathcal{C})( \mathbf{S}_{\mathbf{Z}'\mathbf{Z}})^{-1} \mathbf{R}(\mathcal{C})' \right]^{-1}\bm{\delta}
		\end{split}
	\end{align}
	With this, we can show that the procedure described in Algorithm \ref{algo:F-test} retrieves the club structure as $\xi_{n} \rightarrow \infty \text{, and } \xi_{n}=o(n)\text{ when } n \rightarrow \infty\text{.}$ The condition on the critical values is the same as in \citet{Andrews1999Consistent} and \citet{Windmeijer2021Confidence}. Let the number of clusters formed at some step of Algorithm \ref{algo:ward} be $K$ and the true number of clubs be $K^0$. The F-test is applied to the model selected by the largest cluster at each step in the following situations:
	\begin{enumerate}
		\item $1 \leq K < K^0$. For each of these steps, at least one cluster must be a mixture of different clubs.
		By Property \ref{prop:Fstatistic} and $\xi_{n}\rightarrow\infty$, we have 
		\begin{equation*}
			\underset{n \rightarrow \infty}{lim} P ((J-K)F(\hat{\mathbf{p}},{\mathcal{C}}) < \xi_{n}) = 0 \text{.}
		\end{equation*}
		In this case, the F-test would be rejected, and Algorithm \ref{algo:F-test} moves to the next step.
		\item $K = K^0$. By Corollary \ref{coro:OnPath}, we know that the $K$ clusters are associated with the $K^0$ clubs. Then, applying the F-test at this step would test the equality of propensity scores inside each club, hence, we have that 
		\begin{equation}\label{eq:Fpass}
			\underset{n \rightarrow \infty}{lim} P ((J-K)F(\hat{\mathbf{p}},{\mathcal{C}}) < \xi_{n}) = 1 \text{.}
		\end{equation}
		and Algorithm \ref{algo:F-test} selects the partition $\mathcal{C}^0$.
	\end{enumerate}
	To summarize,  as $n \rightarrow \infty$, at steps $1 \leq K < K^0$, Algorithm \ref{algo:F-test} moves towards step $K = K^0$ and, as it reaches that step, selects the correct partition. Combining \textit{Part 1} and \textit{Part 2}, we prove Theorem \ref{th:ConsistentClassification}.\end{proof}

		\subsubsection{Proof of Lemma \ref{lem:mu_asymnorm} \label{app:oracle_distribution}}
			
			\noindent To show the proof, we need to first introduce some additional notation and an assumption. 
			
			For each observation \(i\), construct the \(3K^0\)-vector\[\mathbf{x}_i = \bigl( Y_i V_{1i},\; D_i V_{1i},\; V_{1i},\; Y_i V_{2i},\; D_i V_{2i},\; V_{2i},\; \dots,\; Y_i V_{K^0 i},\; D_i V_{K^0 i},\; V_{K^0 i} \bigr)'\]\noindent the mean of which is\[\boldsymbol{\mu}_X = \bigl(r_{1}\pi_1,\; p_1\pi_1,\; \pi_1,\; r_2\pi_2,\; p_{2}\pi_2,\; \pi_2,\; \dots,\; r_{K^0}\pi_{K^0},\; p_{K^0}\pi_{K^0},\; \pi_{K^0} \bigr)',\]Let \(\bar{\mathbf{X}}_n = \frac1n\sum_{i=1}^n \mathbf{x}_i\) be the vector of full-sample averages, with components \[A_{Yk} = \frac1n\sum_{i=1}^n Y_i V_{ki},\quad A_{Dk} = \frac1n\sum_{i=1}^n D_i V_{ki},\quad B_k = \frac1n\sum_{i=1}^n V_{ki}.\]Thus \(\bar{\mathbf{X}}_n = (A_{Y1},A_{D1},B_1,\dots,A_{YK^0},A_{DK^0},B_{K^0})'\).\begin{assumption}\label{ass:multiCLT} (Multivariate Central Limit Theorem) The following holds:\[\sqrt{n}\bigl( \bar{\mathbf{X}}_n - \boldsymbol{\mu}_X \bigr) \xrightarrow{d} N(\mathbf{0},\mathbf{\Sigma}_X),\]where \(\mathbf{\Sigma}_X = \operatorname{Cov}(\mathbf{x}_i)\) is a finite covariance matrix.\end{assumption}\noindent This ensures that the building blocks of the ratios in $\tilde{\bm{\mu}}$ used further on are well-behaved. Note that the subsample means are ratios of the full-sample averages:\[\tilde{r}_k = \frac{A_{Yk}}{B_k},\qquad \tilde{p}_k = \frac{A_{Dk}}{B_k}.\]
			
							\begin{proof}
			Consider that \(\tilde{\boldsymbol{\mu}} = f(\bar{\mathbf{X}}_n)\) where \(f:\mathbb{R}^{3K^0}\to\mathbb{R}^{2K^0}\) applies the ratio of the first and third as well as the second and third element to each triple \((A_{Yk},A_{Dk},B_k)\). 
			The Jacobian matrix \(\mathbf{J}_f\) evaluated at \(\boldsymbol{\mu}_X\) is block‑diagonal:
			\[
			\mathbf{J}_f = \begin{pmatrix}
				\mathbf{J}_1 & \mathbf{0}_{2\times 3} & \cdots & \mathbf{0}_{2\times 3} \\
				\mathbf{0}_{2\times 3} & \mathbf{J}_2 & \cdots & \mathbf{0}_{2\times 3} \\
				\vdots & \vdots & \ddots & \vdots \\
				\mathbf{0}_{2\times 3} & \mathbf{0}_{2\times 3} & \cdots & \mathbf{J}_{K^0}
			\end{pmatrix} = \operatorname{diag}(\mathbf{J}_1,\mathbf{J}_2,\dots,\mathbf{J}_{K^0}),
			\]
			with each \(\mathbf{J}_k\) a \(2\times3\) matrix: $
			\mathbf{J}_k = \begin{pmatrix}
				\frac1{\pi_k} & 0 & -\frac{r_k}{\pi_k} \\[4pt]
				0 & \frac1{\pi_k} & -\frac{p_k}{\pi_k}
			\end{pmatrix}.
			$
			
			\noindent By the delta method,
			
			\[
			\sqrt{n}\bigl( \tilde{\boldsymbol{\mu}} - \boldsymbol{\mu} \bigr) \xrightarrow{d} N(0,\mathbf{\Sigma}_{\boldsymbol{\mu}}),
			\quad\text{where}\quad
			\mathbf{\Sigma}_{\boldsymbol{\mu}} = \mathbf{J}_f \mathbf{\Sigma}_X \mathbf{J}_f'.
			\]
			
			%	\paragraph{Structure of \(\Sigma_{\boldsymbol{\mu}}\)} 
			
			\noindent Partition \(\mathbf{\Sigma}_X\) into $k$-specific blocks so that for groups \(k\) and \(\ell\), \(\mathbf{\Sigma}_{k\ell}\) is the \(3\times3\) covariance matrix between \((Y_iV_{ki}, D_iV_{ki}, V_{ki})\) and \((Y_iV_{\ell i}, D_iV_{\ell i}, V_{\ell i})\). We will now show that $\mathbf{\Sigma}_\mu$ is block-diagonal by considering the blocks of $	\mathbf{\Sigma}_{\boldsymbol{\mu}}$, i.e. 
			$
			\mathbf{\Sigma}_{\boldsymbol{\mu},kl} = \bigl( \mathbf{J}_k \mathbf{\Sigma}_{k\ell} \mathbf{J}_\ell' \bigr).
			$
			\(\mathbf{J}_k\mathbf{\Sigma}_{k\ell}\mathbf{J}_\ell' = \mathbf{0}\) for \(k\neq\ell\), and for \(k=\ell\), the blocks $\mathbf{J}_k \mathbf{\Sigma}_{kk} \mathbf{J}_\ell'$ are equal to those given in the Lemma.
			
			\underline{\textit{1. Diagonal blocks (within a group)}}
			
			For a fixed group $k$, %let $V_i = \mathbbm{1}(Z_i\in V_{ki})$, $\pi = \mathbb{P}(V_{ki}=1)$, 	$r = \mathbb{E}[Y_i\mid V_{ki}=1]$, $p = \mathbb{E}[D_i\mid V_{ki}=1]$, 
			let $\sigma_{Yk}^2 = \operatorname{Var}(Y_i\mid V_{ki}=1)$, $\sigma_{Dk}^2 = \operatorname{Var}(D_i\mid V_{ki}=1)$, 
			$\sigma_{YDk} = \operatorname{Cov}(Y_i,D_i\mid V_{ki}=1)$. 
			Using the law of total expectation,
			\[
			\begin{aligned}
				\mathbb{E}[Y_iV_{ki}] &= r_k\pi_k,\quad 	\mathbb{E}[D_iV_{ki}]  = p_k\pi_k,\quad \\
				\mathbb{E}[Y_i^2V_{ki}] &= \mathbb{E}[Y_i^2\mid V_{ki}=1]\pi_k = (\sigma_{Yk}^2+r_k^2)\pi_k,\quad \\
				\mathbb{E}[Y_iD_iV_{ki}] &= \mathbb{E}[Y_iD_i\mid V_{ki}=1]\pi_k = (\sigma_{YDk}+r_kp_k)\pi_k.
			\end{aligned}
			\]
			Define the \(3\times3\) covariance matrix \(\mathbf{\Sigma}_{kk} = \operatorname{Cov}((Y_iV_{ki}, D_iV_{ki}, V_{ki})')\). Its entries are:
			\[
			\begin{aligned}
				\operatorname{Var}(Y_iV_{ki}) &= \mathbb{E}[Y_i^2V_{ki}] - (\mathbb{E}[Y_iV_{ki}])^2 = (\sigma_{Yk}^2+r_k^2)\pi_k - r_k^2\pi_k^2 \\
				&= \pi_k\sigma_{Yk}^2 + r_k^2\pi_k(1-\pi_k),\\[4pt]
				\operatorname{Var}(D_iV_{ki}) &= \pi_k\sigma_{Dk}^2 + p_k^2\pi_k(1-\pi_k),\\
				\operatorname{Var}(V_{ki}) &= \pi_k(1-\pi_k),\\[4pt]
				\operatorname{Cov}(Y_iV_{ki},D_iV_{ki}) &= \mathbb{E}[Y_iD_iV_{ki}] - \mathbb{E}[Y_iV_{ki}]\mathbb{E}[D_iV_{ki}] \\
				&= (\sigma_{YDk}+r_kp_k)\pi_k - (r_k\pi_k)(p_k\pi_k) = \pi_k\sigma_{YDk} + r_kp_k\pi_k(1-\pi_k),\\[4pt]
				\operatorname{Cov}(Y_iV_{ki},V_{ki}) &= \mathbb{E}[Y_iV_{ki}] - \mathbb{E}[Y_iV_{ki}]\mathbb{E}[V_{ki}] = r_k\pi_k - r_k\pi_k^2 = r_k\pi_k(1-\pi_k),\\[4pt]
				\operatorname{Cov}(D_iV_{ki},V_{ki}) &= p_k\pi_k(1-\pi_k).
			\end{aligned}
			\]
			\noindent	Now compute \(\mathbf{J}_k \mathbf{\Sigma}_{kk} \mathbf{J}'_k\). With \(\mathbf{J}_k = \begin{pmatrix} \mathbf{j}_Y' \\ \mathbf{j}_D' \end{pmatrix}\) where \(\mathbf{j}_Y' = (\frac{1}{\pi_k}, 0, -\frac{r_k}{\pi_k})\), \(\mathbf{j}_D' = (0, \frac{1}{\pi_k}, -\frac{p_k}{\pi_k})\):
			\[
			\begin{aligned}
				\mathbf{j}_Y' \mathbf{\Sigma}_{kk} \mathbf{j}_Y &= \frac{1}{\pi_k^2}\operatorname{Var}(Y_iV_{ki}) - 2\frac{r_k}{\pi_k^2}\operatorname{Cov}(Y_iV_{ki},V_{ki}) + \frac{r_k^2}{\pi_k^2}\operatorname{Var}(V_{ki}) \\
				&= \frac{1}{\pi_k^2}\bigl(\pi_k\sigma_{Yk}^2 + r_k^2\pi_k(1-\pi_k)\bigr) 
				- 2\frac{r_k}{\pi_k^2}\bigl(r_k\pi_k(1-\pi_k)\bigr)
				+ \frac{r_k^2}{\pi_k^2}\bigl(\pi_k(1-\pi_k)\bigr) \\
				&= \frac{\sigma_{Yk}^2}{\pi_k} + \frac{r_k^2(1-\pi_k)}{\pi_k}
				- 2\frac{r_k^2(1-\pi_k)}{\pi_k} + \frac{r_k^2(1-\pi_k)}{\pi_k} = \frac{\sigma_{Yk}^2}{\pi_k}.
			\end{aligned}
			\]
			Similarly,
			\[
			\mathbf{j}_D' \mathbf{\Sigma}_{kk} \mathbf{j}_D = \frac{\sigma_{Dk}^2}{\pi_k},\qquad
			\mathbf{j}_Y'\mathbf{ \Sigma}_{kk} \mathbf{j}_D = \frac{\sigma_{YDk}}{\pi_k}.
			\]
			Thus the diagonal block is exactly \(\frac{1}{\pi_k}\mathbf{\Sigma}_{YD\mid V_{k}}\).
			
			\underline{\textit{2. Off‑diagonal blocks (between groups)}} 	
			Take two distinct groups \(k\) and \(\ell\) (\(k\neq\ell\)). %Define \(V = V_{ki}\), \(W = V_{\ell i}\). 	
			Let \(\mathbf{\Sigma}_{k\ell} = \operatorname{Cov}\bigl((Y_iV_{ki}, D_iV_{ki}, V_{ki})',\,(Y_iV_{\ell i}, D_i V_{\ell i}, V_{\ell i})'\bigr)\).  
			Because \(V_{ki}V_{\ell i}=0\) for every observation, every product between a component from group \(k\) and a component from group \(\ell\) that contains the factor \(V_{k i} V_{\ell i}=0\) is zero. Hence  
			\[
			\mathbb{E}\bigl[(Y_iV_{ki}, D_iV_{ki}, V_{ki})'\,(Y_iV_{\ell i}, D_iV_{\ell i}, V_{\ell i})\bigr] = \mathbf{0}_{3\times3}.
			\]
			Consequently, the covariance matrix is  
			$
			\mathbf{\Sigma}_{k\ell} = -\,\mathbf{a}_k\,\mathbf{a}_\ell',
			$
			where  
			\[
			\mathbf{a}_k = \bigl(\mathbb{E}[Y_iV_{ki}],\; \mathbb{E}[D_iV_{ki}],\; \mathbb{E}[V_{ki}]\bigr)' = (r_k\pi_k,\; p_k\pi_k,\; \pi_k)',\,\,
			\text{ and }	\mathbf{a}_\ell = (r_\ell\pi_\ell,\; p_\ell\pi_\ell,\; \pi_\ell)'.
			\]
			Now compute the contribution to the asymptotic covariance matrix of \(\hat{\boldsymbol{\mu}}\):
			\[
			\mathbf{J}_k\mathbf{\Sigma}_{k\ell}\mathbf{J}_\ell' = -\, (\mathbf{J}_k\mathbf{a}_k)(\mathbf{a}_\ell' \mathbf{J}_\ell').
			\]
			Using the definition of \(\mathbf{J}_k\),
			\[
			\mathbf{J}_k\mathbf{a}_k = \begin{pmatrix}
				\frac1{\pi_k} & 0 & -\frac{r_k}{\pi_k} \\[4pt]
				0 & \frac1{\pi_k} & -\frac{p_k}{\pi_k}
			\end{pmatrix}
			\begin{pmatrix}
				r_k\pi_k \\ p_k\pi_k \\ \pi_k
			\end{pmatrix}
			= \begin{pmatrix}
				r_k - r_k \\ p_k - p_k
			\end{pmatrix}
			= \mathbf{0}_{2\times1}.
			\]
			Similarly,  
			$
			\mathbf{a}_\ell' \mathbf{J}_\ell' = \mathbf{0}_{1\times2}.
			$
			Thus \(\mathbf{J}_k\mathbf{\Sigma}_{k\ell}\mathbf{J}_\ell' = \mathbf{0}_{2\times2}\) for every \(k\neq\ell\). The lower‑left block \(\mathbf{J}_\ell\mathbf{\Sigma}_{\ell k}\mathbf{J}_k'\) is zero by symmetry. 
			Hence, the off‑diagonal blocks vanish 
			and \(\mathbf{\Sigma}_{\boldsymbol{\mu}}\) is block‑diagonal:
			\[
			\mathbf{\Sigma}_{\boldsymbol{\mu}} = \operatorname{diag}\left( \frac{1}{\pi_1}\mathbf{\Sigma}_{YD\mid V_1},\;
			\frac{1}{\pi_2}\mathbf{\Sigma}_{YD\mid V_2},\; \dots,\;
			\frac{1}{\pi_{K^0}}\mathbf{\Sigma}_{YD\mid V_{K^0}} \right),
			\]
			matching the expression in equation \ref{eq:symmetricmu}.
		\end{proof}	
	
	\subsubsection{Proof of Proposition \ref{prop:oracle_asymnorm}}\label{app:oracle_asymnorm}

	\begin{proof} We are considering the pairs \((k,\ell)\)  with $k<\ell$, so that the vector $\hat{\bm{\beta}}^{or}$ has dimension \(Q= \binom{K^0}{2}\). 
	The function \(\mathbf{h}:\mathbb{R}^{2K^0}\to\mathbb{R}^{Q}\) is defined by stacking the Wald ratios. Thus
	\[
	\mathbf{h}(\boldsymbol{\mu}) = \bigl( h_{12}(\boldsymbol{\mu}),\; h_{13}(\boldsymbol{\mu}),\; \dots,\; h_{1K^0}(\boldsymbol{\mu}),\; h_{23}(\boldsymbol{\mu}),\; \dots,\; h_{(K^0-1)K^0}(\boldsymbol{\mu}) \bigr)',
	\]
	where each component is
	$
	h_{k\ell}(\boldsymbol{\mu}) = \frac{r_{k} - r_{\ell}}{p_{k} - p_{\ell}} =\beta_{k\ell}, \text{ with } 1\le k<\ell\le K^0.
	$
	%	In particular, the first element is \(h_{12}(\boldsymbol{\mu}) = \dfrac{\mu_{Y1} - \mu_{Y2}}{\mu_{D1} - \mu_{D2}}\) 	and the \(M\)-th (last) element is \(h_{(K^0-1)K^0}(\boldsymbol{\mu}) = \dfrac{\mu_{Y,K^0-1} - \mu_{Y,K^0}}{\mu_{D,K^0-1} - \mu_{D,K^0}}\).
	
	\noindent The partial derivatives are:
	\begin{equation}\label{eq:partialderivatives}
		\frac{\partial h_{k\ell}}{\partial r_{k}} = \frac{1}{\Delta_{k\ell}},\qquad
		\frac{\partial h_{k\ell}}{\partial p_{k}} = -\frac{\beta_{k\ell}}{\Delta_{k\ell}},\qquad
		\frac{\partial h_{k\ell}}{\partial r_{\ell}} = -\frac{1}{\Delta_{k\ell}},\qquad
		\frac{\partial h_{k\ell}}{\partial p_{\ell}} = \frac{\beta_{k\ell}}{\Delta_{k\ell}}.
	\end{equation}
	where \(\Delta_{k\ell} = p_{k}-p_{\ell} > 0\). All other partial derivatives of the Wald ratios with respect to \(r_{j}\) or \(p_{j}\) for \(j\notin\{k,\ell\}\) are zero. 
	For example, the Jacobian matrix \(\nabla\mathbf{h}(\boldsymbol{\mu})\) (\(Q\times 2K^0\)) when $K^0=4$, is:
	
	% Explicit example for K^0=4 (no dots, full matrix)
	\begin{equation}\label{eq:examplematrix}
\nabla\mathbf{h}(\boldsymbol{\mu}) =
	\begin{pmatrix}
		\frac{1}{\Delta_{12}} & -\frac{\beta_{12}}{\Delta_{12}} & -\frac{1}{\Delta_{12}} & \frac{\beta_{12}}{\Delta_{12}} & 0 & 0 & 0 & 0 \\[6pt]
		\frac{1}{\Delta_{13}} & -\frac{\beta_{13}}{\Delta_{13}} & 0 & 0 & -\frac{1}{\Delta_{13}} & \frac{\beta_{13}}{\Delta_{13}} & 0 & 0 \\[6pt]
		\frac{1}{\Delta_{14}} & -\frac{\beta_{14}}{\Delta_{14}} & 0 & 0 & 0 & 0 & -\frac{1}{\Delta_{14}} & \frac{\beta_{14}}{\Delta_{14}} \\[6pt]
		0 & 0 & \frac{1}{\Delta_{23}} & -\frac{\beta_{23}}{\Delta_{23}} & -\frac{1}{\Delta_{23}} & \frac{\beta_{23}}{\Delta_{23}} & 0 & 0 \\[6pt]
		0 & 0 & \frac{1}{\Delta_{24}} & -\frac{\beta_{24}}{\Delta_{24}} & 0 & 0 & -\frac{1}{\Delta_{24}} & \frac{\beta_{24}}{\Delta_{24}} \\[6pt]
		0 & 0 & 0 & 0 & \frac{1}{\Delta_{34}} & -\frac{\beta_{34}}{\Delta_{34}} & -\frac{1}{\Delta_{34}} & \frac{\beta_{34}}{\Delta_{34}}
	\end{pmatrix}.
	\end{equation}
	By the multivariate delta method, and by Lemma \ref{lem:mu_asymnorm}
	\[
	\sqrt{n}\bigl( \hat{\boldsymbol{\beta}}^{\text{or}} - \boldsymbol{\beta} \bigr) \xrightarrow{d} N(0,\mathbf{\Omega}_{\text{or}}),
	\]
	where \(\mathbf{\Omega}_{\text{or}} = \nabla\mathbf{h}(\boldsymbol{\mu}) \mathbf{\Sigma}_{\boldsymbol{\mu}}  \nabla\mathbf{h}(\boldsymbol{\mu})'\).% and \(\nabla\mathbf{h}(\boldsymbol{\mu})\) is the Jacobian evaluated at \(\boldsymbol{\mu}\). 
	%\paragraph{Explicit form of \(\Omega_{\text{or}}\)}
	Now we  want to look at the explicit entries of the asymptotic covariance matrix \(\mathbf{\Omega}_{\text{or}} = \nabla\mathbf{h}(\boldsymbol{\mu})\,\mathbf{\Sigma}_{\boldsymbol{\mu}}\,\nabla\mathbf{h}(\boldsymbol{\mu})'\). We can use Lemma \ref{lem:mu_asymnorm} to retrieve the specific shape of $\bm\Sigma_{\mu}$.
	
	\underline{\textit{1. Variance of $\hat{\beta}_{k,\ell}$:}}
	We start by considering the variances of single estimators. 
	Since \(\nabla h_{k\ell}\) (representing a row of $\mathbf{\nabla h}$) only has support for the \(2\times2\) blocks for groups \(k\) and \(\ell\), we can write
	\begin{equation}\label{eq:avar-deriv}
		\operatorname{AVar}\bigl(\sqrt{n}\,\hat\beta_{k\ell}^{\text{or}}\bigr) = \mathbf{v}_k' \left(\frac{1}{\pi_k}\mathbf{\Sigma}_{YD\mid V_k}\right) \mathbf{v}_k + \mathbf{v}_\ell' \left(\frac{1}{\pi_\ell}\mathbf{\Sigma}_{YD\mid V_\ell}\right) \mathbf{v}_\ell,
	\end{equation}
	where
	$
	\mathbf{v}_k = \left(1/\Delta_{k\ell}, \, -\beta_{k\ell}/\Delta_{k\ell}\right)'$ and 
	$\mathbf{v}_\ell =\left( -1/\Delta_{k\ell}, \, \beta_{k\ell}/\Delta_{k\ell}\right)'$.
	
	\noindent Now compute each quadratic form. For group \(k\):
	\[
	\mathbf{v}_k' \mathbf{\Sigma}_{YD\mid V_k} \mathbf{v}_k = \frac{1}{\Delta_{k\ell}^2}\operatorname{Var}(Y\mid V_k) - 2\frac{\beta_{k\ell}}{\Delta_{k\ell}^2}\operatorname{Cov}(Y,D\mid V_k) + \frac{\beta_{k\ell}^2}{\Delta_{k\ell}^2}\operatorname{Var}(D\mid V_k).
	\]
	This equals \(\frac{1}{\Delta_{k\ell}^2}\operatorname{Var}(Y - \beta_{k\ell}D\mid V_k)\). We get the same result for group \(\ell\), changing the subscript. 
	Thus, plugging into \ref{eq:avar-deriv}, we get
	\[
	\operatorname{AVar}\bigl(\sqrt{n}\,\hat\beta_{k\ell}^{\text{or}}\bigr)
	= \frac{1}{\Delta_{k\ell}^2}\left( \frac{\operatorname{Var}(Y-\beta_{k\ell}D\mid V_k)}{\pi_k} + \frac{\operatorname{Var}(Y-\beta_{k\ell}D\mid V_\ell)}{\pi_\ell} \right).
	\]

\underline{	\textit{2. Covariance between two estimators sharing one club:} }
	Consider two ordered pairs \((k,\ell)\) and \((k',\ell')\) where one group index is the same, i.e. $k=k'$ or $\ell=\ell'$; i.e. one club is shared and it is in the same position. %Remember that $\sigma(a,b)$ is the sign function of the difference $b-a$.
	%	so that \(\sigma(a,b)=+1\) if \(a<b\) and \(\sigma(a,b)=-1\) if \(a>b\). 
	%For the pair \((k,\ell)\), the gradient of \(\hat\beta_{k\ell}^{\text{or}}\) with respect to \(r_k\), \(p_k\), \(r_\ell\) and \(p_\ell\) is
	%\[\frac{\partial h_{k\ell}}{\partial r_k} = \frac{1}{\Delta_{k\ell}},\qquad\frac{\partial h_{k\ell}}{\partial p_k} = -\frac{\beta_{k\ell}}{\Delta_{k\ell}}, \qquad 	\frac{\partial h_{k\ell}}{\partial r_\ell} = -\frac{1}{\Delta_{k\ell}},\qquad	\frac{\partial h_{k\ell}}{\partial p_l} = \frac{\beta_{k\ell}}{\Delta_{k\ell}}.\]
	%Analogous expressions hold for \(h_{k\ell'}\). 
	%Because \(\mathbf{\Sigma}_{\boldsymbol{\mu}}\) is block‑diagonal, and the $\nabla h$ vectors only have support for blocks associated with their indices, 
	The asymptotic covariance then is
	\begin{align}\label{eq:asycov}
		\operatorname{ACov}\bigl(\sqrt{n}\hat\beta_{k\ell}^{\text{or}},\sqrt{n}\,\hat\beta_{k\ell'}^{\text{or}}\bigr)
		& = \nabla h_{k\ell} \;\mathbf{\Sigma}_{\boldsymbol{\mu}}\; \nabla h_{k\ell'}' \\ \nonumber
		& =  \left( \frac{1}{\Delta_{k\ell}},\; -\frac{\beta_{k\ell}}{\Delta_{k\ell}} \right)
		\left( \frac{1}{\pi_k}\mathbf{\Sigma}_{YD\mid V_k} \right)
		\left( \frac{1}{\Delta_{k\ell'}},\; -\frac{\beta_{k\ell'}}{\Delta_{k\ell'}} \right)'\\ \nonumber
		& 	= \frac{1 }{\pi_k \Delta_{k\ell} \Delta_{k\ell'}} \;
		\begin{pmatrix} 1 & -\beta_{k\ell} \end{pmatrix}
		\mathbf{\Sigma}_{YD\mid V_k}
		\begin{pmatrix} 1 & -\beta_{k\ell'} \end{pmatrix}'\text{.}
	\end{align}
	The quadratic form equals \(\operatorname{Cov}\bigl(Y - \beta_{k\ell}D,\; Y - \beta_{k\ell'}D \mid V_k\bigr)\). Therefore,
	\[
	\operatorname{ACov}\bigl(\sqrt{n}\hat\beta_{k\ell}^{\text{or}},\, \sqrt{n}\hat\beta_{k\ell'}^{\text{or}}\bigr)
	= \frac{1}{\Delta_{k\ell}\Delta_{k\ell'}} \cdot
	\frac{\operatorname{Cov}\bigl(Y - \beta_{k\ell}D,\; Y - \beta_{k\ell'}D \mid V_k\bigr)}{\pi_k}.
	\]
	The same expression multiplied with $-1$ holds for the case where the shared club is not in the same position, e.g. $k=\ell'$, because one of the vectors appearing in the multiplication in \ref{eq:asycov} containing partial derivatives appears with flipped signs.\footnote{For example, referring to  the example matrix in \ref{eq:examplematrix}, such a case would occur when we would consider the covariance between the $\hat{\beta}_{12}$ and $\hat{\beta}_{23}$, and we would need to consider the first and fourth row of that matrix in \ref{eq:asycov}.}
	%	The product \(\sigma(k,\ell)\sigma(k,\ell')\) equals \(+1\) if both \(\ell>k\) and \(\ell'>k\) or both \(\ell<k\) and \(\ell'<k\)), and \(-1\) otherwise. This means that the shared club occupies the same position in both pairs. 
	
\underline{	\textit{3. Covariance between two estimators that share no clubs:} }
	Consider the case where there is no shared club, $s=\emptyset$, i.e. all indices $k,\,\ell, \, k', \ell'$ are distinct. The gradient rows \(\nabla h_{k\ell}\) and \(\nabla h_{k'\ell'}\) have non‑zero entries corresponding to the blocks for groups \(\{k,\ell\}\) and \(\{k',\ell'\}\), respectively. Because these sets are disjoint, the product \(\nabla h_{k\ell} \,\mathbf{\Sigma}_{\boldsymbol{\mu}}\, \nabla h_{k'\ell'}'\) always involves multiplication with zeroes and hence 
	%involves only off‑diagonal blocks of \(\mathbf{\Sigma}_{\boldsymbol{\mu}}\) between distinct groups. Since \(\mathbf{\Sigma}_{\boldsymbol{\mu}}\) is block‑diagonal (off‑diagonal blocks are zero), this product equals zero. Hence
	\[
	\operatorname{ACov}\bigl(\sqrt{n}\hat\beta_{k\ell}^{\text{or}},\,\sqrt{n}\hat\beta_{k'\ell'}^{\text{or}}\bigr)=0.
	\]
	Hence, we have shown the shape of the entries of the asymptotic variance-covariance matrix and this concludes the proof.
\end{proof}

\subsubsection{Proof of Theorem \ref{th:gpiv_asymnorm}}\label{app:gpiv_asymnorm}

\begin{proof}
	\noindent Define the event
	$
	\mathcal{E}_n = \bigl\{ \hat{\mathcal{C}} = \mathcal{C}^0,\; \hat{\mathcal{R}}^{M} = \mathcal{V}\ \bigr\}.
	$
	By consistent selection, \(\mathbb{P}(\mathcal{E}_n) \to 1\). For the event \(\mathcal{E}_n\), the GPIV estimator coincides with the oracle estimator $\tilde{\bm{\beta}}^{or}$ as defined in the text. %which for a specific group comparison is
	%	\[
	%		\hat{\beta}_{k\ell}^{\text{or}} = \frac{\tilde{r}_k - \tilde{r}_{\ell}}{\tilde{p}_k - \tilde{p}_{\ell}} \, ,
	%		\]
	%and the vector of oracle Wald estimators, for the set of ordered pairs $(k,\ell)$ with $k<\ell$ is 
	%\begin{equation}\label{eq:oracle}
	%	\hat{\bm{\beta}}^{or} = \left( \hat{\beta}^{or}_{1 2},\hat{\beta}^{or}_{13}, ..., \hat{\beta}^{or}_{(K^0-1)K^0} \right)' \, .
	%\end{equation}
	On the complement \(\mathcal{E}_n^c\) we may define \(\hat{\boldsymbol{\beta}}^{\text{GPIV}}\) arbitrarily (e.g., as the zero vector); this choice does not affect the asymptotic distribution because \(\mathbb{P}(\mathcal{E}_n^c)\to 0\).
	
	\noindent Decompose the scaled difference as
	\begin{equation}\label{eq:decomposegpiv}
		\sqrt{n}\bigl( \hat{\boldsymbol{\beta}}^{\text{GPIV}} - \boldsymbol{\beta} \bigr)
		= \sqrt{n}\bigl( \tilde{\boldsymbol{\beta}}^{\text{or}} - \boldsymbol{\beta} \bigr) \mathbf{1}_{\mathcal{E}_n}
		\;+\; \sqrt{n}\bigl( \hat{\boldsymbol{\beta}}^{\text{GPIV}} - \boldsymbol{\beta} \bigr) \mathbf{1}_{\mathcal{E}_n^c} \, .
	\end{equation}
	
	\noindent The first term: \(\sqrt{n}(\tilde{\boldsymbol{\beta}}^{\text{or}} - \boldsymbol{\beta}) \xrightarrow{d} N(0,\mathbf{\Omega}_{\text{or}})\) by Proposition \ref{prop:oracle_asymnorm} in Appendix \ref{app:oracle_distribution}, and \(\mathbf{1}_{\mathcal{E}_n} \xrightarrow{p} 1\). By Slutsky’s theorem, $\sqrt{n}\bigl( \tilde{\boldsymbol{\beta}}^{\text{or}} - \boldsymbol{\beta} \bigr) \mathbf{1}_{\mathcal{E}_n} \xrightarrow{d} N(0,\mathbf{\Omega}_{\text{or}})$.
	
	\noindent Due to consistent selection, the second term is $o_p(1)$, because for any \(\epsilon>0\),
	\[
	\mathbb{P}\bigl( \|\sqrt{n}(\hat{\boldsymbol{\beta}}^{\text{GPIV}} - \boldsymbol{\beta})\mathbf{1}_{\mathcal{E}_n^c}\| > \epsilon \bigr)
	\le \mathbb{P}(\mathcal{E}_n^c) \to 0 \, .
	\]
	%The first inequality holds because the first event (in the first probability) can only occur when \(\mathbf{1}_{\mathcal{E}_n^c}=1\), i.e. when $\mathcal{E}_n^c$ is true. 
	Therefore,
	\[
	\sqrt{n}\bigl( \hat{\boldsymbol{\beta}}^{\text{GPIV}} - \boldsymbol{\beta} \bigr) \xrightarrow{d} N(0,\mathbf{\Omega}_{\text{or}}) \, .
	\]
	
\end{proof}

\subsection{Identical complier effects} \label{app:identical_compliers}

%We point out that methodologically, our approach is somewhat different from existing studies aiming at selecting valid instruments, which pre-select relevant instruments and where the first-stage relationship affects group membership. In contrast, our approach relies on first forming clubs with a homogeneous first-stage association in order to detect violations by investigating the heterogeneity in reduced form parameters within clubs. Comparing between clubs, we automatically have a difference in propensity scores and hence relevance is given. Moreover, the first-stage does not affect group membership, which is solely defined in terms of the reduced form parameter, in contrast to studies invo\king parametric assumptions when selecting valid instruments; such as \citet{Kang2016Instrumental}, \citet{Guo2018Confidence} and \citet{Windmeijer2021Confidence}.

The identification of an identical complier effect under either pair of instrument values is driven by the fact that the satisfaction of $\psi(z)=\psi(z^*)$ for any pair $z,z^*$ implies that $\Pr(D=1| Z=z)=\Pr(D=1| Z=z^*)$. This is the case because for any value $z$, $\Pr(D=1| Z=z)$ is the probability of the event $(\psi(z)\geq V)$: $\Pr(D=1| Z=z)=\Pr(\psi(Z)\geq V|Z=z)=\Pr(\psi(z)\geq V)$, where the first equation follows from equation \eqref{eq:vyt_equivalence} and the second from the independence of $D(z)$ and $Z$ implied by Assumption \ref{ass:Validity}. The converse holds as well, i.e., $\Pr(D=1| Z=z)=\Pr(D=1| Z=z^*)$ implies that $\psi(z)=\psi(z^*)$. To see this, let us normalize $V$ to only take values between 0 and 1. Rather than considering the unobservable $V$, we take its cumulative distribution function (cdf), denoted by $F_V$, which is bounded between 0 and 1. Formally, $F_V\sim Unif[0,1]$. As noticed in the literature on marginal treatment effects, see, e.g.,\ \citet{HeckVytlacil00} and \citet{HeVy05}, this normalization is innocuous in the sense that it does not affect the results of our treatment model postulated in Equation \eqref{eq:vyt_equivalence}. The latter can, without loss of generality, be reparametrized in the following way when expressing the treatment as a function of $F_V$ rather than $V$:
\begin{eqnarray}\label{eq:FirstStage}
	D=\mathbbm{1}(F_V(\psi(Z))\geq F_V)),
\end{eqnarray}
where $F_V(\psi(Z))$ is the cdf of $V$ evaluated at the value $\psi(Z)$. By the definition of a cdf and treatment equation \eqref{eq:vyt_equivalence}, $F_V(\psi(Z))=\Pr(V\leq \psi(Z))=\Pr(D=1|Z)$, which is the treatment propensity score. For this reason, there exists a one-to-one correspondence between $\psi(z)=\psi(z^*)$ and $\Pr(D=1|Z=z)=\Pr(D=1|Z=z^*)$. This implies that distinct pairs of instruments entailing identical pairs of upper and lower propensity scores necessarily identify the LATE for the very same complier group (i.e., with the same distribution of unobserved characteristics), if Assumptions \ref{ass:Validity} to \ref{ass:FirstStage} hold.

\subsection{Violations of the LATE assumptions - bias of Wald estimand} \label{app:violation_bias}

\noindent To refine our approach such that it allows for the possibility that some instruments violate any of assumptions \ref{ass:Validity} and \ref{ass:Monotonicity} (henceforth ``the LATE assumptions''), based on the previous insight that all IVs in the same club-pair which satisfy the LATE assumptions yield the same LATE. To this end, we exploit that for any pairs of IVs with the same propensity scores, the denominator in \eqref{Wald} is equal. This, in turn, implies that by Theorem \ref{th:MainResult}, the reduced form effect in the numerator in \eqref{Wald} is equal, too, if the instruments satisfy the LATE assumptions and thus, have the same Wald estimand. More formally, for any  pairs of instruments $z, z'$ and $z^*, z''$ satisfying the LATE assumptions and $p_z=p_{z^*}$ and $p_{z'}=p_{z''}$ such that $p_z-p_{z'}=p_{z^*}-p_{z''}=c$, with some abuse of notation, it holds that
\begin{align}\label{eq:ViolationsViaRF}
	&&\frac{E(Y | Z=z) - E(Y| Z=z')}{c} = \frac{E(Y | Z=z^*) - E(Y| Z=z'')}{c} \\ \nonumber
	&\Leftrightarrow& E(Y | Z=z) - E(Y| Z=z') = E(Y | Z=z^*) - E(Y| Z=z'')
\end{align}
\noindent Hence, using any of the judge-specific means of the outcome $E(Y | Z=z) := r_z$ within the same pair of clubs must yield the same reduced form effect. We maintain relevance, given that existence of the first stage ($c\neq0$) can be tested. 

%\iffalse
%To illustrate this argument, let us consider direct effects of the judges on the outcomes, such that the exclusion restriction is violated. Our argument also holds for other violations of the LATE Assumptions.
%\begin{equation}\label{eq:LATEInvalid}
%	Y = \eta(D,Z_j,U)
%\end{equation}
%which means that a judge $j$ has a direct effect on the outcome (or also through an unobservable, where $U$ is a function of $Z$).
%\fi

%First, we want to better understand what happens in presence of violations of the LATE assumptions.
\noindent In many empirical applications, however, the assumption that all IVs satisfy the LATE assumptions might be challenged. Let us, for instance, assume that a pair of judges violates the exclusion restriction postulated in Assumption \ref{ass:Validity}. In this case, one judge directly affects recidivism relative to another, for instance by using distinct alternative measures to imprisonment, such as electronic monitoring. This generally implies for a pair of judges $z, z'$  that the average direct effect of the instrument conditional on the compliance type $(D(z)=d,D(z')=d')$, denoted by $\gamma_{zz'}^{dd'}$, is non-zero:
\iffalse
\begin{equation}
	\gamma_{zz'}^{dd'}:=E[Y(d)|D(z)=d,D(z')=d',Z=z]-E[Y(d)|D(z)=d,D(z')=d',Z=z']\neq 0\textrm{ for }d,d' \in \{0,1\}.
\end{equation}
\fi

\begin{align}\label{eq:LATEInvalidity}
	\gamma_{zz'}^{dd'} & :=E[Y(d)|D(z)=d,D(z')=d',Z=z]\\ &-E[Y(d)|D(z)=d,D(z')=d',Z=z']\neq 0 \\ \nonumber
	&\textrm{ for }d,d' \in \{0,1\}.
\end{align}
\noindent In this case, the Wald estimand defined in \eqref{Wald} no longer identifies the LATE but is biased, as discussed in \citet{Huber2014Sensitivity}, who demonstrates that under a violation of the exclusion restriction, the LATE corresponds to
\begin{align}
	\Delta_{z,z'} = \frac{E(Y | Z=z) - E(Y| Z=z') - p_{z'}\cdot \gamma_{zz'}^{11}-(1-p_{z})\cdot\gamma_{zz'}^{00}}{p_{z} - p_{z'}} - \gamma_{zz'}^{10} \text{.}
\end{align}
Therefore,
\begin{equation}
	\frac{E(Y | Z=z) - E(Y| Z=z')}{p_z - p_{z'} } = \Delta_{z,z'}  + \frac{p_{z'} \cdot \gamma_{zz'}^{11} + (1-p_{z})\cdot\gamma_{zz'}^{00}}{p_z - p_{z'}} + \gamma_{zz'}^{10},
\end{equation}
where the expression to the right of the LATE $\Delta_{z,z'}$ characterizes the bias of the Wald estimand. 
Such biases are not constrained to cases with violations of the exclusion restriction, but they generally arise under violations of any of the LATE assumptions. For instance, the bias occurring under a violation of the monotonicity assumption has been characterized in \citet{Angrist+96}.

\newpage
\subsection{Example: within-club plurality} \label{app:example}
%
%To better understand plurality in the context of our paper, assume there are eleven judges with different imprisonment and recidivism rates in the cases that they consider: the four judges 1 to 4 have imprisonment rates of 10 percent, judges 5 to 8 have an imprisonment rate of 40 percent and the remaining judges are strict judges with 80 percent imprisonment rate. Hence, there are three clubs. In club 1, two judges have a recidivism rate of 20 percent, one has a rate of 40 percent and one 50 percent. Hence, this club’s largest group of judges has the same recidivism rate. We might now also consider the largest groups inside the other two clubs. The plurality assumption says that judges who are part of the largest group inside of each club belong to the validity set, i.e., if we pair them together, they create valid IVs. Table \ref{tab:withinplurality} illustrates the within-club plurality with this example. The bold and underlined IV values in the leftmost column indicate the validity set, and those in the grey area belong to the largest sets inside each club. Hence, the plurality assumption holds when the set of bold and underlined values and the grey area are the same. It would not hold, for example, if some of the values in the grey area were not part of the validity set (not bold and underlined).
%

\begin{table}[ht!]
	
	\caption{Hypothetical example to illustrate within-club plurality\label{tab:withinplurality}}
	
	\begin{center}
		\begin{tabular}{ccc|c}
			\hline
			z & $p_z$ & c & $r_z$ \\
			\hline
			\rowcolor{Gray}
			\underline{\textbf{1}} & 0.1 & 1 & 0.2 \\
			\rowcolor{Gray}
						\underline{\textbf{2}} & 0.1 & 1 & 0.2 \\
			3 & 0.1 & 1 & 0.4 \\
			4 & 0.1 & 1 & 0.5 \\
			\hline
			5 & 0.4 & 2 & 0.2 \\
			\rowcolor{Gray}
			\underline{\textbf{6}} & 0.4 & 2 & 0.6 \\
			\rowcolor{Gray}
			\underline{\textbf{7}} & 0.4 & 2 & 0.6 \\
			8 & 0.4 & 2 & 0.7 \\
			\hline
			\rowcolor{Gray}
			\underline{\textbf{9}} & 0.8 & 3 & 0.4 \\
			\rowcolor{Gray}
			\underline{\textbf{10}} & 0.8 & 3 & 0.4 \\
			11 & 0.8 & 3 & 0.5 \\
			\hline
		\end{tabular}
	\end{center}
	
	\textit{\scriptsize Note: z stands for IV values, $p_z$ for imprisonment rate, $c$ for club-identifier and $r_z$ for recidivism rate. Underlined and bold values in $z$ belong to validity set, gray set shows the largest groups in terms of recidivism rates within each club.}
\end{table}

\clearpage
\subsection{Empirical method - estimation approach}

\begin{figure}[ht!]
	\caption{Illustration of the estimation approach\label{app:fig:Illustration}}
	\begin{center}
		\includegraphics[scale=0.23]{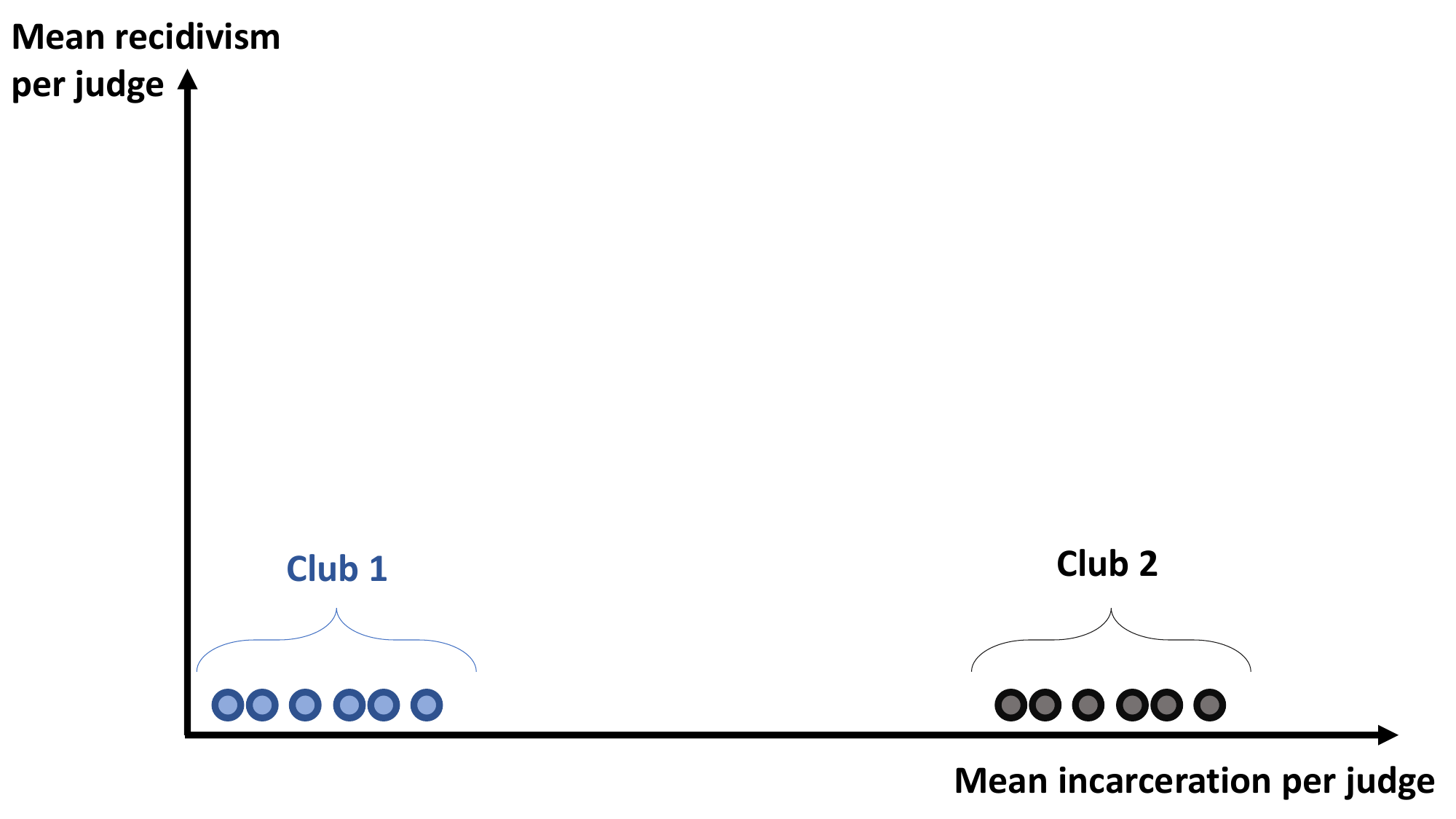}
		\includegraphics[scale=0.23]{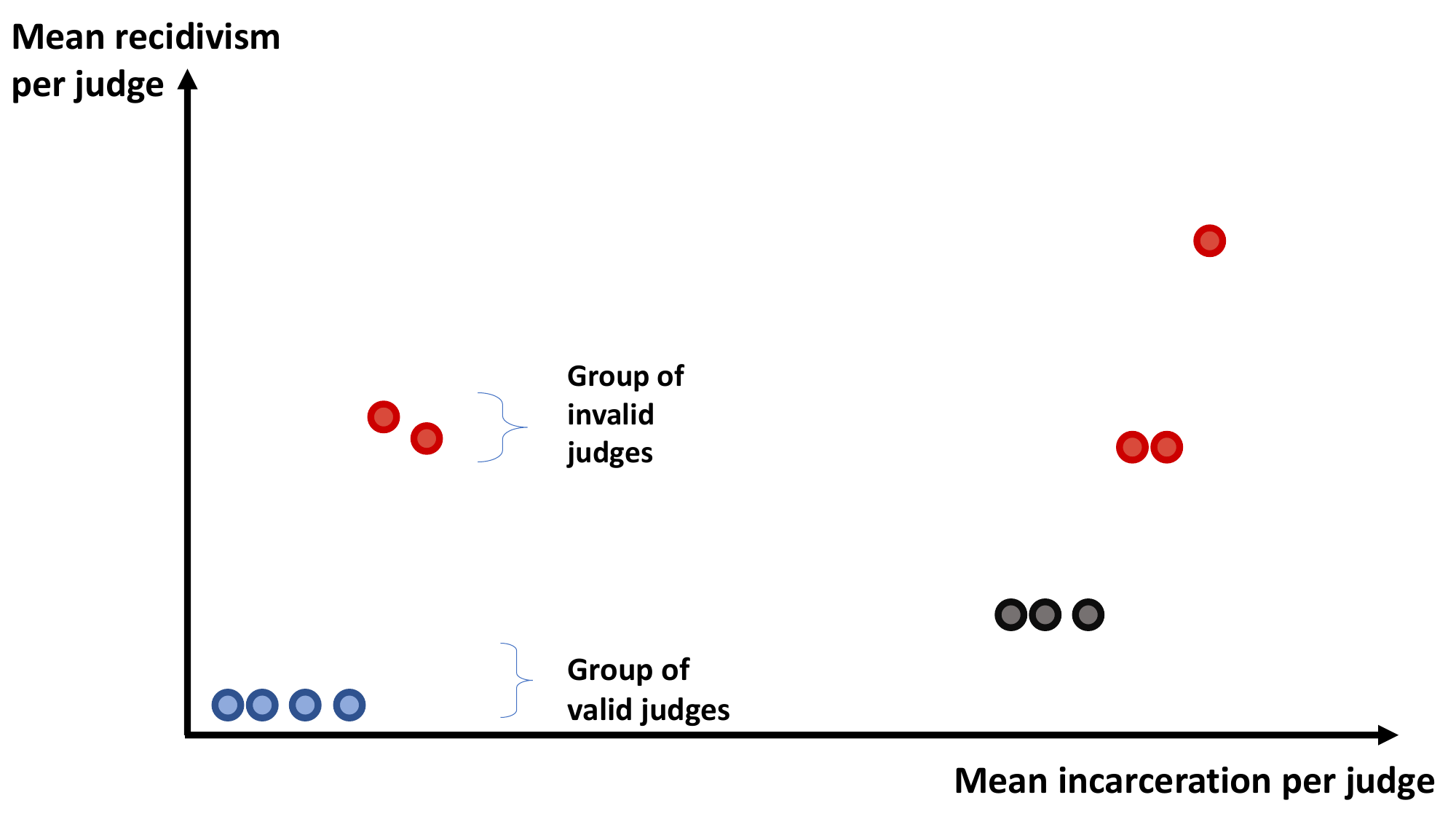}
	\end{center}
	
	\scriptsize
	\textit{Note:} The first step of the procedure (left graph) detects clubs with comparable propensity scores. This provides us with two clubs in the given example. The second step detects groups with comparable reduced form estimates of judge- (or IV-) specific mean outcomes and selects the largest groups within clubs as the validity set (right graph). After this two-step classification, pairing the largest groups across different clubs (or propensity score values) permits estimating the LATE.
\end{figure}

\newpage

	\section{CARDS}

\subsection{Improving finite-sample clustering via CARDS}

% Idea
% Estimation - formula
% LLA
% Key condition

\noindent The arguments that justify AHC to find clubs and groups to estimate treatment effects via our GPIV are valid in large samples. However, mistakes might occur in finite samples when classifying clubs and groups. We, therefore, propose to apply CARDS \citep*{Ke2015Homogeneity} to improve the finite sample behavior of our procedure. 
CARDS uses a pre-classification and a penalized regression with two penalty terms to find an optimal grouping. The advanced version of CARDS requires a preliminary segmentation.

\noindent Denote $L_{N T}(\boldsymbol{p})$ the quadratic loss function. 
The penalty function consists of within and between penalty
\begin{equation}
P_{B, \lambda_1, \lambda_2}(\boldsymbol{p})=\sum_{l=1}^{L-1} \sum_{i \in B_l, j \in B_{l+1}} scad_{\lambda_1}\left(\left\|\boldsymbol{p}_i-\boldsymbol{p}_j\right\|_1\right)+\sum_{l=1}^L \sum_{i \in B_l, j \in B_l} scad_{\lambda_2}\left(\left\|\boldsymbol{p}_i-\boldsymbol{p}_j\right\|_1\right)
\end{equation}
where $\mathcal{B} = \{B_1, ..., B_L\}$ is the preliminary segmentation and $scad_{\lambda}$ is the SCAD penalty function \citep*{Fan2001Variable}. We use AHC to find the preliminary segmentation. The overall goal function then is the sum of loss and penalty

\begin{equation}
Q_{N}(\boldsymbol{p})=L_{N}(\boldsymbol{p})+P_{B, \lambda_1, \lambda_2}(\boldsymbol{p})
\end{equation}

\noindent We minimize this function with respect to $\boldsymbol{p}$.\footnote{
To find the minimizer of this function, we apply the local linear approximation algorithm, as proposed in \citet*{Wang2018Homogeneity}.}
\iffalse
:

\begin{equation}
\hat{\boldsymbol{p}}^{(s+1)}=\arg \min \left\{L_{N}(\boldsymbol{p})+R\left(\hat{\boldsymbol{p}}^{(s)} ; \boldsymbol{p}\right)\right\}
\end{equation}

where 
\begin{align}
R\left(\hat{\boldsymbol{p}}^{(s)} ; \boldsymbol{p}\right)& =\sum_{l=1}^{L-1} \sum_{i \in B_l, j \in B_{l'}} p_{\lambda_1}^{\prime}\left(\left\|\hat{\boldsymbol{p}}_i^{(s)}-\hat{\boldsymbol{p}}_j^{(s)}\right\|_1\right)\left\|\boldsymbol{p}_i-\boldsymbol{p}_j\right\|_1 \\
& +\sum_{l=1}^L \sum_{i \in B_l, j \in B_{l}}, p_{\lambda_2}^{\prime}\left(\left\|\hat{\boldsymbol{p}}_i^{(s)}-\hat{\boldsymbol{p}}_j^{(s)}\right\|_1\right)\left\|\boldsymbol{p}_i-\boldsymbol{p}_j\right\|_1
\end{align}

where the first line denotes the between-segment penalty and the second line denotes the within-segment penalty, and $p'(\cdot)$ is the derivative of the SCAD function.
\fi
\noindent We then evaluate the information criterion 
$IC(\lambda)=\ln \left(\sigma_{N}^2(\lambda)\right)+ 0.5 \hat{K}(\lambda) (N)^{-1/2}$ at a fine grid of values $\boldsymbol{\lambda} = (\lambda_1, \lambda_2)$ and choose the combination of tuning parameters with the lowest value of $IC$. As an alternative procedure, we evaluate the F-test as in (\ref{eq:Ftest}) and choose the model that passes the test at a pre-specified significance level, with the minimal number of clubs. 
\begin{equation}
\boldsymbol{\lambda}_{F} =\{\boldsymbol{\lambda}: \mathcal{C}|(\boldsymbol{\lambda}_{pass})|=\underset{\boldsymbol{\lambda}_{pass}}{min}\mathcal{C}|(\boldsymbol{\lambda}_{pass})|\}  
\end{equation}
where 
$\boldsymbol{\lambda}_{pass} = \{\boldsymbol{\lambda}: p_F(\boldsymbol{\lambda}) > \alpha_F \}$. 

\noindent Theorem 6 in \citet*{Ke2015Homogeneity} and Theorem 2 in \citet*{Wang2018Homogeneity} state that if the preliminary segmentation is such that the classification algorithm wrongly assigns members to adjacent clubs, so that misclassification is not severe in this sense, and regularity conditions hold, the clustering algorithm finds the correct clubs with high probability. We expect that finite sample performance will improve through the combined use of AHC and CARDS. 

% But this argument is asymptotic as well!

%Again, we can test the strength of these unions in each first-stage regression. The estimates associated with club pairs that pass the first-stage test are then reported, along with their confidence intervals, the estimates of club pairs that do not pass the first-stage test can be discarded. In this way, we get a map of heterogeneous, club pair-specific first-stage coefficients and estimates.

% What to do to deal with invalidity

%To summarize, Stage 1 of Procedure 1 works as follows: We use the IVs, the treatment, outcome and controls as input. Then, we find the clubs of judges for which the propensity score is the same. We compute all possible pairs of clubs. We construct single IVs or unions of IVs from each club-pair. We compute the first-stage F-statistic and check if it exceeds a conventional threshold. Next, we can discard club-pairs which do not pass the first-stage test. Finally, report IV or 2SLS estimates for the remaining unions of IVs.
\newpage
\subsection{Simulation results - CARDS}\label{app:CARDS:sim}

\begin{table}[!htbp]
	\caption{Simulation: LATEs.}
	\label{tab:sim_lates_CARDS}
\fontsize{9}{11}\selectfont
\begin{tabularx}{\textwidth}{XXXXcccccc}
\hline
Setting & Inv. & Method  & Step & 1-2 & SE & 1-3 & SE & 2-3 & SE \\
\midrule
\multirow{2}{*}{20} & \multirow{6}{*}{no} & Oracle & \multirow{6}{*}{first} & 30.43 & 4.83 & 26.28 & 2.86 & 20.48 & 5.53 \\ %
& & CARDS & & 29.40 & 8.84 & 26.34 & 4.84 & 22.44 & 36.32 \\ %
\multirow{2}{*}{60} & & Oracle & & 30.28 & 2.75 & 26.19 & 1.62 & 20.23 & 2.75 \\
& & CARDS & & 30.71 & 7.18 & 26.46 & 2.43 & 20.17 & 7.01  \\
\multirow{2}{*}{100} & & Oracle & & 30.26 & 2.12 & 26.21 & 1.26 & 20.15 & 2.23 \\
& & CARDS & & 30.23 & 2.99 & 26.28 & 1.81 & 20.23 & 3.35 \\
\hline
\multirow{2}{*}{20} & \multirow{6}{*}{yes} & Oracle & \multirow{6}{*}{first} & 30.33 & 6.09 & 26.12 & 3.13 & 20.67 & 6.94 \\ %
& & CARDS & & 28.19 & 12.47 & 37.67 & 7.57 & 37.76 & 372.15  \\ %
\multirow{2}{*}{60} & & Oracle & & 30.10 & 3.43 & 26.03 & 1.80 & 20.14 & 3.57 \\
& & CARDS & & 28.10 & 6.60 & 35.96 & 3.31 & 52.27 & 11.19 \\
\multirow{2}{*}{100} & & Oracle & & 30.22 & 2.65 & 26.10 & 1.39 & 20.08 & 2.72 \\
& & CARDS & & 27.58 & 4.08 & 36.33 & 2.45 & 51.40 & 7.18 \\
\hline
\multirow{2}{*}{20} & \multirow{6}{*}{yes} & Oracle & \multirow{6}{*}{second} & 30.33 & 6.09 & 26.12 & 3.13 & 20.67 & 6.94 \\ %
& & CARDS & & 37.35 & 1092.56 & 21.4 & 38.6 & 34.76 & 262.40 \\ %
\multirow{2}{*}{60} & & Oracle & & 30.10 & 3.43 & 26.03 & 1.80 & 20.14 & 3.57 \\
& & CARDS & & 26.76 & 3.83 & 25.71 & 2.59 & 28.04 & 6.59 \\
\multirow{2}{*}{100} & & Oracle & & 30.22 & 2.65 & 26.10 & 1.39 & 20.08 & 2.72 \\
& & CARDS & & 28.31 & 3.35 & 25.94 & 1.99 & 23.02 & 4.37 \\
   \bottomrule
\end{tabularx}
\footnotesize \textit{Note:} Setting: number of cases per judge ($Unif(3,5) \times (20,60,100)$); Invalidity: no: $\gamma =  (0i_{10})$, yes: $\gamma =  (0.5,  0i_{3},  0i_{2},  0.4, 0.6,  0i_{2})$; Method: Oracle: estimator that uses group and club identity information, CARDS: AHC combined with a clustering algorithm in regression via data-driven segmentation; Stage: results after the first (detecting clubs with comparable propensity scores) or second (detecting clubs with comparable propensity scores and eliminating invalid judges) stage; 1-2, 1-3, 2-3: Group-wise LATE estimates of imprisonment on recidivism, using judge IVs. SE: robust standard errors
\end{table}

	\newpage
\section{Simulations}  \label{sim}

\noindent We investigate the finite sample properties of our clustering and testing procedure in a simulation study consisting of two settings, with and without invalid judge IVs, where invalidity refers to a violation of the IV exclusion restriction. We consider three different sample sizes to analyze consistency.

\subsection{Simulation design}

\noindent We analyze settings with unbalanced panels, with $J=10$ judges. To determine the number of cases per judge, we draw from a uniform distribution $Unif(3, 5)$, multiply this value by 20, 60, or 100, respectively, and round the number to the next integer. We repeat this approach for all $J$ judges and  construct dummy variables for each judge.  %: $Z_{ji} = 1$ if $j=k$,  otherwise $0$ for $k \in \{1, 2, ..., 30\}$.
Furthermore, we define the first-stage model for the treatment of an observation $i$ in the following way: 
$$D_i = \mathbbm{1}(\mathbf{z}_i\cdot \bm{\pi} > V_i)$$
where $\mathbf{Z}$ is the $(\sum_{z=1}^J n_z) \times J$ matrix of judge dummies and $\mathbf{z}_i$ is a row vector. $\bm{\pi} = (0.8 \,\, \mathbf{i}_4, 0.5 \,\, \mathbf{i}_4, 0.3 \,\, \mathbf{i}_2)'$ is the vector of first-stage coefficients (or propensity scores) of the judge dummies, where $\mathbf{i}_k$ is a $1\times k$ vector of ones. $V_i$ is the first-stage error and follows a uniform distribution, $V_i \sim Unif(0,1)$. 
In this setup, we create three clubs with 4, 4, and 2 different judges per club, respectively. 

\noindent To avoid the correlation between clustering and estimating the LATEs, we apply sample splitting. We divide the cases per judge into a training and a test set. We use the training set to cluster the judges into clubs and apply the second step of the procedure to detect invalid judges. We use the groupings from this step to estimate the group-pair-wise LATEs.
% \textcolor{red}{We split the cases per judge randomly, applying the clustering algorithms to one half of the cases and estimate the LATE based on the other half.}  

\noindent The outcome model is given by the nonlinear function
$$Y_i = (D_i \cdot 0.5 + D_i \cdot U_i +  \mathbf{Z} \gamma +  U_i)^4.$$ 
$U_i$ is the error in the outcome equation and modelled as $U_i = 0.5\cdot V_i + W_i$, where $W_i = Unif(0,1)$ is a uniform random variable which is independent of $V_i$. As the first-stage error affects the error in the outcome equation, the treatment is endogenous, which motivates our IV approach. $\bm{\gamma}$ is a vector of direct effects of the instruments on the outcome. Therefore, a nonzero entry $\bm{\gamma}$ violates the exclusion restriction for the related judge dummy and, thus, of one of the LATE assumptions. We define two different settings regarding the invalidity of the judges to identify the accuracy of the LATEs in both stages, before and after identifying the invalid IVs. In the first setting we choose $\bm{\gamma} = (0\bm{i}_{10})$. This allows us to investigate the club assignment in detail. In the second setting, we set $\bm{\gamma} =  (0.5,  0\bm{i}_{3},  0\bm{i}_{2},  0.4, 0.6,  0\bm{i}_{2})$, so that in club 1, three judges are valid and one is invalid and hence majority is still fulfilled, in club 2, two judges are invalid and two valid. Hence, plurality is fulfilled, but majority is violated, and in club 3, all judges are valid. While we focus on violations of the exclusion restriction to investigate the finite sample performance of our method, we notice that one could also consider violations of other LATE assumptions, namely treatment monotonicity or random IV assignment.

\noindent We report the following statistics for the simulations. Under $\#clubs$, we report the mean number of clubs, under $\#corr$ we report the fraction of times the correct number of clubs has been selected. We report the estimated LATEs for each pair comparison when the number of clubs selected is the true number of clubs ($1-2$, $1-3$, $2-3$). For comparison, we also report the true (or oracle) LATE. This fully informed estimator can be computed directly from our simulation’s data-generating process. We take the average of Hansen p-values for each repetition (over pair comparisons) and then average these means over repetitions ($Hansen$ $p$). We also calculate the probability ($Power$) of detecting a non-zero effect by testing the beta coefficients against zero. Respectively, we calculate the average coverage rate ($Cover$) by testing whether the estimated 95 percent confidence intervals contain the true LATE. We simulate the true LATE by calculating the mean of 5000 oracle estimates, estimated using data independent from the one used to illustrate the selection and estimation procedures. 

\noindent We report the mean normalized mutual information (NMI), which is an indicator for the quality of the clustering, as used in \citet{Ke2015Homogeneity} and \citet{Ana2003Robust}. The NMI is defined for the comparison of a clustering $\mathcal{C}$ and the oracle clustering $\mathcal{C}_0$:
$$\mathrm{NMI}(\mathcal{C}, \mathcal{C}_0)=\frac{I(\mathcal{C}; \mathcal{C}_0)}{[H(\mathcal{C})+H(\mathcal{C}_0)] / 2}$$
\noindent where $I(\mathcal{C} ; \mathcal{C}_0)=\sum_{k, j}\left(\left|C_k \cap C_{0j}\right| / \right) \log \left(\mid C_k \cap\right.\left.C_{0j}|/| C_k|| C_{0j} \mid\right)$ is the mutual information between the two clusterings and $H(\mathcal{C})=-\sum_k\left(\left|C_k\right| / \right) \log \left(\left|C_k\right| \right)$
is the entropy of $\mathcal{C}$. The NMI takes values between 0 and 1, with larger NMIs indicating more similar clusterings and a value of 1, meaning that they are equal. 

\noindent In a separate table, we report the results for the second-step AHC: the average of the fraction of valid IVs detected ($ValDet$) and invalid IVs detected ($InvDet$). Further, we also report the percentage of repetitions in which \textit{all} valid and invalid IVs were correctly identified ($CorVal$) and again $Hansen$ $p$, $Power$, $Cover$, and the estimated LATEs.

\subsection{Simulation results}

\noindent We first consider a setting without invalid IV dummies, implying that all entries of $\gamma$ in the outcome equation are equal to zero, and run 1000 Monte Carlo simulations. Table \ref{tab:2SLS_OLS} reports the mean OLS and 2SLS coefficient estimates and standard error for the three sample size settings. In the smallest sample size setting (20), the OLS estimate is 18.87, with 1.02 as a standard error. The 2SLS coefficient estimate is slightly higher at 20.31 and 1.48 as standard error. We observe almost the same coefficient estimates for the two larger sample settings, with considerably lower standard errors for OLS and 2SLS. The F-statistic on the instruments, in all settings, indicates that judge dummies are strong instruments for incarceration.

% Table 1: Mean OLS and 2SLS 
\begin{table}[ht!]
\caption{Mean OLS and 2SLS}
\label{tab:2SLS_OLS}
\begin{center}
\begin{tabular}{cccccc}
\hline
Setting & Method & Est & SE & F-statistic \\ \hline
\multirow{2}{*}{20} & OLS & 18.87 & 1.02 &  \\
 & 2SLS & 20.31 & 1.48 & 74.63 \\
\hline
\multirow{2}{*}{60} & OLS & 18.91 & 0.59 &  \\
 & 2SLS & 20.38 & 0.86 & 219.13 \\
 \hline
\multirow{2}{*}{100} & OLS & 18.90 & 0.46 &  \\
 & 2SLS & 20.38 & 0.67 & 363.87 \\
   \bottomrule

\end{tabular}
\end{center}
\footnotesize \textit{Note:} Setting: number of cases per judge ($Unif(3,5) \times (20,60,100)$); Method: OLS regression, two-stage least-squares regression; Est: estimates of incarceration on recidivism, using judge dummies for 2SLS; SE: robust standard errors; F-statistic: First-stage F-statistic, testing joint relevance of the instruments. 
\end{table}

\noindent Table \ref{tab:sim_first_step_class} shows the classification results after the first step of our method for both settings, with and without invalidity. In the first step, we aim at detecting clubs of treatment propensity scores based on Assumption \ref{ass:Clubs}. We estimate club membership based on our algorithms \ref{algo:ward} and \ref{algo:F-test}. For comparison, we also report the oracle LATE estimator. 
In Table \ref{tab:sim_first_step_class}, we first look at the setting without invalid IVs. Column 5 ($\#corr$) shows that using AHC, we find the correct number of clubs, respectively, in 33 percent, 95 percent, and 99 percent of the repetitions, based on the different sample sizes. Increasing the sample size greatly improves the clustering performance. The average of the Hansen p-value is considerably above any conventional significance level, indicating we cannot find evidence for invalid instruments. Therefore, the second classification step (based on the reduced form parameters) is unnecessary. The CI coverage is close to its nominal level for all sample sizes.

\noindent In the setting with invalid IVs, we observe a comparable performance when identifying the correct number of clubs. However, the Hansen p-value consistently falls below the 0.05 significance level, indicating a high frequency of rejections of the null hypothesis. Two possible reasons for rejecting the Hansen test are effect heterogeneity and invalidity. By classifying the judges into clubs with the same incarceration rate, we can rule out effect heterogeneity as a potential cause of a rejection of the Hansen test. Therefore, if it still rejects after classifying judges into groups, the only reason left for rejecting is the presence of invalid judges.

\begin{table}[!htbp]
	\caption{Simulation: First step - club allocation.}
	\label{tab:sim_first_step_class}
%\fontsize{7}{9}\selectfont
\begin{tabularx}{\textwidth}{ccXXXXXXX}
\hline
Setting & Inv. & Method & \#clubs & \#corr & Hansen p & Power & Cover & NMI \\
\midrule
\multirow{2}{*}{20} & \multirow{6}{*}{no} & Oracle & 3 & 1 & 0.50 & 0.99 & 0.95 & 1 \\ %
& & AHC & 2.33 & 0.33 & 0.49 & 0.87 & 0.94 & 0.66 \\ %
\multirow{2}{*}{60} & & Oracle & 3 & 1 & 0.49 & 1 & 0.95 & 1 \\
& & AHC & 2.97 & 0.95 & 0.50 & 0.99 & 0.95 & 0.95 \\
\multirow{2}{*}{100} & & Oracle & 3 & 1 & 0.49 & 1 & 0.95 & 1 \\
& & AHC & 3.01 & 0.99 & 0.49 & 1 & 0.95 & 0.99 \\
\hline
\multirow{2}{*}{20} & \multirow{6}{*}{yes} & Oracle & 3 & 1 & 0.01 & 0.99 & 0.38 %0.93 
& 1 \\
& & AHC & 2.34 & 0.34 & 0.05 & 0.82 & 0.70 & 0.66 \\
\multirow{2}{*}{60} & & Oracle & 3 & 1 & 0.00 & 1 & 0.27 %0.91 
& 1 \\
& & AHC & 2.98 & 0.96 & 0.00 & 0.99 & 0.33 & 0.96 \\
\multirow{2}{*}{100} & & Oracle & 3 & 1 & 0.00 & 1 & 0.24 %0.89 
& 1 \\
& & AHC & 3.02 & 0.98 & 0.00 & 1 & 0.29 & 0.99 \\
   \bottomrule

\end{tabularx}
\footnotesize \textit{Note:} Mean results after the first step of the procedure (detecting clubs with comparable propensity scores). Setting: number of cases per judge ($Unif(3,5) \times (20,60,100)$); Invalidity: no: $\gamma =  (0i_{10})$, yes: $\gamma =  (0.5,  0i_{3},  0i_{2},  0.4, 0.6,  0i_{2})$; Method: Oracle: estimator that uses group and club identity information, AHC: agglomerative hierarchical clustering; $\#clubs$: mean number of clubs; $\#corr$ percentage of iterations in which the correct number of clubs has been selected; Hansen p: mean of p-values of Hansen-test from an over-identified specification; Power: fraction of confidence intervals that do not include zero; Cover: fraction of confidence intervals containing the true effect (in iterations in which the correct number of clubs was found); NMI: normalized mutual information. 
\end{table}

\noindent To address invalidity, we apply the second step of our IV selection procedure. In Table \ref{tab:sim_second_step_class}, we present the classification results after our procedure's second stage, which is applied to identify and eliminate invalid judges from the estimation. We apply AHC in both procedure steps, using the classification results in Table \ref{tab:sim_first_step_class}. 

\noindent After applying the second step, Table \ref{tab:sim_second_step_class} shows that the average of the Hansen p-value is considerably above any conventional significance level for all methods, indicating we no longer find evidence of invalid instruments. The power is close to 1, depending on the sample sizes, while the Coverage is between 0.76 and 0.91 for AHC. Respectively, 97 to 99 percent of the valid IVs have been correctly detected ($ValDet$). The fraction of correctly identified invalid IVs ($InvDet$) is slightly lower at 69, 87, and 96 percent, respectively. In total, again, respectively, for the sample sizes, in 33, 68, and 90 percent of the repetitions, AHC identified all valid and invalid IVs correctly ($CorVal$). An NMI close to 1 also indicates that the correct grouping has been retrieved in many cases.

\begin{table}[!htbp]
	\caption{Simulation: First and second step - club allocation and identification of invalid judges.}
	\label{tab:sim_second_step_class}
%\fontsize{7}{9}\selectfont
\begin{tabularx}{\textwidth}{XXXXXXXX}
\hline
Setting & Method & Hansen p & Power & Cover & ValDet & InvDet & CorVal \\
\midrule
\multirow{2}{*}{20} & Oracle & 0.51 & 0.96 & 0.94 & 1 & 1 & 1 \\
& AHC & 0.30 & 0.72 & 0.76 & 0.97 & 0.69 & 0.33 \\
\multirow{2}{*}{60} & Oracle & 0.49 & 1 & 0.95 & 1 & 1 & 1 \\
& AHC & 0.42 & 0.96 & 0.82 & 0.97 & 0.87 & 0.68 \\
\multirow{2}{*}{100} & Oracle & 0.50 & 1 & 0.95 & 1 & 1 & 1 \\
& AHC & 0.48 & 0.99 & 0.91 & 0.99 & 0.96 & 0.90 \\
   \bottomrule
\end{tabularx}
\footnotesize \textit{Note:} Mean results after the first and second step of the procedure (detecting clubs with comparable propensity scores and eliminating invalid judges) for $\gamma =  (0.5,  0i_{3},  0i_{2},  0.4, 0.6,  0i_{2})$. Setting: number of cases per judge ($Unif(3,5) \times (20,60,100)$); Method: Oracle: estimator that uses group and club identity information, AHC: agglomerative hierarchical clustering; Hansen p: mean of p-values of Hansen-test from an over-identified specification; Power: fraction of confidence intervals that do not include zero; Cover: fraction of confidence intervals containing the true effect (in iterations in which the correct number of clubs was found); ValDet: fraction of valid IVs selected as valid; InvDet: fraction of detected, invalid IVs; CorVal: percentage of iterations in which all valid and invalid IVs were identified correctly. 
\end{table}

\noindent In Table \ref{tab:sim_lates}, we report the respective LATE estimates, averaged over all repetitions (in which the correct number of clubs was found). In the setting without invalidity, the estimated LATEs, using AHC, are extremely close to the oracle LATEs, and the SE decrease with increasing sample size. In the setting with invalidity, we first evaluate the performance of LATE estimates after only the first (club assignment) step without trying to detect invalid judges. As expected, given that invalid judges are still present in the estimation, mean coefficient estimates are far from the oracle estimates, irrespective of sample size. Applying the second group selection step of the method, however, dramatically improves our estimation, and now the estimates from our method and the oracle estimates are very close. With a small sample, when there are only 60 to 100 observations per judge, the variance can be large. However, after inspection of the single simulation results we find that this is mostly due to single outliers in the simulation. The performance improves in the medium sample setting, with 180 to 300 observations, and it is best in the largest sample setting.

\begin{table}[!htbp]
	\caption{Simulation: LATEs.}
	\label{tab:sim_lates}
\fontsize{9}{11}\selectfont
\begin{tabularx}{\textwidth}{XXXXcccccc}
\hline
Setting & Inv. & Method  & Step & 1-2 & SE & 1-3 & SE & 2-3 & SE \\
\midrule
\multirow{2}{*}{20} & \multirow{6}{*}{no} & Oracle & \multirow{6}{*}{first} & 30.43 & 4.83 & 26.28 & 2.86 & 20.48 & 5.53 \\ %
& & AHC & & 29.56 & 9.04 & 26.45 & 4.83 & 22.62 & 35.73 \\ %
\multirow{2}{*}{60} & & Oracle & & 30.28 & 2.75 & 26.19 & 1.62 & 20.23 & 2.75 \\
& & AHC & & 30.36 & 3.89 & 26.47 & 2.40 & 20.16 & 6.93  \\
\multirow{2}{*}{100} & & Oracle & & 30.26 & 2.12 & 26.21 & 1.26 & 20.15 & 2.23 \\
& & AHC & & 30.26 & 3.00 & 26.27 & 1.80 & 20.25 & 3.29 \\
\hline
% Oracle group, AHC 1 
\multirow{2}{*}{20} & \multirow{6}{*}{yes} & Oracle & \multirow{6}{*}{first} & 30.33 & 6.09 & 26.12 & 3.13 & 20.67 & 6.94 \\ %
& & AHC & & 28.27 & 12.34 & 36.65 & 9.29 & 39.76 & 355.23  \\ %
\multirow{2}{*}{60} & & Oracle & & 30.10 & 3.43 & 26.03 & 1.80 & 20.14 & 3.57 \\
& & AHC & & 27.87 & 5.26 & 35.94 & 3.27 & 52.59 & 10.98 \\
\multirow{2}{*}{100} & & Oracle & & 30.22 & 2.65 & 26.10 & 1.39 & 20.08 & 2.72 \\
& & AHC & & 27.45 & 4.10 & 36.46 & 2.43 & 51.15 & 6.97 \\
\hline
\multirow{2}{*}{20} & \multirow{6}{*}{yes} & Oracle & \multirow{6}{*}{second} & 30.33 & 6.09 & 26.12 & 3.13 & 20.67 & 6.94 \\ %
& & AHC & & 33.09 & 1057.28 & 21.21 & 44.26 & 39.87 & 207.54 \\ %
\multirow{2}{*}{60} & & Oracle & & 30.10 & 3.43 & 26.03 & 1.80 & 20.14 & 3.57 \\
& & AHC & & 25.62 & 5.02 & 25.08 & 2.73 & 31.09 & 8.04 \\
\multirow{2}{*}{100} & & Oracle & & 30.22 & 2.65 & 26.10 & 1.39 & 20.08 & 2.72 \\
& & AHC & & 28.79 & 3.86 & 25.98 & 2.00 & 22.95 & 4.35 \\
   \bottomrule
\end{tabularx}
\footnotesize \textit{Note:} Setting: number of cases per judge ($Unif(3,5) \times (20,60,100)$); Invalidity: no: $\gamma =  (0i_{10})$, yes: $\gamma =  (0.5,  0i_{3},  0i_{2},  0.4, 0.6,  0i_{2})$; Method: Oracle: estimator that uses group and club identity information, AHC: agglomerative hierarchical clustering; Stage: results after the first (detecting clubs with comparable propensity scores) or second (detecting clubs with comparable propensity scores and eliminating invalid judges) stage; 1-2, 1-3, 2-3: Group-wise LATE estimates of incarceration on recidivism, using judge IVs. SE: robust standard errors
\end{table}

\noindent Overall, these simulations indicate that our method delivers reliable results, retrieving clubs and groups that fulfill the LATE assumptions, and they suggest that the oracle properties shown earlier in the paper indeed hold. Additionally, the methods can be expected to perform well already with samples of medium size. We expect that performance not only improves with the number of observations but also when making the violations more pronounced, increasing the degree of separation among propensity scores, and decreasing error variance.

\subsection{Sensitivity analysis: Local violations to club assumptions}

We check how violations of Assumption \ref{ass:Clubs} impact our simulation results by introducing random perturbations to the propensity scores of the judges. Table \ref{tab:club_violations} shows again the classification results after the first step of our method without invalidity. We create club violations by adding Gaussian noise, sampled as $N(0, 0.01^2)$ in Panel A and $N(0, 0.04^2)$ in Panel B. In each iteration, random noise is drawn and added to the original propensity scores of the judges. Our procedure in Panel A assigns the clubs correctly with high frequency and achieves correct coverage despite the minor local deviations. We see only slight deviations from the results in Table \ref{tab:sim_first_step_class}. However, when the deviations become larger, as in Panel B, the accuracy of the allocations deteriorates while coverage stays at the nominal level. The deviations in Panel B are quite large relative to the propensity scores, which are $\pi = (0.8, 0.5, 0.3)’$. Overall, this exercise has illustrated that our procedure still can produce satisfactory assignment, even in presence of slight deviations.   

\begin{table}[!htbp]
	\caption{Simulation: Local violations to club assumptions \\ First step - club allocation}
	\label{tab:club_violations}
%\fontsize{9}{11}\selectfont
\begin{tabularx}{\textwidth}{XXXXXXXX}
\hline
Setting & Method & \#clubs & \#corr & Hansen p & Power & Cover & NMI \\
\midrule 
\multicolumn{8}{c}{Panel A: Local violations sampled as $N(0, 0.01^2)$} \\
\midrule
\multirow{2}{*}{20} & Oracle & 3 & 1 & 0.52 & 0.99 & 0.96 & 1 \\ %check 
                    & AHC    & 2.36 & 0.36 & 0.51 & 0.88 & 0.96 & 0.35 \\ %check 
\multirow{2}{*}{60} & Oracle & 3 & 1 & 0.50 & 1 & 0.95 & 1 \\ %
                    & AHC    & 2.96 & 0.95 & 0.49 & 0.99 & 0.95 & 0.45 \\ %check
\multirow{2}{*}{100}& Oracle & 3 & 1 & 0.49 & 1 & 0.94 & 1 \\ %check
                    & AHC    & 3.02 & 0.97 & 0.49 & 1 & 0.95 & 0.46 \\ %check
\midrule
\multicolumn{8}{c}{Panel B: Local violations sampled as $N(0, 0.04^2)$} \\
\midrule
\multirow{2}{*}{20} & Oracle & 3 & 1 & 0.49 & 0.98 & 0.95 & 1 \\ %check
                    & AHC & 2.44 & 0.43 & 0.5 & 0.88 & 0.94 & 0.36 \\ %check
\multirow{2}{*}{60} & Oracle & 3 & 1 & 0.5 & 1 & 0.95 & 1 \\ %check
                    & AHC & 3.11 & 0.82 & 0.49 & 0.99 & 0.95 & 0.47 \\ %check
\multirow{2}{*}{100}& Oracle & 3 & 1 & 0.45 & 1 & 0.94 & 1 \\ %   
                    & AHC & 3.37 & 0.65 & 0.47 & 1 & 0.94 & 0.50 \\ % 
\bottomrule

\end{tabularx}
\footnotesize \textit{Note:} Mean results after the first step of the procedure (detecting clubs with comparable propensity scores) after introducing local violations to the club assumptions. No invalidity. Panel A: Adding mean-zero Gaussian noise to the propensity scores, sampled as $N(0, 0.01^2)$; Panel B: Adding mean-zero Gaussian noise to the propensity scores, sampled as $N(0, 0.04^2)$; Setting: number of cases per judge ($Unif(3,5) \times (20,60,100)$); Method: Oracle: estimator that uses group and club identity information, AHC: agglomerative hierarchical clustering; $\#clubs$: mean number of clubs; $\#corr$ percentage of iterations in which the correct number of clubs has been selected; Hansen p: mean of p-values of Hansen-test from an over-identified specification; Power: fraction of confidence intervals that do not include zero; Cover: fraction of confidence intervals containing the true effect (in iterations in which the correct number of clubs was found); NMI: normalized mutual information. 
\end{table}

\iffalse
\subsection{Figures}

\begin{figure}[ht!]
	\begin{center}
		\includegraphics[scale=0.7]{tab3_hist_equal_3_large.pdf}
	\end{center}
\caption{Group-pair-LATEs: First and second step - club allocation and identification of invalid judges.}\label{fig:sim_inv_hist_equal_3_2_small}
\footnotesize \textit{Note:} Group-pair LATES after the first and second step of the procedure (detecting clubs with comparable propensity scores and eliminating invalid judges) for $\gamma =  (0.5,  0i_{3},  0i_{2},  0.4, 0.6,  0i_{2})$. Number of cases per judge: $Unif(3,5) \times 100$, $c= 3$. Oracle: fully informed estimator, AHC: agglomerative hierarchical clustering, CARDS: AHC combined with a clustering algorithm in regression via data-driven segmentation.
\end{figure}
\fi

\FloatBarrier
	\newpage
\section{Application}\label{app:application}
\subsection{Data}  \label{app:desc}

In Table \ref{app:tab:desc}, we provide some descriptive statistics for our data, namely the mean of outcome and covariates in the total sample, as well as among those offenses resulting in a prison or jail sentence ($D = 1$) and those that are not sanctioned with a prison or jail sentence ($D = 0$). The descriptive statistics suggest that the distribution of prison/jail sentences differs in crime type and the offender's race and gender. The proportion of cases that resulted in a prison or jail sentence is lower among white or Hispanic offenders than among offenders of other races and higher among men than among women. The share of property crimes sanctioned with incarceration is smaller than that of crimes against persons, drug crimes, weapon offenses, and sex offenses. The table also shows that the severity of crimes and the likelihood of incarceration are positively correlated, where the variable ``Severity'' is an indicator of the seriousness of a crime as defined by the Minnesota Sentencing Guidelines Commission, ranging from low (Severity = 1) to high (Severity = 11)\footnote{The Minnesota Sentencing Guidelines Commission defines a different severity grid for sex offenders. The severity levels of the sex offenders in our sample are translated into the standard severity level according to the sentence lengths the Minnesota Sentencing Guidelines Commission suggests for each severity level.}.

\begin{table}[ht]
	\caption{Descriptive Statistics}
	\label{app:tab:desc}
		\centering
	% latex table generated in R 4.0.3 by xtable 1.8-4 package
% Mon Nov 01 18:14:53 2021
\begin{tabular}{llllll}
  \toprule
Variable & Overall & $D = 1$ & $D = 0$ & Diff & pval \\ 
  \midrule
Recidivism & 0.292 & 0.3 & 0.219 & 0.08 & 4.4e-37 \\ 
  Female & 0.183 & 0.173 & 0.276 & -0.1 & 4.3e-53 \\ 
  Age at Sentence & 33.04 & 33.02 & 33.2 & -0.18 & 0.28 \\ 
  Race &&&&&\\
  \ \ \ \ \ \ \ \ \ \ White & 0.59 & 0.585 & 0.638 & -0.05 & 6.9e-13 \\ 
  \ \ \ \ \ \ \ \ \ \ Black & 0.261 & 0.267 & 0.204 & 0.06 & 6.3e-24 \\ 
  \ \ \ \ \ \ \ \ \ \ Amerindian & 0.073 & 0.074 & 0.065 & 0.01 & 0.021 \\ 
  \ \ \ \ \ \ \ \ \ \ Hispanic & 0.05 & 0.047 & 0.074 & -0.03 & 5.6e-12 \\ 
  \ \ \ \ \ \ \ \ \ \ Asian & 0.026 & 0.027 & 0.018 & 0.01 & 5e-05 \\ 
  \ \ \ \ \ \ \ \ \ \ Unknown & 0 & 0 & 0 & 0 & 0.73 \\ 
  Crime Type &&&&&\\
  \ \ \ \ \ \ \ \ \ \ Property Crime & 0.332 & 0.324 & 0.412 & -0.09 & 9.7e-32 \\ 
  \ \ \ \ \ \ \ \ \ \ Crime against a Person & 0.299 & 0.306 & 0.235 & 0.07 & 8e-28 \\ 
  \ \ \ \ \ \ \ \ \ \ Drug Crime & 0.238 & 0.242 & 0.2 & 0.04 & 7.5e-12 \\ 
  \ \ \ \ \ \ \ \ \ \ Sex Offenses & 0.056 & 0.056 & 0.054 & 0 & 0.49 \\ 
  \ \ \ \ \ \ \ \ \ \ Weapons Offense & 0.007 & 0.007 & 0.009 & 0 & 0.19 \\ 
  \ \ \ \ \ \ \ \ \ \ Other & 0.068 & 0.065 & 0.091 & -0.03 & 1.3e-09 \\ 
  Severity & 3.44 & 3.46 & 3.25 & 0.21 & 6.1e-09 \\ 
   \bottomrule
\end{tabular}

	   \begin{minipage}{\textwidth}
\textit{\footnotesize Note: `Overall', `D = 1' and `D = 0', report the mean of the respective variable in the total
			sample, among the treated and the non-treated in the baseline sample of 2009-14 criminal cases. `Diff' and `p-val' provide the mean difference (across treatment states) and the p-value of a two-sample t-test.}
		\end{minipage}
\end{table}

\newgeometry{margin=0.5in,top=0.5in}
\vspace{-3cm}
\begin{table}[ht]
	\caption{Crimes considered in baseline sample}
	\label{app:crimelist}
	{\notsotiny
	\centering
	\begin{tabularx}{\textwidth}{brs}
		\toprule
		Crime Type & Crime Category & N  \\ 
		\midrule
		Theft  & Property  &  3416 \\
		Theft of Firearm   &  Property &    89 \\
		Theft Over 35K  & Property  & 144 \\
		Possession of Shoplifting Gear   & Property  & 85 \\
		Theft From Pers  & Property  & 301 \\
		Theft of Motor Vehicle   & Property  & 159 \\
		Motor Vehicle Use Without Consent  & Property  & 1341 \\
		Receiving Stolen Property   & Property  & 1330 \\
		Arson, 1st degree  & Property  &  10 \\
		Arson, 2nd degree  & Property  &  64 \\
		Arson, 3rd degree  & Property  &  25 \\
		Burglary, 1st degree (Occupied Dwelling)  & Property  & 483 \\
		Burglary, 2nd degree (Dwelling/Bank/Government Building/  & Property  & 1330 \\
		Religious Est./ Historic Property/School Building)& & \\
		Burglary, 2nd degree (Pharmacy/Tool)  & Property  & 212 \\
		Burglary 3rd degree (Non Residential)   & Property  & 1954 \\
		Possession of Burglary Tools  & Property  & 552 \\
		Criminal Damage  & Property  & 617 \\
		Check Forgery (Over \$35,000)  & Property  &   8 \\
		Check Forgery (Over \$2,500)  & Property  & 276 \\
		Check Forgery (\$251-\$2,500)  & Property  & 1144 \\
		Check Forgery (\$250 or Less)  & Property  &  81 \\
		Other Forgery  & Property  & 306 \\
		Issuance of Dishonored  Checks  & Property  & 422 \\
		Financial Transaction Card Fraud   & Property  & 989 \\
		Welfare/Unemployment Benefits/Food Stamp Fraud  & Property  & 303 \\
		Identity Theft  & Property  & 177 \\
		Possession or Sale of Counterfeit Check   & Property  & 133 \\
		Mail Theft  & Property  &  48 \\
		Other Property Crimes  & Property  &  238 \\

		Criminal Vehicular Homicide or Injury, severity=5  & Person  &  92 \\
		Criminal Vehicular Homicide or Injury, severity=3  & Person  &  270 \\
		Assault, 2nd degree &  Person &   970 \\
		Assault, 3rd degree &  Person &   1637 \\
		Assault, 4th degree &  Person &  705 \\
		Assault, 5th degree &  Person &    253 \\
		Domestic Assault & Person  &  2280 \\
		Domestic Assault by Strangulation  &  Person &  1147 \\
		Simple Robbery  & Person  &   420 \\
		Aggravated Robbery, 1st degree  & Person  &    88 \\
		Aggravated Robbery, 2nd degree  & Person  &    102 \\
		Kidnapping (Safe Release/No Great Bodily Harm)   &  Person &  17 \\
		Kidnapping (Unsafe Release/Great Bodily Harm/Victim Under 16)  &  Person &   1 \\
		False Imprisonment  &  Person &   65 \\
		Depriving Another of Cust. or Parental Rights   & Person  & 37 \\
		Coercion  & Person  &    13 \\
		Accidents   &  Person &  9 \\
		Malicious Punishment of Child  &  Person &  101 \\
		Threats of Violence (Terror/Evacuation)  &  Person & 2900 \\
		Threats of Violence (Replica Firearm/Bomb Threat)   & Person  &  87 \\
		Stalking, severity=4  & Person  &   121 \\
		Stalking, severity=5  & Person  &  160 \\
		Violation of Harassment Restraining Order &  Person & 2941 \\
		Tampering with a Witness   &  Person & 24 \\
		Burglary 1st Degree (w/Weapon or Assault)  &  Person &   8 \\
		Prostitution  &  Person &  31 \\
		Other Person & Person  &   133 \\
		
		Controlled Substance Crime, 2nd degree   & Drug & 18 \\
		Controlled Substance Crime, 3rd degree & Drug & 1454 \\
		Controlled Substance Crime, 4th degree  & Drug & 584 \\
		Controlled Substance Crime, 5th degree & Drug &  9322 \\
		Possession of Substances with Intent to Manufacture Methamphetamine  & Drug &  46 \\
		Other Drug  Offense & Drug & 206 \\
				
		Drive-by Shooting (Unoccupied Motor Vehicle/Building) & Weapon &   9 \\
		Discharge of Firearm & Weapon & 139 \\
		Other Weapon Related Crimes & Weapon & 204 \\
		
		Criminal Sexual Conduct, 2nd degree & Sex &  240  \\
		Criminal Sexual Conduct, 3rd degree  & Sex & 373 \\
		Criminal Sexual Conduct, 4th degree  & Sex & 283 \\
		Criminal Sexual Conduct, 5th degree  & Sex &   4 \\
		Solicitation of Minors to Engage in Sexual Conduct  & Sex & 94 \\
		Failure to Register as Predator  & Sex & 1380 \\
		Possession of Child Pornography  & Sex &  334 \\
		Other Sex Crimes & Sex &   12 \\ 
	
		Bribery & Other &  13 \\
		Perjury & Other &  35 \\
		Escape, severity=3 & Other & 240 \\
		Fleeing a Police Officer & Other & 1660 \\
		Aiding Offender to Avoid Arrest & Other & 124 \\
		Accomplice After the Fact  & Other & 54 \\
		Obstruction of Legal Procedure & Other &  26 \\
		Lottery Fraud  & Other & 38 \\
		Felony Driving while Intoxicated  & Other & 530 \\
		Failure to Appear in Court  & Other &  65 \\
		Tax Offenses  & Other & 72 \\
		Other & Other & 441 \\
		
		\bottomrule
	\end{tabularx}}
	\textit{\footnotesize Note: Overview of the offenses considered in the application, the broader crime category to which each offense belongs, and the number of observations per offense in the baseline sample of 2009-14 criminal cases.}
\end{table}
\restoregeometry

\begin{table}[ht]
	\centering
	\caption{Club-level: Mean values for each club, including SEs\label{tab:club_balance}}

\begin{tabular}{l
		S[table-format=2.3] S[table-format=1.3]
		S[table-format=2.3] S[table-format=1.3]
		S[table-format=2.3] S[table-format=1.3]}
	\toprule
	& \multicolumn{2}{c}{Club 1} & \multicolumn{2}{c}{Club 2} & \multicolumn{2}{c}{Club 3} \\
	\cmidrule(lr){2-3}\cmidrule(lr){4-5}\cmidrule(lr){6-7}
	{Variable} & {Mean} & {SE} & {Mean} & {SE} & {Mean} & {SE} \\
	\midrule
	female     & 0.168 & 0.007 & 0.164 & 0.005 & 0.177 & 0.011 \\
	agesent    & 34.204 & 0.193 & 33.543 & 0.159 & 33.163 & 0.323 \\
	White      & 0.398 & 0.009 & 0.403 & 0.007 & 0.546 & 0.015 \\
	Black      & 0.506 & 0.009 & 0.450 & 0.008 & 0.348 & 0.014 \\
	AmInd      & 0.044 & 0.004 & 0.031 & 0.003 & 0.040 & 0.006 \\
	Hispanic   & 0.027 & 0.003 & 0.045 & 0.003 & 0.032 & 0.005 \\
	Asian      & 0.025 & 0.003 & 0.072 & 0.004 & 0.033 & 0.005 \\
	Unk        & 0.001 & 0.000 & 0.000 & 0.000 & 0.001 & 0.001 \\
	person     & 0.265 & 0.008 & 0.411 & 0.007 & 0.291 & 0.013 \\
	property   & 0.380 & 0.009 & 0.317 & 0.007 & 0.307 & 0.013 \\
	drugs      & 0.239 & 0.008 & 0.164 & 0.005 & 0.263 & 0.013 \\
	noncscSex  & 0.050 & 0.004 & 0.043 & 0.003 & 0.058 & 0.007 \\
	Weapons    & 0.010 & 0.002 & 0.007 & 0.001 & 0.004 & 0.002 \\
	Other      & 0.055 & 0.004 & 0.057 & 0.003 & 0.077 & 0.008 \\
	severity   & 3.333 & 0.037 & 3.463 & 0.027 & 3.449 & 0.068 \\
	\bottomrule
\end{tabular}
\begin{tablenotes}
	\small
	\item Notes: Club-level means with heteroskedasticity robust standard errors (clustered by OffenderID), estimated from a regression of each variable on club dummies (no intercept).
\end{tablenotes}
\end{table}

\begin{table}
	\centering
	\caption{Judge validity frequency over 10000 resamples \label{tab:app_resamples}}
	
\begin{tabular}{lc}
\hline
Judge ID & Frequency \\
\hline
00037K73 & 0.9943 \\
00059M82 & 0.6534 \\
00076M70 & 0.9984 \\
00481B73 & 0.9423 \\
01000F9  & 0.6410 \\
02006M27 & 0.9950 \\
02011G27 & 0.9720 \\
02012E27 & 0.9938 \\
02018D27 & 0.9828 \\
02021D27 & 0.9840 \\
02160A27 & 0.9959 \\
02215B27 & 0.9932 \\
02219E27 & 0.9979 \\
02987D27 & 0.9996 \\
03114C62 & 0.7858 \\
03120H62 & 0.9471 \\
03197F62 & 0.9353 \\
03210G62 & 0.8861 \\
03372C62 & 0.9469 \\
03669B62 & 0.9309 \\
03710J62 & 0.9552 \\
03762A62 & 0.7991 \\
03868G62 & 0.9313 \\
\hline
\end{tabular}
\vspace{0.5em}
\begin{minipage}{\linewidth}
	\small
	\textit{Notes: Stability of judge validity classification across resampling of the second-step procedure. Judges are held fixed in their original first-step club assignment (based on propensity score clustering); in each of 10{,}000 iterations, the sample is randomly split in half and the second-step (relevant/valid judge) classification is re-estimated within each fixed club on the subsample. Reported frequencies indicate the share of iterations in which each judge is classified as valid.}
\end{minipage}

\end{table}

\begin{table}
	\caption{Placebo test: judge and club assignment do not predict pre-index criminal history\label{tab:placebo}}
	\centering

\begin{tabular}{lcc}
	\toprule
	& Judge-level & Club-level \\
	& (1)         & (2)        \\
	\midrule
	Joint $F$-statistic      & 1.57        & 0.08       \\
	$p$-value                & 0.047       & 0.926      \\
	\midrule
	Number of judges / clubs & 22          & 3          \\
	\bottomrule
\end{tabular}
\begin{tablenotes}\small
	\item \textit{Notes: Joint $F$-test of the judge (column 1) or club (column 2)
	fixed effects in a regression of the number of an offender's prior cases on those fixed effects. All specifications include year fixed effects. Standard errors are
	clustered by offender-ID; $p$-values are shown below each $F$-statistic. Invalid judges
	and singleton clubs are excluded, and the sample is restricted to judges with at least
	300 cases ($c=300$). Sample with 9763 observations.}
\end{tablenotes}
\end{table}

\FloatBarrier
\subsection{Figures}  \label{app:fig}

\begin{figure}[h]
\includegraphics[scale=0.75]{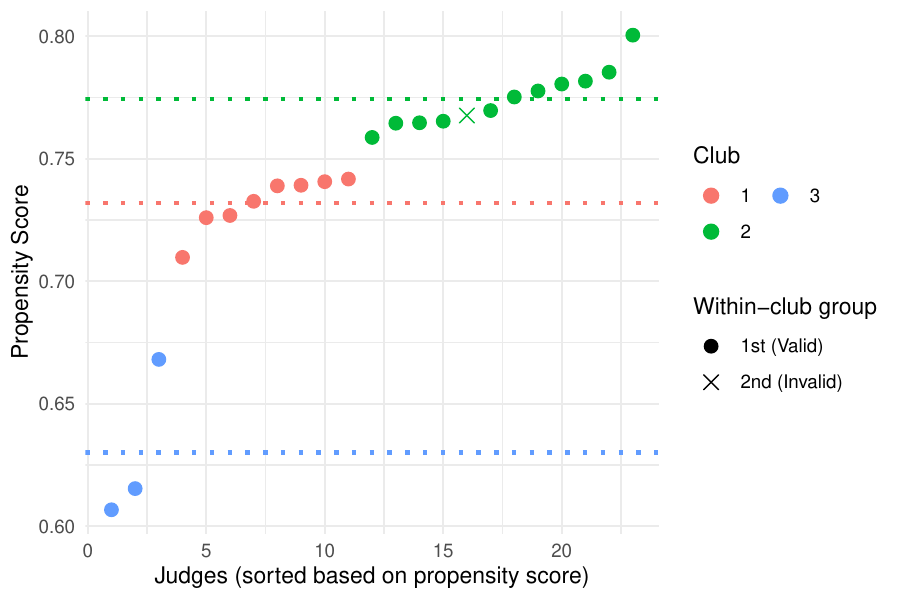}
\caption{First-stage clustering}\label{fig:App_FS_300}
%\\
\textit{Note: Minimum number of cases per judge: 300. On vertical axis: propensity scores with 95 percent confidence intervals. Dotted vertical lines denote cluster mean. Different colors denote different clubs.}
\end{figure}

\FloatBarrier
\subsection{Additional results, c = 200}  \label{app:add}

\begin{table}[ht]
	\begin{center}
	\caption{First-stage clubs, 200}\label{tab:fs-200}
	% latex table generated in R 4.2.2 by xtable 1.8-4 package
% Tue Jun  2 10:39:55 2026
\begin{tabular}{p{1.5cm}p{1.5cm}p{1.5cm}}
  \toprule
Club & Mean & Nr \\ 
  \midrule
1 & 0.84 & 41 \\ 
  2 & 0.60 & 5 \\ 
  3 & 0.79 & 19 \\ 
  4 & 0.73 & 12 \\ 
  5 & 0.17 & 1 \\ 
   \bottomrule
\end{tabular}

	\end{center}
	\textit{Note: Clustering results when minimum case count per judge is 200.}
\end{table}

\begin{table}[ht]\centering
	\caption{Effect of Imprisonment on Recidivism, 200}
	\begin{tabularx}{1.05\textwidth}{cXXXXXXXX}
          &\multicolumn{1}{c}{OLS}&\multicolumn{1}{c}{2SLS}&\multicolumn{1}{c}{1-2}&\multicolumn{1}{c}{1-3}&\multicolumn{1}{c}{1-4}&\multicolumn{1}{c}{2-3}&\multicolumn{1}{c}{2-4}&\multicolumn{1}{c}{3-4}\\

\midrule \multicolumn{9}{p{\textwidth}}{\centering Panel A: Post Club Clustering}\\  \midrule D&    0.061&    0.051&    0.038&     0.51&     0.21&    -0.15&    -0.19&  -0.0038\\
          & (0.0096)&  (0.025)&  (0.078)&   (0.20)&   (0.12)&   (0.11)&   (0.19)&   (0.26)\\

J         &         &         &       46&       60&       53&       24&       17&       31\\
N         &    23958&    23958&     7404&     9221&     8103&     3565&     2447&     4264\\
P         &         &     0.00&     0.00&     0.01&     0.01&     0.05&     0.47&     0.22\\
F         &         &    33.44&   199.51&    81.30&   142.54&   101.11&    36.87&    29.97\\
\end{tabularx}
\\
	\begin{tabularx}{1.05\textwidth}{cXXXXXXXX}

\midrule \multicolumn{9}{p{\textwidth}}{\centering Panel B: Post Valid IV Selection} \\ \midrule D&         &         &     0.12&     0.72&     0.33&    -0.15&    -0.19&  -0.0038\\
          &         &         &  (0.080)&   (0.20)&   (0.12)&   (0.11)&   (0.19)&   (0.26)\\

J         &         &         &       30&       44&       37&       24&       17&       31\\
N         &         &         &     5085&     6902&     5784&     3565&     2447&     4264\\
P         &         &         &     0.17&     0.17&     0.15&     0.05&     0.47&     0.22\\
F         &         &         &    206.6&     95.6&    156.2&    101.1&     36.9&     30.0\\
\hline \multicolumn{9}{p{\textwidth}}{Note: Sample restricted so that minimum number of cases per judge is 200. Standard errors are clustered by Offender-ID. Significance level in testing procedure: 0.01. D: coefficient for LATE estimate when comparing judge clubs or groups as detailed in top line. Panel A: Results after club-clustering, without valid IV selection, Panel B: Results after valid IV selection. J: Number of judges, N: number of observations, P: p-value of Hansen-test from an over-identified specification.} \end{tabularx}

\end{table}

%{ % red
%\arrayrulecolor{red} % red
\begin{table}[htbp]
%	\color{red} % red
\caption{Second-step groups, 200\label{tab:c200groups}}
	\label{tab:second-stage}
	\begin{center}
	\begin{tabular}{cccc}  		\toprule
	Club & Total number of judges & Valid judges & Invalid judges \\ 
	  \midrule
	  1 & 41  & 25  & 16   \\ 
	  2 & 5   & 5  & 0 \\ 
	  3 & 19   & 19  & 0  \\ 
	  4 & 12   & 12  & 0  \\ 
	  5 & 1    & 0   & 0 \\ 
	   \bottomrule
\end{tabular} 	\end{center}
    \textit{Note: Clustering results after the second step, with a minimum case count per judge of 200. Singleton clubs are discarded from further analysis because we cannot verify whether they are valid. }
\end{table}
%\arrayrulecolor{black} % red
%} % red

\begin{figure}[ht]
	\includegraphics[scale=0.75]{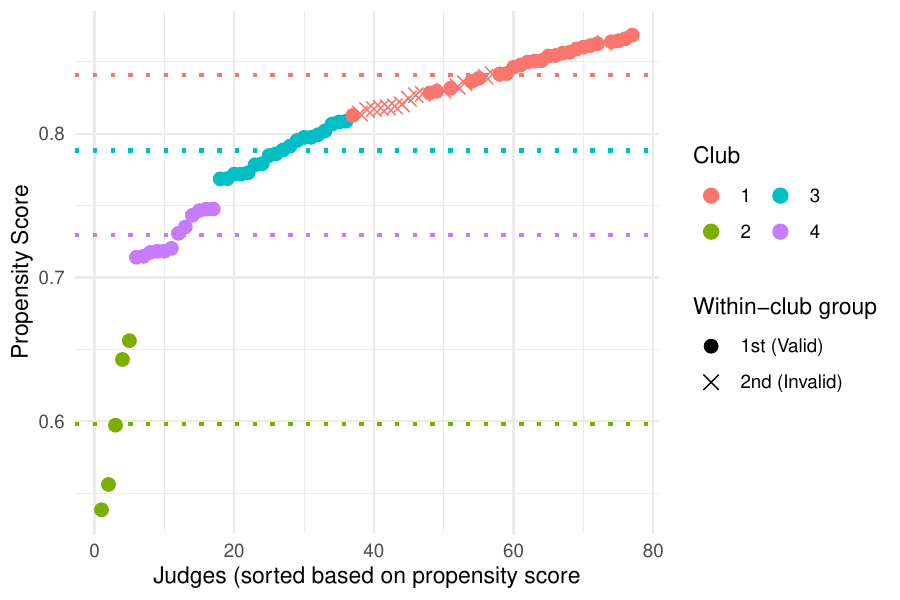}
	\caption{First-stage clustering, min. 200 cases per judge}\label{fig:App_FS_200}
	\textit{Note: Minimum number of cases per judge: 200. On vertical axis: propensity scores with 95 percent confidence intervals. Dotted vertical lines denote cluster mean. Different colors denote different clubs.}
\end{figure}

\begin{figure}[ht]
	\includegraphics[scale=0.75]{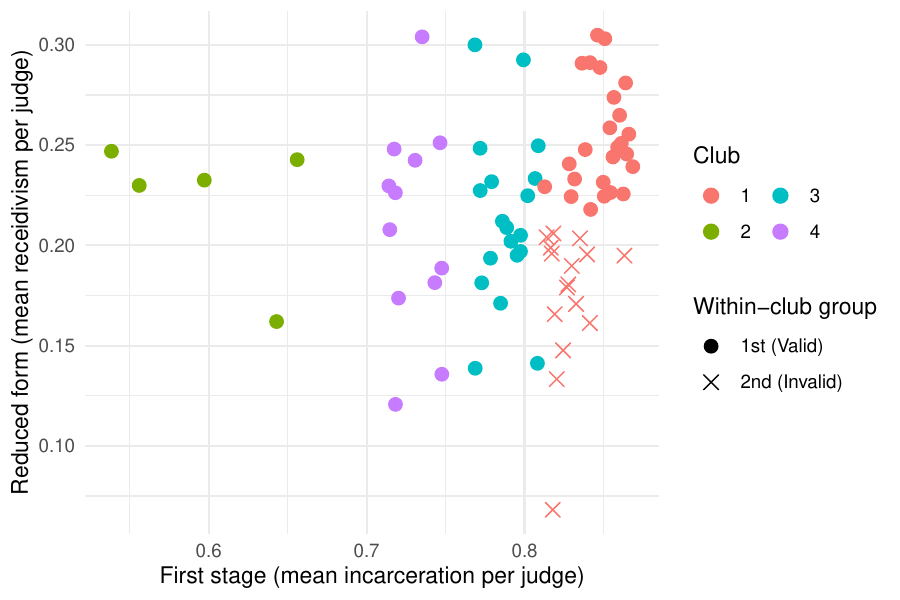}
	\caption{Reduced-form clustering, min. 200 cases per judge}\label{fig:App_RF_200}
	\textit{Note: Minimum number of cases per judge: 200. On vertical axis: mean outcomes per judge. Crosses denote judges selected as invalid, dots denote the selected validity set. Different colors denote different clubs. }
\end{figure}

\clearpage

%\color{red} % red
\subsection{Subset analysis}\label{sec:Subset}

We tackle the problem of accounting for observable controls in an additional way by reducing the data to a subset with similar observables and clustering clubs within this reduced dataset. We limit the data to male, white defendants aged 100 or younger with crimes against persons and a severity index equal to 4. Moreover, we only consider judges with a minimum case count of 20 to ensure the calculation of propensity scores remains reliable.

%{ % red
%\arrayrulecolor{red} % red
\begin{table}[htbp]
%	\color{red} % red
	\caption{First step clustering - Subset analysis}
	\label{tab:subset1}
	\begin{center}
		% latex table generated in R 4.2.2 by xtable 1.8-4 package
% Tue Jun  2 11:35:14 2026
\begin{tabular}{p{1.5cm}p{1.5cm}p{1.5cm}}
  \toprule
Club & Mean & Nr \\ 
  \midrule
1 & 0.14 & 71 \\ 
  2 & -0.03 & 31 \\ 
   \bottomrule
\end{tabular}

	\end{center}	
		\textit{Note: Clustering results when minimum case count per judge is 20.}
\end{table}
%\arrayrulecolor{black} % red
%} % red

\noindent In this case, the OLS estimate is statistically insignificant with a coefficient of 0.05. The 2SLS estimate (0.13) is more than double the size of the OLS estimate. The SE is 0.07 and hence one could be tempted to claim significance at a higher significance level. 

\noindent We find two clubs when using algorithms \ref{algo:ward} and \ref{algo:F-test}. The judge clubs have sizes 71 and 31 (see Table \ref{tab:subset1}). The LATE estimate for the club pair is presented in Table \ref{tab:subset2}. The LATE between Club 1 and 2 is -0.03 and the SE is 0.17, and therefore the effect is statistically insignificant at any conventional significance level. 

\noindent We don't proceed to interpret the LATE based on the invalid IV selection (see Panel B of Table \ref{tab:subset2}), because the Hansen-Sargan test does not reject after the first step. We can see that the first-stage F-statistic for the 2SLS estimate is very low (4.95), indicating weak IVs. However, using our procedure, the F-statistic becomes higher, highlighting one of the advantages of our method, even when invalidity does not seem to be an issue.

%{ % red
%\arrayrulecolor{red} % red
\begin{table}[htbp]
%	\color{red} % red
	\caption{Effect of Imprisonment on Recidivism, Subset analysis}
	\label{tab:subset2}
	\begin{center}
	\newcolumntype{C}[1]{>{\centering\arraybackslash}p{#1}}
\begin{tabular}{p{1cm}C{2cm}C{2cm}C{2cm}}
          & OLS & 2SLS & 1--2 \\
\midrule \multicolumn{4}{p{0.6\textwidth}}{\centering Panel A: Post Club Clustering}\\
\midrule 
D 		  &    0.054&     0.13&   -0.030\\
          &  (0.028)&  (0.071)&   (0.17)\\
J         &         &         &      102\\
N         &     3280&     3280&     1627\\
P         &         &     0.30&     0.26\\
F         &         &     4.95&    63.28\\
\end{tabular}

	\newcolumntype{C}[1]{>{\centering\arraybackslash}p{#1}}
\begin{tabular}{p{1cm}C{2cm}C{2cm}C{2cm}}
\midrule \multicolumn{4}{p{0.6\textwidth}}{\centering Panel B: Post Valid IV Selection} \\ \midrule
D 		  &         &         &   0.0054\\
          &         &         &   (0.18)\\
J         &         &         &       63\\
N         &         &         &     1032\\
P         &         &         &     0.99\\
F         &         &         &    62.69\\
\hline  \end{tabular}

	\end{center}
	\textit{Note: Sample restricted to guarantee similar covariates. Cluster-robust standard errors in parentheses. Significance level in testing procedure: 0.1/log(N). D: coefficient for LATE estimate when comparing judge clubs or groups as detailed in top line. Panel A: Results after club-clustering, without valid IV selection, Panel B: Results after valid IV selection. J: Number of judges, N: number of observations, P: p-value of Hansen-test from an over-identified specification.}
\end{table}
%\arrayrulecolor{black} % red
%} % red

\noindent Table \ref{tab:second-stage_sub} reports the size of the second and third largest groups found within Club 1. The group of valid judges in Club 1 consists of 32 judges, while there are two smaller groups, each consisting of 20 and 19 judges, respectively. Due to the clear difference in group sizes, we refrain from repeating the analyses for the smaller groups. 

%{ % red
%\arrayrulecolor{red} % red
\begin{table}[htbp]
%	\color{red} % red
\caption{Second-step groups, Subset analysis}
	\label{tab:second-stage_sub}
	\begin{center}
	\begin{tabular}{ccccc}  		\toprule
	Club 
& \shortstack{Total number\\of judges}
& \shortstack{Valid judges\\(Group 1)}
& \shortstack{Invalid judges\\(Group 2)}
& \shortstack{Invalid judges\\(Group 3)} \\ 
	  \midrule
	  1 & 71  & 32  & 20 & 19    \\ 
	  2 & 31   & 31  & 0 & 0 \\ 
	   \bottomrule
\end{tabular} 	\end{center}
    \textit{Note: Clustering results after the second step in the subset analysis. Singleton clubs are discarded from further analysis because we cannot verify whether they are valid. }
\end{table}
\arrayrulecolor{black} % red
%} % red

\begin{figure}[ht]
	\includegraphics[scale=0.75]{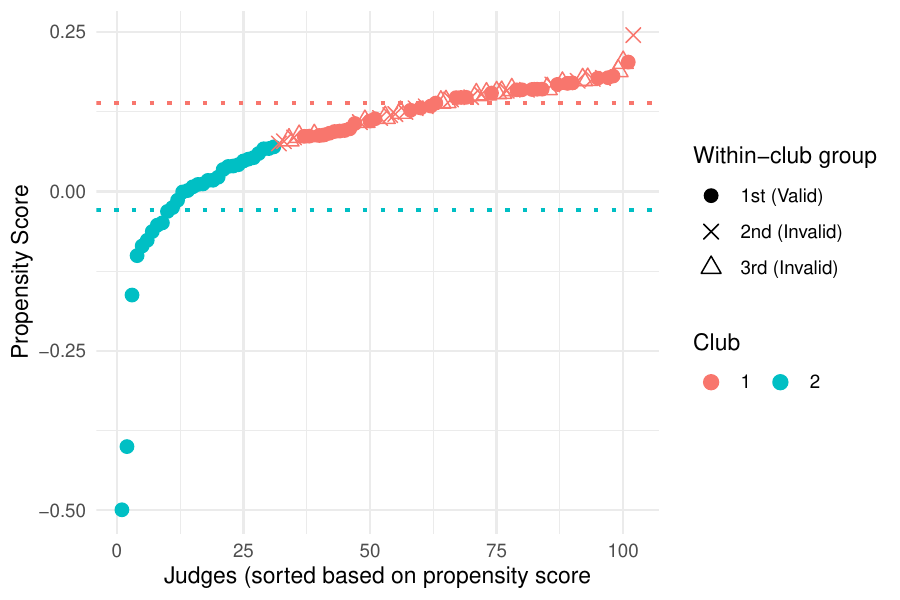}
	\caption{First-stage clustering, subset analysis}\label{fig:App_FS_subset}
	\textit{Note: Subset analysis. On vertical axis: propensity scores with 95 percent confidence intervals. Dotted vertical lines denote cluster mean. Different colors denote different clubs.}
\end{figure}

\begin{figure}[ht]
	\includegraphics[scale=0.75]{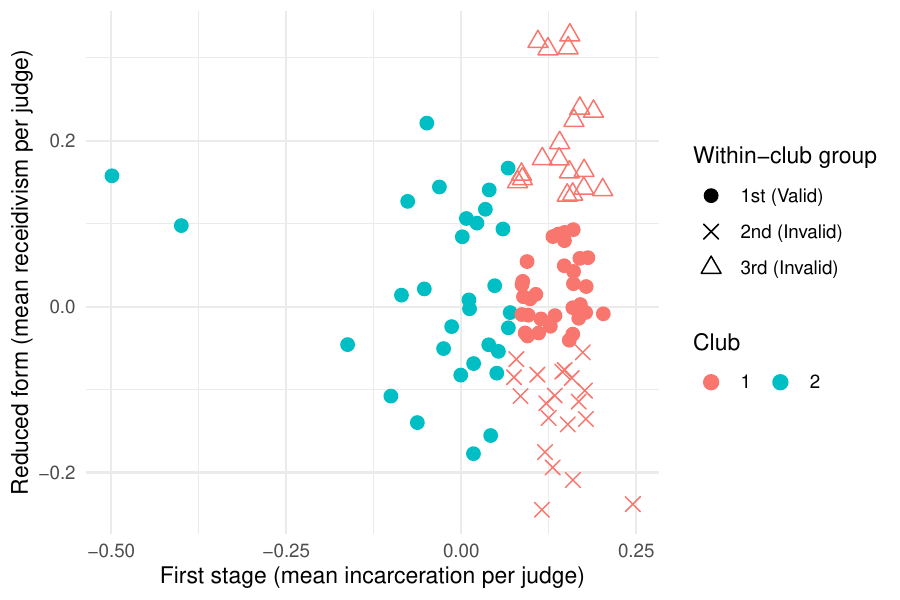}
	\caption{Reduced-form clustering, subset analysis}\label{fig:App_RF_subset}
	\textit{Note: Subset analysis. On vertical axis: mean outcomes per judge. Crosses denote judges selected as invalid, dots denote the selected validity set. Different colors denote different clubs. }
\end{figure}
\FloatBarrier
	\color{black} % red

\end{appendix}		
	
\bibliography{Judges}
\pagebreak

\clearpage

\end{document}